\documentclass[longauth]{aa}  

\usepackage{multirow}
\usepackage{graphicx}
\usepackage{amsmath}    
\newcommand{\mstar}{\rm log(M_{star}/M_{\odot})}
\newcommand{\chireduced}{\rm \chi^2_r}
\usepackage[colorlinks=true,allcolors=blue]{hyperref}
\usepackage{orcidlink}
\usepackage{cancel}
\usepackage{txfonts}
\begin{document}

\title{Value-added catalog of physical properties for more than 1.3 million galaxies from the DESI survey}
\titlerunning{Physical properties of DESI galaxies}

   \author{M. Siudek\inst{1,2}\orcidlink{0000-0002-2949-2155}
          \and
          R. Pucha\inst{3,4}\orcidlink{0000-0002-4940-3009}
          \and
          M.~Mezcua\inst{5,1}\orcidlink{0000-0003-4440-259X}
          \and
          S.~Juneau\inst{6}
          \and
          J.~Aguilar\inst{7}
          \and
          S.~Ahlen\inst{8}\orcidlink{0000-0001-6098-7247}
          \and 
            D.~Brooks\inst{9}
          \and
          C.~Circosta\inst{10}
          T.~Claybaugh\inst{7}
          \and
          S.~Cole\inst{11}\orcidlink{0000-0002-5954-7903}
          \and
          K.~Dawson\inst{3}
          \and
          A.~de la Macorra\inst{12}\orcidlink{0000-0002-1769-1640}
          \and
          Arjun~Dey\inst{6}\orcidlink{0000-0002-4928-4003}
          \and 
          Biprateep~Dey\inst{13}\orcidlink{0000-0002-5665-7912}
          \and
          P.~Doel\inst{9}
          \and
          A.~Font-Ribera\inst{14,15}\orcidlink{0000-0002-3033-7312}
          \and
          J.~E.~Forero-Romero\inst{16,17}\orcidlink{0000-0002-2890-3725}
          \and 
          E.~Gaztañaga\inst{5,18,1}
          \and
          S.~Gontcho A Gontcho\inst{7}\orcidlink{0000-0003-3142-233X}
          \and
          G.~Gutierrez\inst{19}
          \and
          K.~Honscheid\inst{20,21,22}
          \and
          C.~Howlett\inst{23}\orcidlink{0000-0002-1081-9410}
          \and
          M.~Ishak\inst{24}\orcidlink{0000-0002-6024-466X}
          \and
          R.~Kehoe\inst{25}
          \and
          D.~Kirkby\inst{26}\orcidlink{0000-0002-8828-5463}
          \and
          T.~Kisner\inst{7}\orcidlink{0000-0003-3510-7134}
          \and
          A.~Kremin\inst{7}\orcidlink{0000-0001-6356-7424}
          \and
          A.~Lambert\inst{7}
          \and
          M.~Landriau\inst{7}\orcidlink{0000-0003-1838-8528}
          \and
          L.~Le~Guillou\inst{27}\orcidlink{0000-0001-7178-8868}
          \and
          M.~Manera\inst{28,15}\orcidlink{0000-0003-4962-8934}
          \and
          P.~Martini\inst{20,29,22}\orcidlink{0000-0002-4279-4182}
          \and
          A.~Meisner\inst{6}\orcidlink{0000-0002-1125-7384}
          \and
          R.~Miquel\inst{30,15}
          \and
          J.~Moustakas\inst{31}\orcidlink{0000-0002-2733-4559}
          \and
          J.~ A.~Newman\inst{13}\orcidlink{0000-0001-8684-2222}
          \and
          G.~Niz\inst{32,33}\orcidlink{0000-0002-1544-8946}
          \and
          Z.~Pan\inst{34}\orcidlink{0000-0003-0230-6436}
          \and
          W.~J.~Percival\inst{35,36,37}\orcidlink{0000-0002-0644-5727}
          \and
          C.~Poppett\inst{7,38,39}
          \and
          F.~Prada\inst{40}\orcidlink{0000-0001-7145-8674}
          \and
          G.~Rossi\inst{41}
          \and
          A.~Saintonge\inst{9}\orcidlink{0000-0003-4357-3450}
          \and 
          E.~Sanchez\inst{42}\orcidlink{0000-0002-9646-8198}
          \and 
          D.~Schlegel\inst{7}
          \and
          D.~Scholte\inst{9}
          \and
          M.~Schubnell\inst{43,44}
          \and
          H.~Seo\inst{45}\orcidlink{0000-0002-6588-3508}
          \and
          F.~Speranza\inst{9}
          \and
          D.~Sprayberry\inst{6}
          \and
          G.~Tarl\'{e}\inst{44}\orcidlink{0000-0003-1704-0781}
          \and
          B.~A.~Weaver\inst{6}
          \and
          H.~Zou\inst{46}\orcidlink{0000-0002-6684-3997}
          }

   \institute{Institute of Space Sciences, ICE-CSIC, Campus UAB, Carrer de Can Magrans s/n, 08913 Bellaterra, Barcelona, Spain\\
              \email{msiudek@iac.es}
         \and
             Instituto Astrofisica de Canarias, Av. Via Lactea s/n, E38205 La Laguna, Spain\\ 
         \and
                Department of Physics and Astronomy, The University of Utah, 115 South 1400 East, Salt Lake City, UT 84112, USA\\
         \and
                Steward Observatory, University of Arizona, 933 N, Cherry Ave, Tucson, AZ 85721, USA\\
         \and
         Institut d’Estudis Espacials de Catalunya (IEEC), Edifici RDIT, Campus UPC, 08860 Castelldefels (Barcelona), Spain\\
         \and
                 NSF NOIRLab, 950 N. Cherry Ave., Tucson, AZ 85719, USA\\
         \and 
                Lawrence Berkeley National Laboratory, 1 Cyclotron Road, Berkeley, CA 94720, USA\\
          \and
            Physics Dept., Boston University, 590 Commonwealth Avenue, Boston, MA 02215, USA\\
        \and
        Department of Physics \& Astronomy, University College London, Gower Street, London, WC1E 6BT, UK\\
        \and
        Department of Physics \& Astronomy, University College London, Gower Street, London, WC1E 6BT, UK\\
        \and
        Institute for Computational Cosmology, Department of Physics, Durham University, South Road, Durham DH1 3LE, UK\\
        \and
        Instituto de F\'{\i}sica, Universidad Nacional Aut\'{o}noma de M\'{e}xico,  Cd. de M\'{e}xico  C.P. 04510,  M\'{e}xico\\
        \and
        Department of Physics \& Astronomy and Pittsburgh Particle Physics, Astrophysics, and Cosmology Center (PITT PACC), University of Pittsburgh, 3941 O'Hara Street, Pittsburgh, PA 15260, USA\\
        \and
        Department of Physics \& Astronomy, University College London, Gower Street, London, WC1E 6BT, UK\\
        \and
        Institut de F\'{i}sica d’Altes Energies (IFAE), The Barcelona Institute of Science and Technology, Campus UAB, 08193 Bellaterra Barcelona, Spain\\
        \and 
        Departamento de F\'isica, Universidad de los Andes, Cra. 1 No. 18A-10, Edificio Ip, CP 111711, Bogot\'a, Colombia\\
        \and
        Observatorio Astron\'omico, Universidad de los Andes, Cra. 1 No. 18A-10, Edificio H, CP 111711 Bogot\'a, Colombia\\
        \and
        Institute of Cosmology and Gravitation, University of Portsmouth, Dennis Sciama Building, Portsmouth, PO1 3FX, UK\\
        \and
        Fermi National Accelerator Laboratory, PO Box 500, Batavia, IL 60510, USA\\
        \and
        Center for Cosmology and AstroParticle Physics, The Ohio State University, 191 West Woodruff Avenue, Columbus, OH 43210, USA\\
        \and
        Department of Physics, The Ohio State University, 191 West Woodruff Avenue, Columbus, OH 43210, USA\\
        \and
        The Ohio State University, Columbus, 43210 OH, USA\\
        \and
        School of Mathematics and Physics, University of Queensland, 4072, Australia\\
        \and
        Department of Physics, The University of Texas at Dallas, Richardson, TX 75080, USA\\
        \and
        Department of Physics, Southern Methodist University, 3215 Daniel Avenue, Dallas, TX 75275, USA\\
        \and
        Department of Physics and Astronomy, University of California, Irvine, 92697, USA\\
        \and
        Sorbonne Universit\'{e}, CNRS/IN2P3, Laboratoire de Physique Nucl\'{e}aire et de Hautes Energies (LPNHE), FR-75005 Paris, France\\
        \and
        Departament de F\'{i}sica, Serra H\'{u}nter, Universitat Aut\`{o}noma de Barcelona, 08193 Bellaterra (Barcelona), Spain\\
        \and
        Department of Astronomy, The Ohio State University, 4055 McPherson Laboratory, 140 W 18th Avenue, Columbus, OH 43210, USA\\
        \and
        Instituci\'{o} Catalana de Recerca i Estudis Avan\c{c}ats, Passeig de Llu\'{\i}s Companys, 23, 08010 Barcelona, Spain\\
        \and
        Department of Physics and Astronomy, Siena College, 515 Loudon Road, Loudonville, NY 12211, USA\\
        \and
        Departamento de F\'{i}sica, Universidad de Guanajuato - DCI, C.P. 37150, Leon, Guanajuato, M\'{e}xico\\
        \and
        Instituto Avanzado de Cosmolog\'{\i}a A.~C., San Marcos 11 - Atenas 202. Magdalena Contreras, 10720. Ciudad de M\'{e}xico, M\'{e}xico\\
        \and
        Kavli Institute for Astronomy and Astrophysics at Peking University, PKU, 5 Yiheyuan Road, Haidian District, Beijing 100871, P.R. China\\
        \and
        Department of Physics and Astronomy, University of Waterloo, 200 University Ave W, Waterloo, ON N2L 3G1, Canada\\
        \and
        Perimeter Institute for Theoretical Physics, 31 Caroline St. North, Waterloo, ON N2L 2Y5, Canada\\
        \and
        Waterloo Centre for Astrophysics, University of Waterloo, 200 University Ave W, Waterloo, ON N2L 3G1, Canada\\
        \and
        Space Sciences Laboratory, University of California, Berkeley, 7 Gauss Way, Berkeley, CA  94720, USA\\
        \and
        University of California, Berkeley, 110 Sproul Hall \#5800 Berkeley, CA 94720, USA\\
        \and
        Instituto de Astrof\'{i}sica de Andaluc\'{i}a (CSIC), Glorieta de la Astronom\'{i}a, s/n, E-18008 Granada, Spain\\
        \and
        Department of Physics and Astronomy, Sejong University, Seoul, 143-747, Korea\\
        \and
        CIEMAT, Avenida Complutense 40, E-28040 Madrid, Spain\\
        \and
        Department of Physics, University of Michigan, Ann Arbor, MI 48109, USA\\
        \and
        University of Michigan, Ann Arbor, MI 48109, USA\\
        \and
        Department of Physics \& Astronomy, Ohio University, Athens, OH 45701, USA\\
        \and
        National Astronomical Observatories, Chinese Academy of Sciences, A20 Datun Rd., Chaoyang District, Beijing, 100012, P.R. China
             }

   \date{Received September 15, 1996; accepted March 16, 1997}


  \abstract
   {}
   { We present an extensive catalog of the physical properties of more than a million galaxies investigated with the Dark Energy Spectroscopic Instrument (DESI), one of the largest spectroscopic surveys to date. Spanning  a full range of target types, including emission-line galaxies, luminous red galaxies, and quasars, our survey encompasses an unprecedented range of spectroscopic redshifts, all the way from 0 to 6.}
   { The physical properties, such as stellar masses and star formation rates, were derived via the {\tt CIGALE} spectral energy distribution (SED) fitting code accounting for the contribution coming from active galactic nuclei (AGNs). Based on the modeling of the optical-mid-infrared ({\it grz} supplemented with WISE photometry) SEDs, we studied the galaxy properties with respect to their location on the main sequence.}
   { We have revised the dependence of stellar mass estimates on model choices and on the availability of  WISE photometry. Indeed, the WISE data are required to minimize the misclassification of star-forming galaxies as AGNs. The lack of WISE bands in SED fits leads to elevated AGN fractions for 68\% of star-forming galaxies identified using emission line diagnostic diagrams, but this does not significantly affect their stellar mass or star formation estimates.  }
   {}


   \maketitle
%

\section{Introduction}\label{sec:introduction}

The exploration of galaxies has been a focal point of astronomical research for centuries, revealing an astonishing diversity of galaxy types and physical processes. 
Large galaxy photometric catalogs mapping an unprecedented number of galaxies and their properties, such as COSMOS~\citep{Scoville2007, Weaver2022}, CANDELS~\citep{Grogin2011, Koekemoer2011},  UltraVISTA~\citep{McCracken2012, Muzzin2013}, and ZFOURGE~\citep{Straatman2016} supplemented
with spectroscopic datasets, such as those of Sloan Digital Sky Survey~\citep[SDSS;][]{York2000}, Galaxy and Mass Assembly~\citep[GAMA;][]{Baldry2018}, Deep Extragalactic VIsible Legacy Survey~\citep[DEVILS;][]{Davies2021}, and VIMOS Public Extragalactic Redshift Survey~\citep[VIPERS;][]{Scodeggio2018} have allowed us to unlock fundamental galaxy scaling relations and have made significant contributions to our understanding of galaxies and physical processes regulating their formation and evolution.  

Template-based techniques relying on the spectral energy distribution (SED; e.g.,~\citealt{Conroy2013} and references therein) fitting of galaxies are generally used to derive physical properties of galaxies from large-scale sky surveys. 
These methods heavily depend on the physical models of galaxy populations (e.g.,~\citealt{Mitchell2013,Moustakas2013,Lower2020,Pacifici2023}) and a statistical method of finding the best fits (e.g.,~\citealt{Leja2018}). 
The optimization of the parameter space and priors in constructing a model library is vital for finely tuning parameters related to stellar populations, dust content, and other key factors. This precision enables a more accurate characterization of galaxies. Building an extensive variety of models is crucial to encompass the full diversity of galaxy properties, but it introduces challenges such as degeneracies, where different parameter combinations yield similar predictions (e.g.,~\citealt{Lower2020}). An excessively large model library also poses the risk of overfitting, whereby models may fit noise or peculiarities in the data, rather than capturing the genuine underlying physical properties of galaxies. The SED fitting demands a wide wavelength coverage for better tracing the contribution from all stellar types in a galaxy~\citep[e.g.,][]{Maraston2006, Pforr2019} and to break degeneracies between the host galaxy and the parameters related to the active galactic nucleus (AGN)  ~\citep[e.g.,][]{CalistroRivera2016,Thorne2022b}. Accounting for the AGN contribution is one of the most significant sources of uncertainty in estimated physical properties as AGNs are relatively common and their contribution to the mid-infrared (MIR) emission might be significant~\citep{Leja2018}.
 The inferred physical properties also hinge on the accuracy of photometry and redshift of galaxies~\citep[e.g.,][]{Acquaviva2015,Iyer2017,PaulinoAfonso2022}. 
 Thus, a careful consideration of the extensive range of possible model combinations within the parameter space is essential for the robust estimations of physical properties of galaxies~\citep[e.g.,][]{Pforr2012, Johnson2021, Han2023, Pacifici2023}.

 To address the aftermentioned challenges, a collection of panchromatic SED codes has been developed relying on the reduced $\rm \chi^2$ ($\chireduced$) techniques such as {\tt MAGPHYS}~\citep{deCunha2008}, {\tt BEAGLE}~\citep{Chevallard2016}, {\tt Prospector}~\citep{Johnson2021}, {\tt BAGPIPES}~\citep{Carnall2018}, {\tt CIGALE}~\citep{Boquien2019}, and {\tt ProSpect}~\citep{Robotham2020}, along with the most recent SED fitting codes based on the  Markov chain Monte Carlo (MCMC) approach, such as {\tt MCSED}~\citep{Bowman2020}, {\tt piXedfit}~\citep{Abdurrouf2021}, {\tt Lightning}~\citep{Doore2023}, {\tt PROVABGS}~\citep{Hahn2023b}, and {\tt GalaPy}~\citep{Ronconi2024}, among others. 
 Employing diverse forward-modeling frameworks or templates, along with a range of Bayesian methods such as MCMC sampling or on a model grid, these codes provide a comprehensive approach to accurately estimate the physical properties of galaxies. 
 We refer to \cite{Pacifici2023} and \cite{Best2023} for a review of the performance of different SED fitting tools. 

The evolving landscape of astronomy and the state-of-the-art instruments, including Dark Energy Spectroscopic Instrument (DESI; \citealt{DESICollaboration2016, DESICollaboration2022}),  Prime Focus Spectrograph (PFS; \citealt{Takada2014}),  {\it Vera C. Rubin} Observatory (\citealt{Ivezic2019}),  {\it James Webb} Space Telescope (\citealt{Gardner2006}), {\it Euclid}~\citep{Laureijs2010}, and  {\it Nancy Grace Roman} Space Telescope (\citealt{Spergel2015}), demands an even deeper understanding of the physical properties of galaxies. We must also account for various target types, spanning a wider redshift range.
In particular, the advent of large multi-wavelength surveys triggered using SED fitting methodology to constrain the AGN and its host galaxy properties for statistical samples~\citep[e.g.,][]{Walcher2011, Boquien2019, Johnson2021, Yang2020, Yang2022, Thorne2022, Bichang2024}. 
The importance of incorporating AGN templates for reliable estimates of physical properties of galaxies hosting AGNs was already raised in previous works, such as \cite{Ciesla2015}. 
The SED fitting approach has revealed the potential to not only derive reliable properties of AGN and host properties (e.g., \citealt{Marshall2022, Mountrichas2021a, Burke2022, Best2023}) but also to identify AGNs based on their multi-wavelength information (e.g., \citealt{Thorne2022, Best2023, Yang2023, Prathap2024}). The AGN SED modeling techniques are also used as the base for the target selection of forthcoming wide-field spectroscopic surveys such as 4MOST (\citealt{Merloni2019}) and VLT-MOONS (\citealt{Maiolino2020}). 

In this paper, we describe the methodology employed in constructing the Value Added Catalog (VAC) of physical properties of DESI Early Data Release (EDR) galaxies obtained via SED modeling with the Code Investigating GALaxy Emission ({\tt CIGALE}; \citealt{Boquien2019}). This code, based on the energy balance principle, has already proved its efficiency and accuracy in estimating physical properties accounting for the AGN contribution~\citep[e.g.,][]{Ciesla2015, Yang2023}. Its modular framework allows the inclusion of various AGN models, both based on both theoretical approaches~\citep[e.g.,][]{Fritz2006} and observational constraints~\citep[e.g.,][]{Stalevski2012,Stalevski2016}. We demonstrate the utility of our catalog by showcasing its potential for discriminating host galaxy properties, while also investigating the influence of the model assumptions and availability of   photometry data. In the follow-up paper (Siudek et al. {under DESI Collaboration review}), we discuss the ability of the SED fitting approach to distinguish narrowline (NL) and broadline (BL) AGNs based on their physical properties. The structure of the paper is as follows. In Sect. \ref{sec:data}, we provide an overview of the DESI survey and EDR data. In Sect. \ref{sec:cigale}, we describe the SED fitting methodology applied to derive the physical properties of DESI galaxies. The general properties of the catalog are presented in Sect. \ref{sec:VACgeneralProperties}. We compared our sample to existing catalogs to
validate the derived properties of galaxies  in Sect.~\ref{sec:comparion}. In Sect.~\ref{sec:DependenceofPhysProp}, we discuss the dependence of the physical properties on model assumptions and MIR availability. Finally, Sect. \ref{sec:conclusions} summarizes the catalog and our analysis. Throughout this paper, we assume WMAP7 cosmology \citep{Komatsu2011}, with $\Omega_{m}$ = 0.272 and $H_{0}$ = 70.4. We also consider the photometry in AB magnitudes (\citealt{Oke1983}).

\section{DESI data}\label{sec:data}

\subsection{DESI survey}

DESI is a 5000-fiber multiobject spectrograph installed on the Mayall 4-meter telescope at Kitt Peak National Observatory. It covers a spectral range of $3600-9800$ {\rm \AA} with a wavelength-dependent spectral resolution, $\rm R = 2000 - 5500$~\citep{DESICollaboration2016b,DESICollaboration2022, Corrector.Miller.2023, FocalPlane.Silber.2023}. 
It is designed to observe approximately 36 million galaxies~\citep{Hahn2023a, Raichoor2023, Zhou2023} and 3 million quasars~\citep{Chaussidon2023} over a 14,000 $\rm deg^2$ within a five-year period~\citep{DESICollaboration2023} with the aim to determine the nature of dark energy through the most precise measurement of the expansion history of the universe ever obtained~\citep{Levi2013}. 
The DESI dataset will be ten times larger than the  SDSS~\citep{York2000, Almeida2023} sample of extragalactic targets and substantially deeper than prior large-area surveys~\citep{DESICollaboration2023}. 
In December 2020, DESI started a five-month survey validation (SV) before the start of the main survey~\citep{DESI2023a.KP1.SV}. 
The SV campaign consisted of three phases: i) SV1: validating the target selections of the five primary target classes: Milky Way survey ({\tt MWS}; ~\citealt{Cooper2023}) and bright galaxy survey ({\tt BGS}; ~\citealt{Hahn2023a}), along with surveys of luminous red galaxies ({\tt LRG}; ~\citealt{Zhou2023}), emission-line galaxies ({\tt ELG}; ~\citealt{Raichoor2023}), and quasars ({\tt QSO};~\citealt{Chaussidon2023}). These targets were further supplemented
with secondary fiber targets for additional science goals ({\tt SCND}; e.g.,~\citealt{Darragh-Ford2023, Fawcett2023}. More details regarding the DESI targeting are described in \citealt{Myers2023});  Furthermore, 
 ii) SV2: operation developments; and iii) SV3 (One-Percent survey) optimized the observing procedures~\citep{SurveyOps.Schlafly.2023}, with very high fiber assignments completeness over an area of 200 $\rm deg^2$, namely, of 1\% of the final DESI main survey. 
The entire SV data, internally known as {\it Fuji}, is publicly released as the DESI Early Data Release~\citep[EDR;][]{DESICollaboration2023} and is used for generating the VAC of physical properties of DESI galaxies presented in this paper. The First Data Release (DR1, DESI Collaboration et al. in prep.) is planned to be released at the {first half of 2025}. The DR1 already showcases the DESI potential in science Key Papers presenting the two-point clustering measurements and validation (DESI Collaboration et al. in prep.), BAO measurements from galaxies and quasars~\citep{DESICollaboration2024III}, and from the Lya forest~\citep{DESICollaboration2024IV}, as well as a full-shape studies of galaxies and quasars (DESI Collaboration et al. in prep). These are supplemented
with the cosmological results from the BAO measurements~\citep{DESICollaboration2024VI} and the full-shape analysis (DESI Collaboration et al. in prep.), as well as constraints on primordial non-gaussianities (DESI Collaboration et al. in prep.).

The DESI spectra are processed with a fully automatic pipeline~\citep{Guy2023}, followed by the spectral classification and redshift estimation using the {\tt Redrock} pipeline\footnote{\url{https://github.com/desihub/redrock}} (\citealt{Anand2024}, Bailey et al. in prep.). 
This $\chi^2$ method relies on the principal component analysis templates generated from a combination of real and synthetic spectra of astronomical sources using an iterative principal component generator \citep{Bailey2012}, which also takes the uncertainties of the data into account.
Along with the redshift ({\tt Z}), we take the redshift uncertainty ({\tt ZERR}), a redshift warning bitmask ({\tt ZWARN}), {\tt Redrock} also assigns a spectral type ({\tt SPECTYPE}) to every target based on the best fit. The resulting DESI EDR redshift catalog consists of 2,847,435 spectra of 2,757,937 unique sources~\citep{DESICollaboration2023}. For multiply observed targets, we chose the "best" spectrum as the one that has a higher signal-to-noise ratio (S/N) spectrum, along with good fiber and redshift measurements ({\tt ZCAT\_PRIMARY = True}\footnote{\url{https://github.com/desihub/desispec/blob/0.51.13/py/desispec/zcatalog.py\#L13}}). 
Furthermore, we selected sources that have been assigned as {\tt GALAXY} or {\tt QSO} by {\tt Redrock} and that do not have any fiber issues ({\tt COADD\_FIBERSTATUS = 0}\footnote{\url{https://desidatamodel.readthedocs.io/en/latest/bitmasks.html\#spectroscopic-reduction-masks}}) , but do have
 a reliable redshift ({\tt ZWARN = 0 or 4}\footnote{\url{https://desidatamodel.readthedocs.io/en/latest/bitmasks.html\#redshift-fitting-redrock-masks}}). We refer  to~\citealt{DESICollaboration2023} for more details about our selection choices. 
After applying all these cuts, we were left with a sample of 1,345,137 objects spanning a redshift range of $\rm 0.001 \le z \le 5.968$ {over $\rm \sim1,100~deg^2$}. Our analysis requires photometric measurements, which are described in the following subsection.

\subsection{Photometry}\label{sec:photometry}
The DESI primary targets ({\tt MWS}, {\tt BGS}, {\tt LRG}, {\tt ELG}, and {\tt QSO}) were selected from the ninth data release of the Legacy imaging surveys\footnote{\url{https://www.legacysurvey.org/dr9/description/}}~\citep[LS/DR9;][]{Dey2019}, which is a combination of three public projects. 
The northern sky is covered by the Beijing-Arizona Sky Survey (BASS; \citealt{Zou2017}) in the {\it g} and {\it r} band and by the Mayall {\it z}-band Legacy Survey (MzLS) in the {\it z} band with a 5$\sigma$ detection limit of {\it g} = 23.48, {\it r} = 22.87, and {\it z} = 22.29 AB magnitude. The south LS footprint is mapped by the Dark Energy Camera Legacy Survey (DECaLS) in all three bands ({\it g}, {\it r}, and {\it z}) with a 5$\sigma$ detection of {\it g} = 23.72, {\it r} = 23.27, and {\it z} = 22.22 AB magnitude. The detection limits are found for a fiducial galaxy size of $\rm 0.45\arcsec$.  

The ground-based optical and near-infrared (NIR) photometry (i.e., {\it grz} photometry, which we hereafter shortly refer to as optical) is supplemented
with observations from MIR bands at $\rm 3.4, 4.6, 12$ and 22 $\mu$m provided by the Wide-field Infrared Survey Explorer (WISE;~\citealt{Wright2010}) and a mission extension NEOWISE-Reactivation forced-photometry~\citep{Mainzer2014} in the unWISE maps at the DESI footprint~\citep{Meisner2017, Schlafly2019, Meisner2021b}.  
The WISE photometry is matched to optical imaging using the {\tt Tractor} package~\citep{Lang2016} based on seven-year custom stacks of WISE/NEOWISE exposures, called unWISE coadds, reaching $5\sigma$ limiting magnitudes of 21.7 and 20.9 AB mag in {\it W1} and {\it W2}~\citep{Meisner2019, Meisner2021}. {Table~\ref{tab:PhotometrySummary} summarizes the information about photometric bands used in this analysis. }

\begin{table}[ht]
\centering
\caption{Summary of the LS9 photometry used for SED fitting.}
\label{tab:PhotometrySummary}
\footnotesize
\begin{tabular}{c|c|c}
    Band & Survey & Depth (5$\sigma$, AB mag)  \\
    \hline\hline
    \multirow{2}{*}{g} & DECaLS & 23.72  \\
                       & BASS   & 23.48 \\
    \hline
    \multirow{2}{*}{r} & DECaLS & 23.27 \\
                       & BASS   & 22.87  \\
    \hline
    \multirow{2}{*}{z} & DECaLS & 22.22  \\
                       & MzLS   & 22.29 \\
    \hline
    W1 & unWISE & 21.7  \\
    W2 & unWISE & 20.9  \\
    W3 & unWISE & 16.7  \\
    W4 & unWISE & 14.5 \\
\end{tabular}
\tablefoot{
We report the 5$\sigma$ detection limits for {\it g}, {\it r}, {\it z} for a fiducial galaxy size of 0.45~arcseconds from~\cite{Dey2019}.
The 5$\sigma$ detection limits for WISE/NEOWISE exposures (unWISE) are reported by~\cite{Meisner2019, Meisner2021}.
}
\end{table}

{\tt Tractor} is a pioneered Python tool based on the statistically rigorous forward-modeling approach to perform source extraction on the pixel data. 
It is designed to fit images and photometry to estimate source shapes and brightness properties taking into account their different point spread function (PSF) and different band sensitivities. 
This approach is particularly useful to process LS sources given their wide range of PSF shapes and sizes: the optical data have a typical PSF of $\sim 1$ arcsec; and the WISE PSF full width at half maximum (FWHM) is $\sim 6$ arcsec in {\it W1–W3} and $\sim 12$ arcsec in {\it W4}~\citep{Dey2019}. 
The pixels associated with each detection (called blob) are fitted with models of surface brightness, including the S\'ersic profile, and the best fit is chosen as the one which minimizes the $\chireduced$. 
The fits are performed separately on each photometric band ({\it g}, {\it r}, and {\it z} bands) accounting for different PSF and sensitivity of each image. 
In DR9, the light profiles are fitted with four models: point source ({\tt PSF}), round exponential galaxy model ({\tt REX}), de Vaucouleurs model ({\tt DEV}), exponential model ({\tt EXP}), and a S\'ersic model ({\tt SER})\footnote{Note: the S\'ersic model replaced the composite model used in the previous, DR8, release.}, in that order. 
However, {\tt Tractor} models do not include more complex structures and the resulting models may not always be ideal. 
The best-fit model is determined by convolving each model with the specific PSF for each exposure, fitting it to each image, and minimizing the residuals for all images. 

\begin{table}
\centering
        \caption{Summary of DESI main target classes used throughout the paper.}
        \label{tab:TargetsDemographics}
        \footnotesize
        \begin{tabular}{r | r | r | r  }
        target & {\tt N} & {\tt \%} & {\tt Z} \\
        \hline \hline
        {\tt BGS} & 435,685 & 34 & $0.01 < z < 0.6$ \\
        {\tt LRG} & 229,347 & 18 & $0.4 < z < 1.1$ \\
        {\tt ELG} & 555,221 & 43 & $0.6 < z < 1.6$ \\
        {\tt QSO} & 163,380 & 13 & $0.6 < z < 3.5$  \\
        \hline 
        {\tt EDR} & 1,286,124 & 100 & $0. < z < 6.0$  

\\
\end{tabular}
\tablefoot{For each class, we provide the number of objects ({\tt N}),  the percentage of the total sample (\%), and the redshift range ({\tt Z}; \citealt{DESICollaboration2023}).}
\end{table}  

The {\tt Tractor} model fits are determined using only the optical ({\it grz}) data. The MIR photometry for each optically detected source is then determined by forcing the location and shape of the model, convolving with the WISE PSF, and fitting to the WISE stacked image. 
The advantage of the "forced photometry" is the ability to deblend any confused WISE sources by using the higher-spatial-resolution optical data and detecting fainter sources than traditional approaches, while preserving the photometric reliability. However, this approach limits the LS catalog to strictly WISE photometry for sources detected at optical wavelengths~\citep{Dey2019}. 

{\tt Tractor} returns the object position, fluxes, galactic extinctions, shape, and morphological parameters, given by the S\'ersic index (among others). 
The Tractor catalog also contains a set of quality measures, such as {\tt FRACMASKED}, {\tt FRACFLUX}, and {\tt FRACIN}, which offer a qualitative calculation of the data in each profile fit and can be used to preselect high-quality samples (see Sect.~\ref{sec:CleaningVAC}). 
More detail about the data reduction can be found in \cite{Dey2019} and Schlegel et al. in prep.

\begin{figure}
    \centerline{\includegraphics[width=0.49\textwidth]{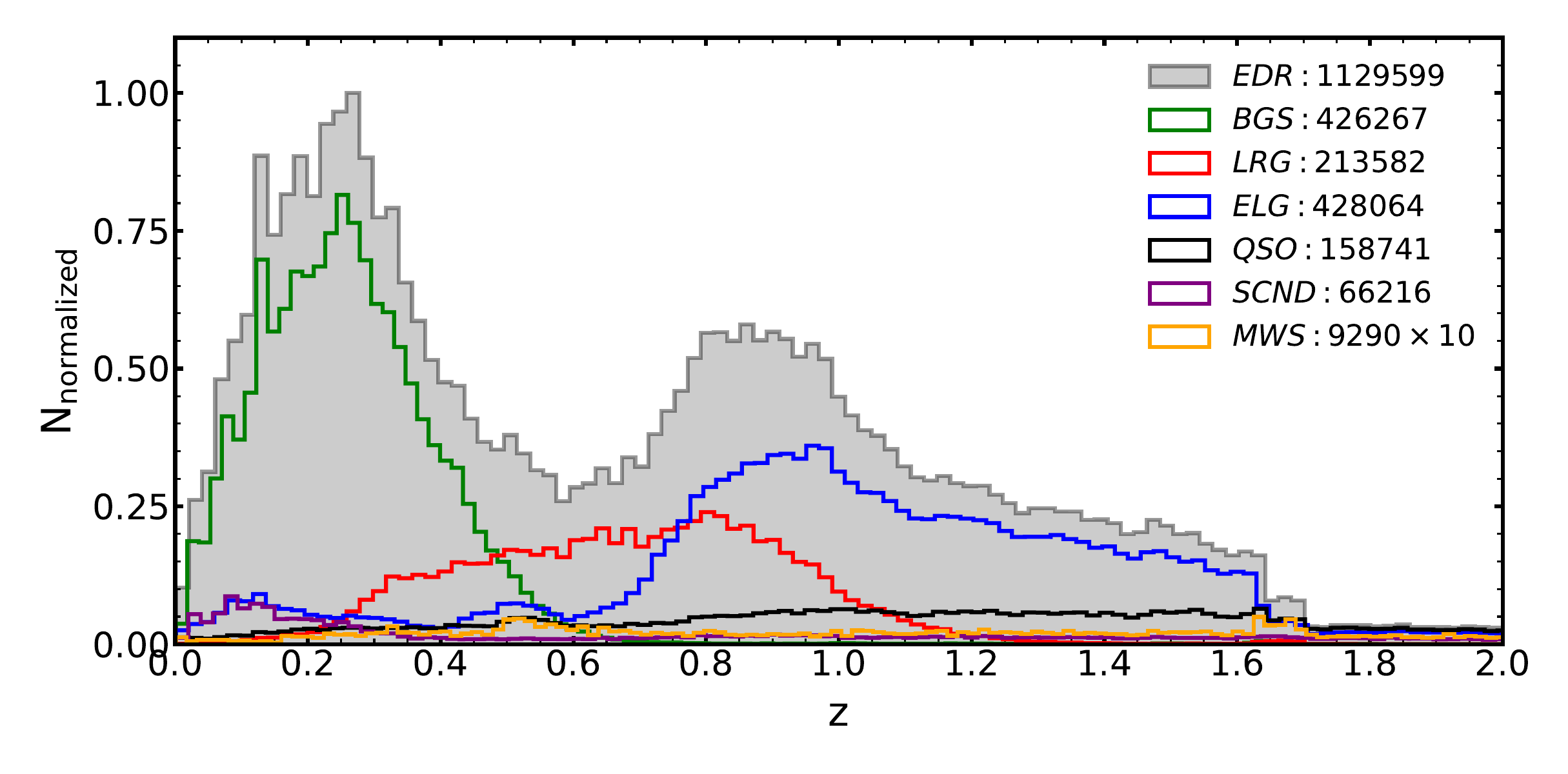}}
    \caption{Redshift distribution of the DESI EDR sources with reliable redshift and photometry estimates (see Sect.~\ref{sec:data}). The catalog includes all galaxies and quasars observed within the DESI target classes: {\tt BGS}, {\tt LRG}, {\tt ELG}, {\tt QSO}, {\tt SCND,} and {\tt MWS}. The redshift distribution of {\tt MWS} targets is scaled up by a factor of 10. Some of the sources are targeted by multiple target classes. The numbers shown in the figure also contain duplicates, so this total is higher than the number of unique targets. }
    \label{fig:z_distribution}
\end{figure}

In addition to the primary targets, DESI EDR also includes some secondary targets and targets of opportunity that do not have LS/DR9 photometry. Given that our analysis requires photometric measurements, we only consider the DESI sources that have LS/DR9 photometry ({\tt RELEASE = 9010 or 9011 or 9012}). This leads to a final sample of 1,286,124 unique objects, spanning a redshift range of $\rm 0.001 \le z \le 5.968$. 
{The demographics of the the DESI primary targets ({\tt MWS}, {\tt BGS}, {\tt LRG}, {\tt ELG}, and {\tt QSO}) is summarized in Table~\ref{tab:TargetsDemographics}.  }
The redshift distribution of all these sources, along with their distributions from different targeting types is shown in Fig.~\ref{fig:z_distribution}, with the distribution of {\tt MWS} scaled up by a factor of ten to more easily compare the shape of each distribution. The DESI VAC covers a wide redshift range spanning over $\rm z = 0-6$ targeting {\tt BGS} galaxies and AGNs at lower redshift ($\rm z < 0.6$; \citealt{Hahn2023a}; \citealt{Juneau2024})  and {\tt ELGs} at higher redshift ($\rm 0.6 < z < 1.6$; \citealt{Raichoor2023}), with {\tt QSOs} spanning out to $\rm z \sim 6$ (\citealt{Chaussidon2023}); {\tt LRGs} extend out to $\rm z\sim 1$ (\citealt{Zhou2023}) and {\tt SCND} sources cover the entire redshift range, incorporating low-z targets~\citep[e.g.,][]{Darragh-Ford2023} as well as high-z QSOs~\citep[e.g.,][]{Fawcett2023}. The catalog also includes a negligible fraction ($\rm <1\%)$ of galaxies and QSO observed within the {\tt MWS} (\citealt{Cooper2023}) spanning  a wide redshift range; namely, it is composed of 3,238 galaxies (with a mean redshift $\rm z = 0.7$) and 6,052 QSOs (with a mean redshift $\rm z = 1.6$).

\section{Spectral energy distribution fitting}\label{sec:cigale}

\begin{table*}
\centering
        \caption{Default parameters used in SED fitting with {\tt CIGALE}.}
        \label{tab:SEDParameters}
        \footnotesize
        \begin{tabular}{r r r}
\hline
\hline
Parameter & Symbol &  Values\\
\hline
\multicolumn{3}{c}{Stellar population models:  \cite{Bruzual2003}}\\
\hline \hline
Initial mass function & IMF & \cite{Chabrier2003}\\
Metallicity & $\rm Z$ & 0.02\\
\hline
\multicolumn{3}{c}{SFH: Double exponentially decreasing}\\
\hline \hline
$\tau$ of the main stellar population (Gyr) & $\rm \tau_{main}$ & {0.1, 0.5, 1, 3, 5, 8}\\
Age of the main stellar population (Gyr) & $\rm t_1$ & 0.5, 1, 3, 4.5, 6, 8, 10, 13 \\
$\tau$ of the burst stellar population (Gyr) & $\rm \tau_{burst}$ & 10\\
Age of the burst stellar population (Gyr) & $\rm t_{burst}$ & 0.05\\
Mass fraction of young stellar population & $\rm f_{ySP}$ & 0, 0.01, 0.1, 0.2\\
\hline
\multicolumn{3}{c}{Nebular emission}\\
\hline \hline
Ionization parameter & $\rm logU$ & -2\\
Gas metallicity & $\rm Z_{gas}$ & 0.02\\

\hline
\multicolumn{3}{c}{Dust attenuation: \cite{Calzetti2000}}\\
\hline \hline
Color excess of the nebular emission & $\rm E(B-V)_{line}$ & 0, 0.05, 0.15, 0.3, 0.5, 0.75, 0.9, 1.1, 1.3, 1.6\\
Reduction factor to apply on $\rm E(B-V)_{line}$ & $\rm E(B-V)_{star}/E(B-V)_{line}$ & 0.44\\
\hline
\hline
\multicolumn{3}{c}{Dust emission: \cite{Draine2014}}\\
\hline \hline
Mass fraction of PAHs  & $\rm q_{PAH}$ &  0.47, 1.12, 2.5, 3.19\\
Minimum radiation field & $\rm U_{min}$ & 15\\
Power law slope of the radiation field & $\rm \alpha$ & 2.0\\
Fraction illuminated from $\rm U_{min}$ to $\rm U_{max}$& $\gamma$ & 0.02 \\
\hline
\multicolumn{3}{c}{AGN: \cite{Fritz2006}}\\
\hline \hline
The angle between the equatorial axis and line-of-sight & AGNPSY [deg] & 0.001, 20.100, 40.1, 70.100, 89.990 \\
Contribution of the AGN to the total LIR & AGNFRAC &  0, 0.01, 0.1, 0.3, 0.5, 0.7, 0.9 \\
\hline \hline
\end{tabular}
\end{table*}    

Physical properties of DESI galaxies are derived by performing SED (optical and mid-IR photometry) fitting using Code Investigating GALaxy Emission~\citep[{\tt CIGALE} v2022.1;][]{Boquien2019}. 
{\tt CIGALE} is a state-of-the-art Python code based on the principles of the energy balance between the dust-absorbed stellar emission in the ultraviolet (UV) and optical and its re-emission in the infrared (IR). 
Thanks to its efficiency, flexibility, and accuracy, {\tt CIGALE} and its modified version {\tt X-CIGALE}~\citep{Yang2020,Yang2022} are widely used to derive the physical properties of galaxies and AGNs in large galaxy surveys (e.g., \citealt{Ciesla2015,Salim2016,Salim2018A, Malek2018,Barrows2021, Mountrichas2021b,Zou2022, Csizi2024,Osborne2024}) as well as high-z AGNs (e.g., \citealt{Juodzbalis2023,Mezcua2023, Yang2023, Burke2024, Durodola2024}). 
{\tt CIGALE} estimates the physical properties of galaxies using a Bayesian approach by evaluating all the possible combinations of SED models on data to maximize the likelihood distribution. 
CIGALE  takes into account the age of the universe at the redshift of each object when fitting models. It excludes stellar population ages that exceed the age of the universe at the given redshift. This constraint helps us avoid unphysical solutions and ensures the consistency of the fitted parameters with cosmological constraints.
The estimates and errors of the physical properties are then computed as the likelihood-weighted mean and standard deviation, respectively, for all the models~\citep{Boquien2019}. 
To build a library of models, {\tt CIGALE} relies on five main modules: star formation history (SFH), SSP models, dust attenuation and emission, and the AGN component. 
For each module, {\tt CIGALE} includes several possible prescriptions and the flexible parametrization of model parameters allows us to adapt the model complexity (i.e., number of free parameters). 
In the next sections, we describe the key assumptions and parametrization of models used to create our catalog, which are summarized in Table~\ref{tab:SEDParameters}. 
This configuration generates 167,529,600 models spanning over a wide redshift range from 0 to 6 (302,400 per redshift). 
We use a single node of 32 cores on the Cori supercomputer at the National Energy Research Scientific Computing Center (NERSC) to fit all SEDs within 3.5 hours. 

\subsection{Stellar component}\label{sec:StellarComponent}

We use the~\citealt{Bruzual2003} SSP models adopting a ~\cite{Chabrier2003} IMF to build the stellar component. 
We assume solar metallicity and following the analysis by~\cite{Ciesla2015}, we use the delayed SFH with an optional exponential burst. 
This prescription allows us to reproduce the SEDs of both star-forming and passive galaxies with a modest number of free parameters~\citep[e.g.,][]{Ciesla2015, Salim2016}.  
Such two exponentially decreasing star formation rate (SFR) laws with different e-folding times show a good performance in decoupling the long-term SFH from the recent star formation activity~\citep{Ciesla2015, Malek2018}. 
The SFR is therefore defined as the sum of two exponentially decreasing SFRs:
\begin{equation}\rm
\mathrm{SFR}(t) = \mathrm{SFR}_{\mathrm{delayed}}(t) + \mathrm{SFR}_{\mathrm{burst}}(t),
\end{equation}
where:

\begin{equation}\rm
\mathrm{SFR}_{\mathrm{delayed}}(t) \propto \frac{t}{\tau_{\mathrm{main}}^2} \, e^{-t/\tau_{\mathrm{main}}},
\end{equation}
and 
\begin{equation}\rm
\mathrm{SFR}_{\mathrm{burst}}(t) =
\begin{cases}
0, & t < t_{\mathrm{main}} - t_{\mathrm{burst}}, \\
k \, e^{-\left(t-(t_{\mathrm{main}}-t_{\mathrm{burst}})\right)/\tau_{\mathrm{burst}}}, & t \geq t_{\mathrm{main}} - t_{\mathrm{burst}} ,
\end{cases}
\end{equation}

 where t is the time since the onset of star formation, $\rm \tau_{main}$ is the e-folding time of the main (old) stellar population, $\rm \tau_{burst}$ is the e-folding time of the burst (young) stellar population, and k is the normalization constant which depends on the fraction of stars formed in the second burst versus the total stellar mass formed ($\rm f_{ySP}$).  
 The SFH module is described in more detail in~\cite{Ciesla2015, Ciesla2017}. 
 
The effects of the choice of IMF, the SFH prescription, and the solar metallicity assumption on the main galaxy physical properties are discussed in Sect.~\ref{sec:DependenceofPhysProp}. 

\subsection{Nebular emission}\label{sec:nebularemission}

Nebular emission (emission from ionized gas) is an important component to include when considering high-redshift galaxies~\citep[e.g.,][]{Stark2013, Barros2014, Yuan2019} or at lower-redshift starburst dwarf and young star-forming galaxies~\citep[e.g.,][]{Boquien2010} as intense star formation and ionization processes lead to stronger nebular emission signatures.  
Neglecting the nebular emission component may lead to the overestimation of the stellar mass~\citep{Yuan2019}.  
The nebular emission lines are pre-computed with {\tt CLOUDY} 17.01~\citep{Ferland2017} with electron density ($\rm N_e$), gas metallicity ($\rm Z_{gas}$), and ionization parameter (U) as the free parameters and re-scaled with the number of Lyman continuum photons from the stellar emission. 
The \cite{Bruzual2003} SSP models using a constant SFH over 10 Myr is used to generate the photo-ionizing field shape. 
The nebular continuum is scaled directly from the number of ionizing photons. 
{\tt CIGALE} takes into account also the fraction of the Lyman photons escaping galaxies and absorbed by dust. More details about the implementation of the {\tt CLOUDY} into {\tt CIGALE} can be found in~\cite{Villa2021}. We retained the default parameters of this module (see Table~\ref{tab:SEDParameters}). 

\begin{figure}
\centerline{\includegraphics[width=0.49\textwidth]{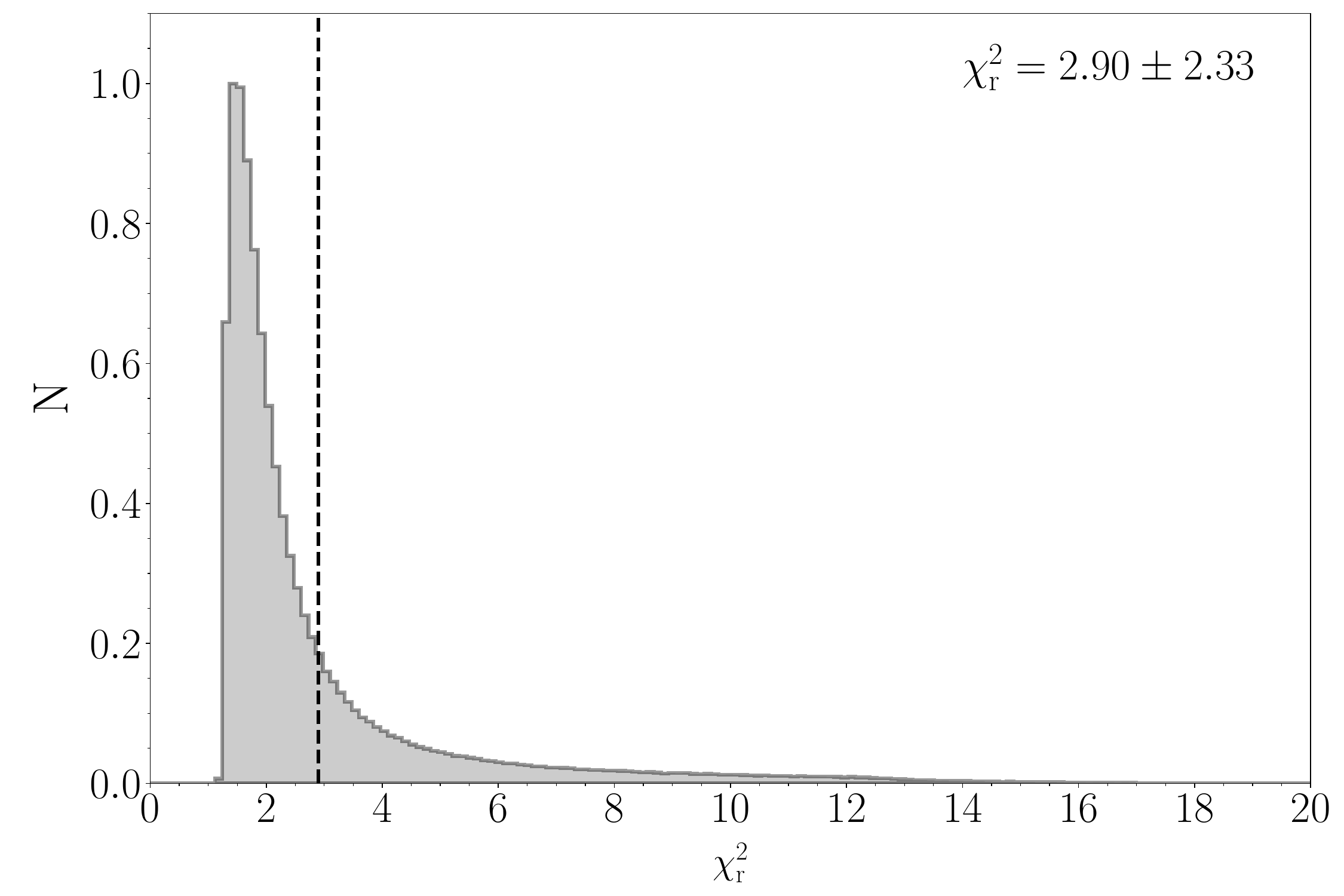}}
        \caption{Distribution of the $\chireduced$ for the DESI EDR SED fits. The mean and standard deviation of $\chireduced$ are reported in the legend.  The dashed line corresponds to the mean $\chireduced$. At least $\sim 88\%$ of the sample is characterized by robust fits (defined as $\chireduced < 5$):}
        \label{fig:chi2}
\end{figure}  

\subsection{Interstellar dust}\label{sec:dust}
Dust is a fundamental component of galaxies that significantly influences their observed SEDs, especially those that are actively star-forming~\citep{Conroy2013}. 
It plays a dual role in stellar population models, as dust absorbs short-wavelength light (from the UV to NIR) and re-emits it at longer wavelengths (from MIR to far-infrared (FIR)). 
SED fitting techniques, such as those employed by {\tt CIGALE}, model the full SED by simultaneously considering both the attenuated light and the re-emitted radiation. This integrated approach ensures that both the absorption and emission by dust are used to constrain the model. This is important because dust obscuration is influenced by the galaxy's geometry, while dust emission is sensitive to the interstellar radiation field (\citealt{Conroy2013}). Dust emission is dominated by: i) polycyclic aromatic hydrocarbon (PAH) bands in the MIR ($\sim 8 \mu m$), ii) very small, warm grains, and iii) big, relatively cold grains ($\gtrsim 100 \mu m$). 
Differences between these dust grains have an impact on the dust SED.  
On the other hand, attenuation also depends on the geometry. 
The simplest way to model dust attenuation is to assume attenuation laws~\citep{Boquien2019}. 
{\tt CIGALE} provides two modules to model attenuation curves: the implementation of the \cite{CharlotFall2000} model and the modified \cite{Calzetti2000} model; for simplicity, we refer to it as the \cite{Calzetti2000} model. 

The starburst model uses the \cite{Calzetti2000} starburst attenuation curve as a baseline, which is extended by a \cite{Leitherer2002} curve from the Lyman break to 150 nm. 
The amount of attenuation is described by the color excess applied to the nebular emission lines, $\rm E(B-V)_{line}$, and the ratio $\rm E(B-V)_{star}/E(B-V)_{line}$, where $\rm E(B-V)_{star}$ is the color excess applied to the whole stellar continuum. 
Following the \cite{Calzetti2000} recommendations, this ratio is fixed to 0.44. 
We used the \cite{Calzetti2000} model to generate our catalog; however, the impact on the physical property estimates when using \cite{CharlotFall2000} is discussed in Sect.~\ref{sec:Dustchoice}.

{\tt CIGALE} provides five modules to describe the IR emission from dust: \cite{Casey2012}, \cite{Dale2014}, \cite{Draine2007} and its updated version of the \cite{Draine2014} model, along with the Themis dust emission models from \cite{Jones2017}. 
To create our catalog, we rely on the \cite{Draine2014} models, which account for very different physical conditions with a variety of radiation fields and a variable PAH emission providing a high flexibility. 
The model assumes that the majority ($\rm 1-\gamma$) of dust mass is heated by a radiation field with an intensity ($\rm U_{min}$), while the remaining fraction ($\gamma$) is exposed to intensities ranging from $\rm U_{min}$ to $\rm U_{max}$ following a power-law index $\rm \alpha$. By default, $\rm \gamma$ and $\rm \alpha$ are fixed values set to 0.02, and 2, respectively. 
We also considered the model from \cite{Dale2014} and describe its influence on the physical properties of galaxies in Sect.~\ref{sec:Dustchoice}.

\subsection{AGN contribution}\label{sec:AGNComponentFritz}

{\tt CIGALE} allows for the separation of the emission from AGN from their host galaxy with several approaches, starting from a simple AGN parameterization by the power slope when using \cite{Casey2012} models to fit IR data. The \cite{Dale2014} module provides simple templates of quasars from UV to IR, with the AGN fraction  left as a free parameter. Those options are fast but do not provide complex AGN SEDs. However, {\tt CIGALE} also incorporates two more flexible models: \cite{Fritz2006} and SKIRTOR~\citep{Stalevski2012, Stalevski2016}. The AGN model from \cite{Fritz2006} covers the UV to IR and assumes that the central engine is surrounded by smoothly distributed dust in the AGN torus \citep[i.e., the AGN unified model; e.g.,][]{Zou2019}, while SKIRTOR models add the possibility that the dust is clumpy \citep[e.g.,][]{Stalevski2012, Assef2013}. However, it is still unclear whether observations can discriminate between these models~\citep{Feltre2012}.

In this work, we apply the \cite{Fritz2006} model, which uses a simple, but realistic torus geometry that relies on the flared disc and a full range of dust grain size.
It allows us to control the geometry and physics of the torus thanks to a flexible ratio of the maximum to minimum radii of the dust torus, the optical
depth at 9.7 $\mu m$, dust density distribution,  opening angle of the torus, and the viewing angle. In particular, viewing from the equatorial direction (with a viewing angle {\tt AGNPSY} $= 0^{\circ}$) leads to the obscuration of the central engine and only the radiation reemitted in IR can be observed (type 2 AGN, i.e., NL AGN). When viewing from the polar direction (with a viewing angle {\tt AGNPSY} $= 90^{\circ}$), the central engine is directly visible (type 1 AGN, i.e., BL AGN). We fixed the parameters to the default, aside from allowing for the flexibility in the viewing angle ({\tt AGNPSY}) and in the AGN fraction ({\tt AGNFRAC}), defined as the ratio of the AGN IR emission to the total IR emission.  Thus, the generated models cover a wide range of objects, including galaxies without any AGN contribution, as well as type 1 and 2 AGNs. This provides flexibility and allows us to build the catalog of physical properties for both AGN and non-AGN host galaxies. The impact of the incorporation or the change of the AGN model is discussed in Sect.~\ref{sec:AGNchoice}.

\begin{figure}
\centerline{\includegraphics[width=0.49\textwidth]{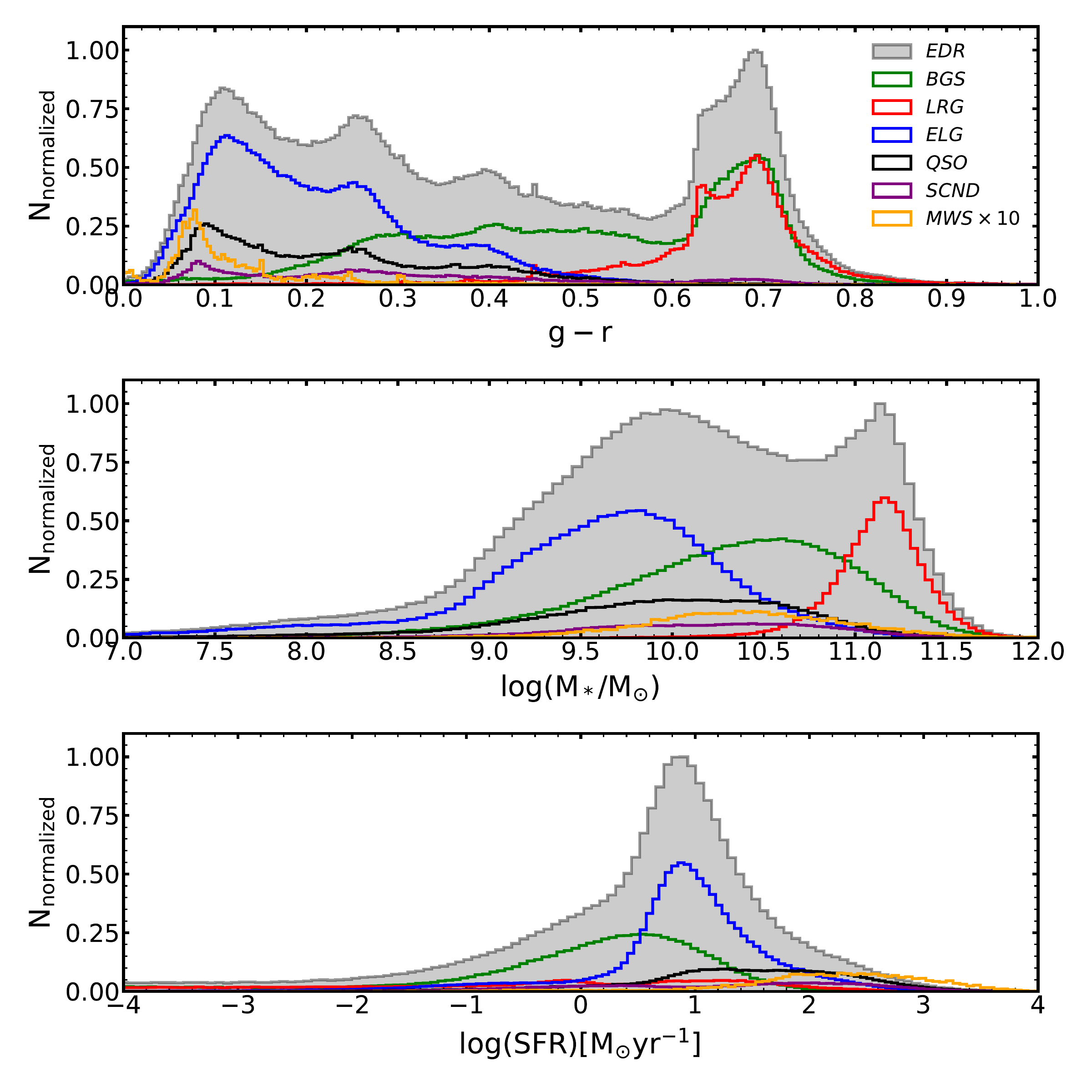}}

        \caption{Distribution of the SED-derived properties: rest-frame {\it g-r} color (top panel), stellar mass ($\mstar$; middle panel), and star formation rate (SFR; bottom panel) of 1,286,124 of DESI EDR galaxies. The MWS target class is scaled up by a factor of ten.    }
        \label{fig:phys_distributio}
\end{figure}

\section{VAC: General properties}\label{sec:VACgeneralProperties}
In this section, we characterize the VAC of physical properties of the DESI galaxies and quasars observed at redshift $\rm 0.001\le z \le 5.968$ (see Sect.~\ref{sec:data} for data description). 
The catalog includes all DESI sources, independent of their target class ({\tt BGS}, {\tt LRG}, {\tt ELG}, {\tt SCND}, {\tt MWS,} and {\tt QSO}) or redshift. 
The physical properties are derived based on the SED fitting using {\tt CIGALE} (see Sect.~\ref{sec:cigale} for the description of the SED procedures and modules and Table~\ref{tab:SEDParameters} for a full list of parameters).  
The catalog includes estimates of stellar masses, SFRs, absolute magnitudes, AGN fractions, and AGN luminosities among others (for the full list, see Table~\ref{tab:CatalogColumns}).

\subsection{Selection of galaxies:\ SEDs with secure fits}\label{sec:CleaningVAC}
The DESI VAC includes all `reliable' sources from the EDR (see Sect.~\ref{sec:data} for the cuts applied for sample selection). The choice of additional cleaning cuts strongly depends on individual scientific goals. 
For general analysis of DESI galaxies, we did not introduce strict cuts to clean the sample, but the cuts can be adjusted as required for the specific science case. 
In the following analysis, we have excluded:
\begin{itemize}
    \item 153,294 (12\%) sources for which $\mstar =0$. 
Almost all sources with $\mstar = 0 $ are characterized by insufficient photometric observations to perform a reliable SED fitting. In particular, half of them are observed in only one optical band with a high signal to noise ratio ({\tt S/N} $\ge 10$); 21\% of them have two or three optical observations with {\tt S/N} $\ge 10$, and only 3\% are observed in two or more WISE bands with {\tt S/N} $\ge 3$).  This suggests that these sources are false detections, faint sources, and other artifacts without valid photometry. 

    \item An additional 3,231 (0.2\%) sources with bad fits characterized by $\chireduced > 17$ (see Appendix~\ref{sec:ReliabilityofFit} for the description of the $\chireduced$ cut). We note that the user may apply a more restrictive cut on $\chireduced$ (e.g., $\chireduced = 5$) or apply further statistical criteria, such as a Bayesian information criterion (BIC; see e.g., \citealt{Masoura2018, Buat2021}). Figure~\ref{fig:chi2} shows the distribution of $\chireduced$ for DESI EDR galaxies. 
\end{itemize}
 
Additionally, the user can introduce several other quality cuts based on additional flags provided in the catalog, namely:
\begin{itemize}
    \item The S/N of the optical photometry. The uncertainties in stellar mass estimates strongly depend on the input photometry (see Appendix~\ref{app:stmass_errors} for details). Sources observed in three optical bands  ({\it grz}) with high S/N ({\tt S/N} $\ge 10$) are characterized by a stellar mass error of $\mstar_{err} \lesssim 0.25$, corresponding to the standard uncertainty of the stellar mass estimates due to model assumptions (see Sect.~\ref{sec:DependenceofPhysProp} and~\citealt{Conroy2013}\footnote{\cite{Conroy2013} showed that different assumptions of the IMF, SSP models, SFH can introduce a systematic uncertainty of $\rm \sim 0.3$ dex.}, \citealt{Pacifici2023}). The {\tt FLAGOPTICAL} defines the number of optical bands with {\tt S/N} $\ge 10$ and can be used to select sources with more reliable photometry and, thus, better estimates of the physical properties. 
    
    \item The S/N of the WISE photometry. The availability of WISE photometry with high S/N ({\tt S/N} $\ge 3$) has an impact not only on the stellar mass estimates (see Sect.~\ref{sec:DependenceonPhootmetry}), but also on stellar mass errors (see App.~\ref{app:stmass_errors}). The {\tt FLAGINFRARED} defines the number of bands ({\it W1-4}) with {\tt S/N} $\ge 3$\footnote{The catalog includes the {\tt S/N} for each band allowing the users to modify the threshold according to their scientific cases.}. 

    \item The probability density function (PDF) of the estimated parameters is asymmetric or multi-peaked. {\tt CIGALE}  introduces two estimates based on the best-fit model ({\tt best}) and the likelihood-weighted mean measured from its PDF marginalized over other parameters ({\tt bayes}). A narrow, one-peak PDF should see very similar values for these estimates; otherwise the PDF would end up asymmetric or multi-peaked. The shape of the PDF for the stellar mass and SFR estimates is expressed as $\tt FLAG\_MASSPDF = logM_{best}/logM_{bayes}$ and $\tt FLAG\_SFRPDF = logSFR_{best}/logSFR_{bayes}$, respectively. To preselect sources with narrow one-peak PDF of stellar mass (SFR), we can consider only the one with values between $\rm 0.2 \le \tt LOGM_{PDF} (LOGSFR_{PDF})\rm \le 5,$ following \cite{Mountrichas2021b}, and \cite{Mountrichas2024}.
    \item Additional cuts based on the {\tt Tractor} photometry information. To reject fragmented sources, we may introduce the cut: {\tt FRACFLUX} $\rm \le 0.25$ as advised by \cite{Juneau2024}, Pucha et al. {under DESI Collaboration review}. 
\end{itemize}

Our final EDR sample after cleaning features 1,129,599 sources (88\% of the whole catalog) and is characterized by a mean  $\overline{\chireduced}=2.9\pm2.3$ (see Fig.~\ref{fig:chi2}). At least $ \sim80\%$ of the sample is characterized by good fits defined with a more strict criterion of $\chireduced \lesssim 5$~\citep{Masoura2018, Buat2021}.

The distributions of SED-derived properties: rest-frame {\it g-r} color, stellar mass, and SFR of the DESI EDR galaxies are shown in Fig.~\ref{fig:phys_distributio}. The distribution shapes are clearly different for each of the main target classes ({\tt MWS} is scaled up by 10 to ease the comparison). The {\tt LRG} are found among the reddest observed galaxies, while {\tt ELG} are among the bluest with {\tt BGS} bridging both the blue and red ends of the distribution (see the top panel in Fig.~\ref{fig:phys_distributio}). {\tt QSO}, {\tt MWS}, and {\tt SCND} are among the blue population. 
The stellar mass distribution follows the distribution of the {\it g-r} color, namely, the redder target class, {\tt LRG}, is also found to be the one covering the high-mass end of the DESI EDR distribution ($\mstar> 11$), while the blue {\tt ELG} peaks at $\mstar \sim 9.5 -10$ forming a long tail towards the low-mass end. The remaining target classes ({\tt BGS}, {\tt QSO}, {\tt SCND}, and {\tt MWS}) are peaking in-between {\tt ELG} and {\tt LRG} covering $\mstar \sim10-11$. 
As expected, {\tt LRG} are characterized by lower SFR ($\rm log(SFR/M_{\odot}yr^{-1})<0$) while {\tt ELG} peak at higher SFR ($\rm log(SFR/M_{\odot}yr^{-1})\sim 1$), with {\tt MWS} found to be among the most active ($\rm log(SFR/M_{\odot}yr^{-1})>2$).

\subsection{Star formation main sequence}\label{sec:MainSequence}

One of the common indicators of the galaxy's current star formation activity is its relation between SFR and stellar mass ($\rm M_{star}-SFR$), commonly known as the star-forming main sequence (MS). 
The position of a galaxy compared to the MS helps in classifying it as either a star-burst galaxy (above the MS), a passive galaxy (below the MS), or as a normal SF galaxy (close to the MS; e.g., \citealt{Elbaz2007, Noeske2007, Whitaker2012, Johnston2015, Davies2016, Siudek2018, Davies2024}). 
The MS relation exists across a range of epochs and environments and is roughly linear, with the normalization increasing with redshift (e.g., \citealt{Schreiber2015, Lee2015, Thorne2021}) suggesting that the majority of star-forming galaxies are in a self-regulated equilibrium state (e.g., \citealt{Bouche2010, Daddi2010, Genzel2010, Lagos2011, Lilly2013, Dave2013, Mitchell2016}).

\begin{figure}
\centerline{\includegraphics[width=0.49\textwidth]{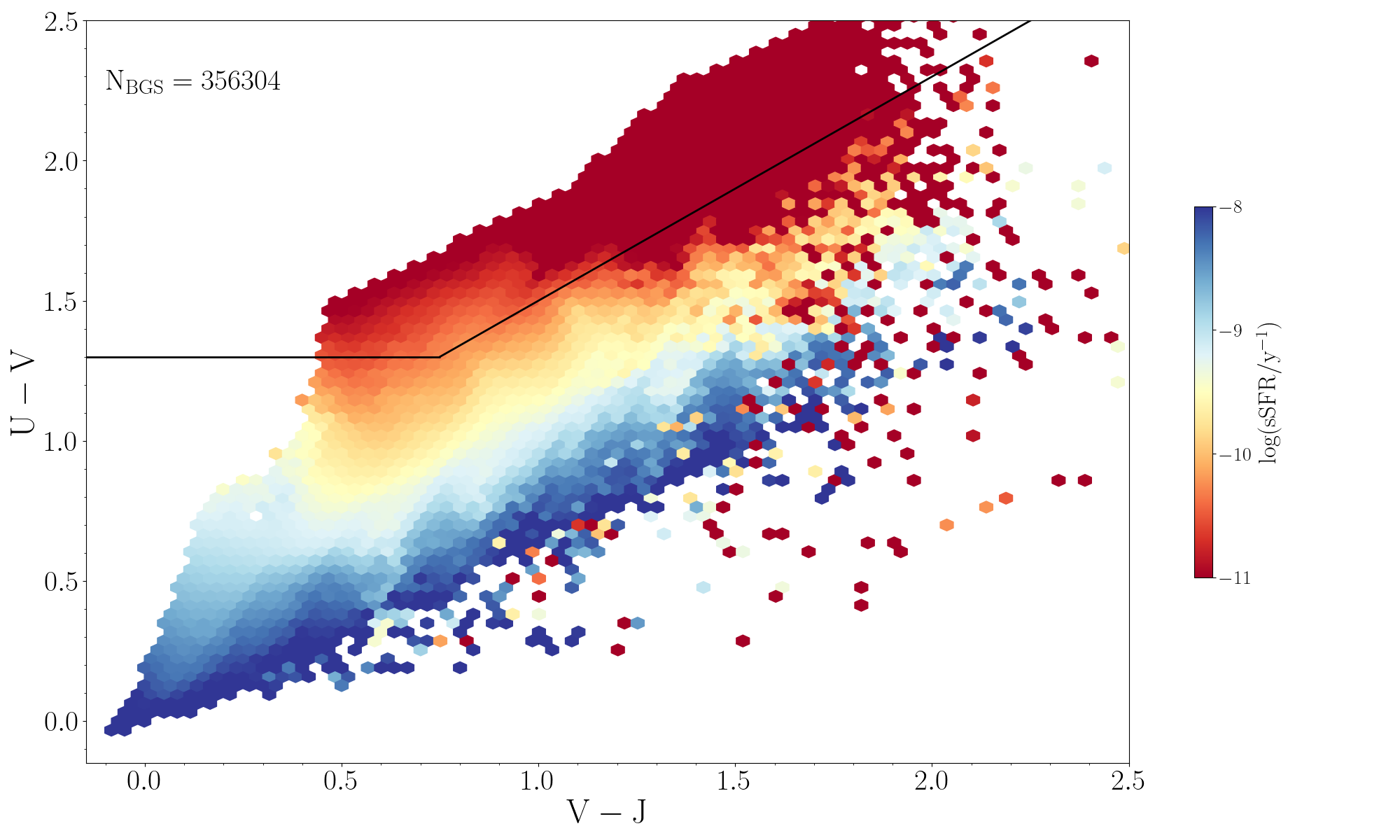}}
        \caption{UVJ diagram for {\tt BGS} galaxies observed at redshift $\rm 0<z<0.6$ color-coded according to their sSFR estimates. Red and blue galaxies are selected by the cut defined by \citet{Whitaker2012} shown in black. }
        \label{fig:UVJ}
\end{figure}

In this paper, we demonstrate the MS relation for {\tt BGS} galaxies. The {\tt BGS} sample is a flux-limited survey at low redshift ($\rm 0 < z < 0.6$) divided into two programs: {\tt BRIGHT} with $r<19.5$ and {\tt DARK} reaching fainter galaxies at $\rm 19.5 < r < 20.175$~\citep{Hahn2023a}. 
We utilized the rest-frame colors derived with {\tt CIGALE}: $\it U-V$ vs. $\it V-J$ to construct the UVJ diagram (\citealt{Williams2009, Whitaker2012}) used to identify red and blue galaxies. 
The UVJ diagram for a sample of 356,304 {\tt BGS} ({\tt BRIGHT} and {\tt DARK}) galaxies\footnote{To select a sample of {\tt BGS} galaxies we implement only the cuts outlined in Sect.~\ref{sec:CleaningVAC}, i.e.,  $\rm \mstar !=0$ and $\chireduced \le 17$.} is shown in Fig.~\ref{fig:UVJ} with a separation cut following the definition given by \cite{Whitaker2012}:  $U - V > 0.8 \times (V - J) + 0.7$, $U - V > 1.3$ and $V - J < 1.5$\footnote{The cut $V-J<1.5$ is no longer implemented, as it is a false upper limit imposed on the quiescent population~\citep{vanderWel2014}.}. 
A gradual change in the specific SFR (sSFR) with colors is evident. Red galaxies are characterized by low sSFR ($\rm log(sSFR/y^{-1}) \lesssim 11$), while blue galaxies are actively star forming ($\rm log(sSFR/y^{-1}) \gtrsim 10$). 

\begin{figure}
\centerline{\includegraphics[width=0.49\textwidth]{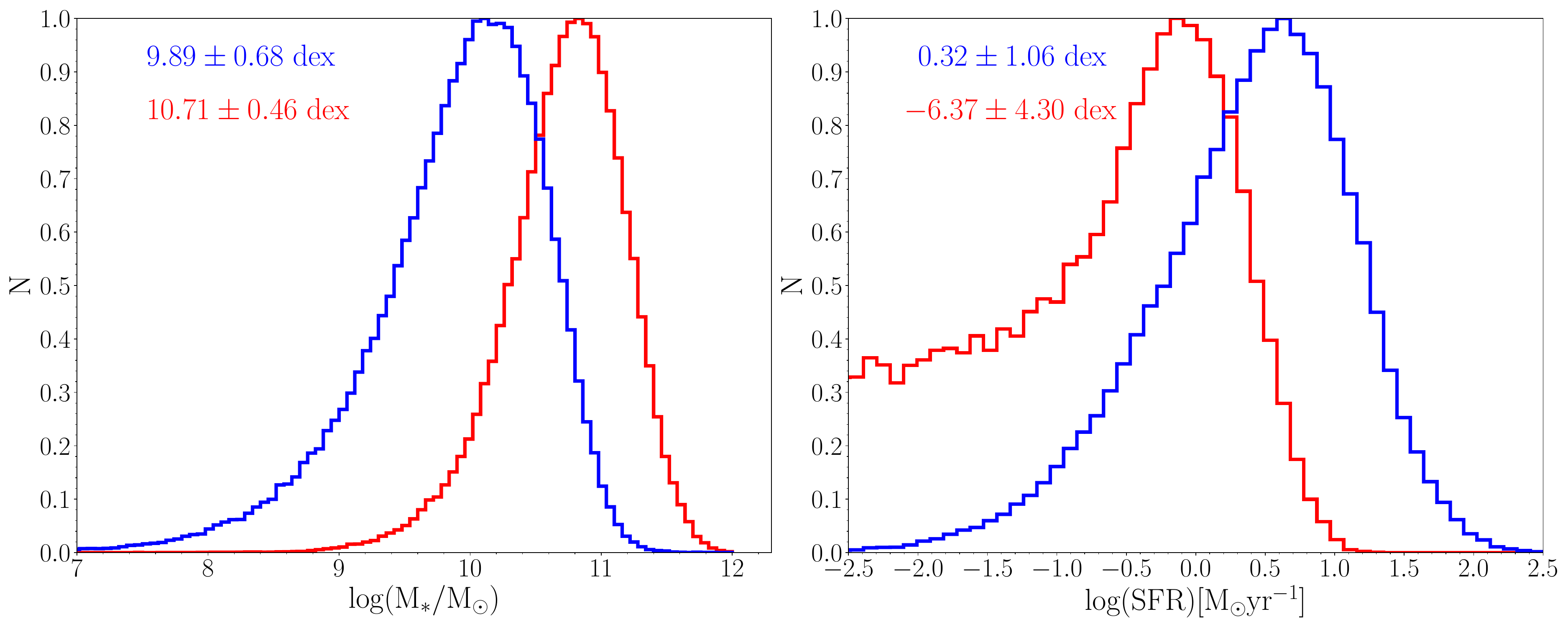}}
        \caption{Stellar mass (left) and SFR (right) distributions for red (in red) and blue (in blue) {\tt BGS} galaxies observed at redshift $\rm 0<z<0.6$ with means and standard deviations reported in the legend. Red and blue galaxies are selected by the cut defined by \citet{Whitaker2012} in the UVJ diagram (see Fig.~\ref{fig:UVJ}). }
        \label{fig:RedBlueDistribution}
\end{figure}

The distribution of stellar mass and SFR for red and blue galaxies is shown in fig.~\ref{fig:RedBlueDistribution}. 
We find a similar number of red and blue galaxies (48\%, and 52\%, respectively) among the flux-limited {\tt BGS} sample, with a mean stellar mass $\mstar = 10.71$ and 9.89 for red and blue galaxies, respectively (see left panel in Fig.~\ref{fig:RedBlueDistribution}). 
As expected, the massive red galaxies are characterized by lower SFRs than low-mass blue galaxies (with a mean SFR  $\rm log(SFR) = -6.4$ and 0.3 $\rm M_{\odot}yr^{-1}$  for red and blue galaxies, respectively;  see right panel in Fig.~\ref{fig:RedBlueDistribution}). 
The blue galaxies tend to follow the MS according to~\cite{Schreiber2015} with slightly higher SFR values possibly due to the choice of the dust attenuation prescription (see e.g.,~\citealt{Siudek2018}). 
The clear MS trend not only validates the SED fitting procedures in recreating the proper physical properties for a population of {\tt BGS} galaxies but also showcases the utility of derived rest-frame colors for galaxy classification purposes. 
We note that to quantitatively compare the fraction and properties of red and blue galaxies one should take into account the selection biases. For example, selecting a mass complete sample (selected following prescription from~\citealt{Pozzetti2010}) changes the fraction of red and blue galaxies to 68\% and 32\%, respectively.  

\begin{figure}
        \centerline{\includegraphics[width=0.49\textwidth]{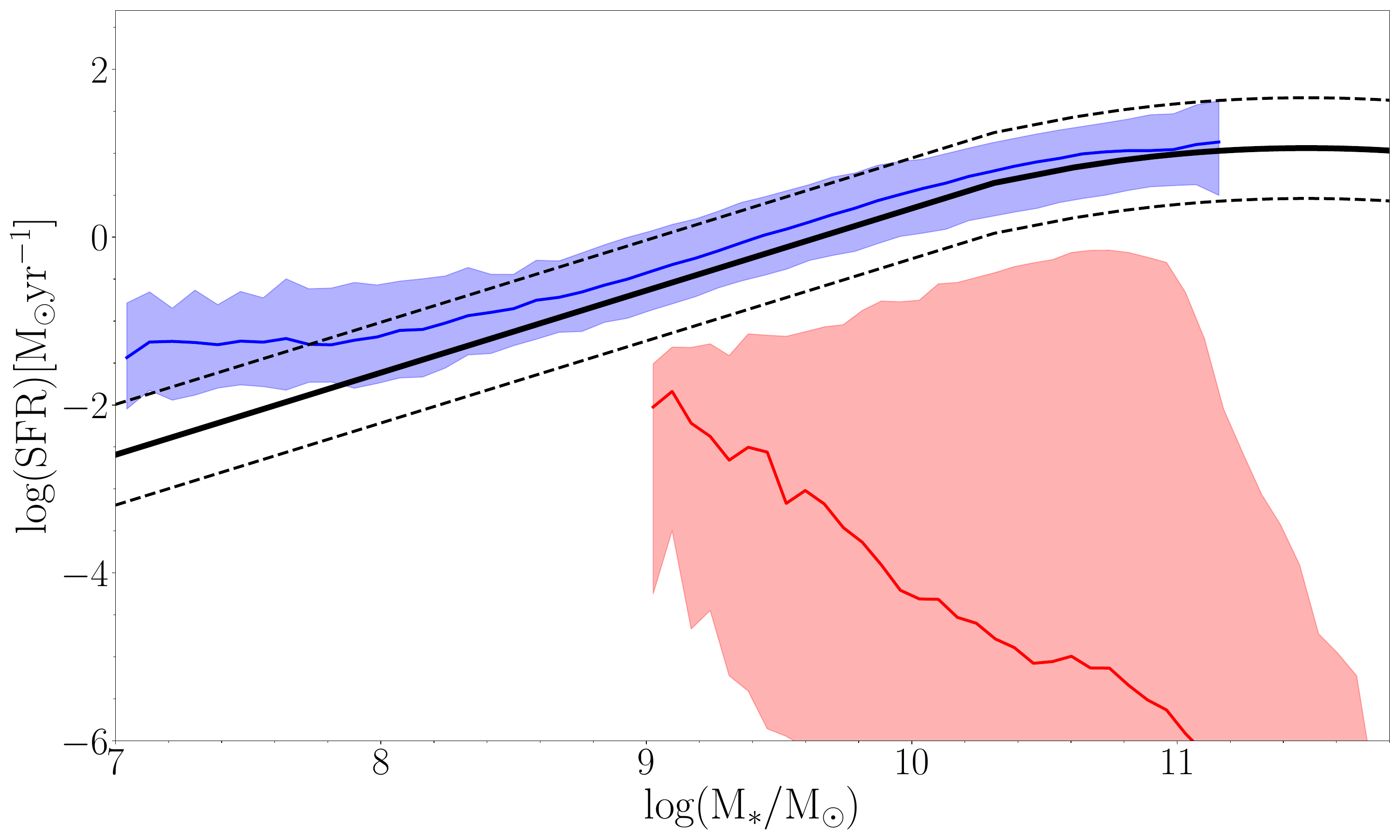}}
        \caption{$\rm M_{star}-SFR$ relation for {\tt BGS} galaxies. The MS at $\rm z\sim 0.24$ (median redshift of the {\tt BGS} sample) according to \cite {Schreiber2015} is shown with a black solid line, while dashed lines correspond to $\rm MS\pm 0.6$ dex to represent starburst and passive galaxies.  The running median and 16th-84th percentile range for red and blue galaxies selected based on the UVJ criterion are marked with red and blue, respectively. }
        \label{fig:MS}
\end{figure}

In Fig.~\ref{fig:MS}, the MS trend of star-forming galaxies observed at a median redshift of the {\tt BGS} sample ($\rm z\sim 0.24$; the median redshift of the {\tt BGS} sample) is reproduced following the prescription given by \cite{Schreiber2015}. 
Starburst and passive galaxies are commonly selected as galaxies that deviate for more than $\rm +0.6$ dex, and $\rm -0.6$ dex from the MS, respectively (e.g., \citealt{Elbaz2007, Rodighiero2011, Sargent2012, Whitaker2012, Buat2019, Donevski2020}). 
The lower limits distinguishing passive galaxies correspond to the transition between red and blue galaxies selected based on the UVJ diagram. 
The bulk of blue {\tt BGS} galaxies are located close to the MS, while the red {\tt BGS} are found under the MS. 

\section{Comparison to other catalogs}\label{sec:comparion}

In this section, we compare our estimates of the physical properties of DESI EDR galaxies with the COSMOS  \citep{Weaver2022},   AGN-COSMOS  (\citealt{Suh2019} and \citealt{Thorne2022}), and DEVILS ~\citep{Thorne2022} catalogs. In Appendix~\ref{sec:DESISTMassCatComparison}, we further compare our estimates with: i) SDSS and Extended Baryon Oscillation Spectroscopic Survey (eBOSS) Firefly VAC ({\tt SDSS(Firefly DR16)}; \citealt{Comparat2017}), ii) SDSS MPA-JHU DR8, after the Max Planck Institute for Astrophysics and Johns Hopkins University ({\tt SDSS(MPA-JHU)}; \citealt{Kauffmann2003b, Brinchmann2004,Tremonti2004}), iii) GALEX-SDSS-WISE Legacy Catalog X2\footnote{\url{https://salims.pages.iu.edu/gswlc/}} ({\tt GSWLC};~\citealt{Salim2016, Salim2018A}), and iv)  DESI VAC of the stellar masses and emission lines presented by~\citealt{Zou2024}. The summary of the comparison of stellar masses from different catalogs (including the ones described in  Appendix~\ref{sec:DESISTMassCatComparison} for simplicity) is presented in Table~\ref{tab:CatalogsComparioson}, using metrics described in Sect.~\ref{sec:metrics_comparison} to quantify the differences. There exist several other DESI VACs with stellar mass estimates, such as i) FastSpecFit Spectral Synthesis and Emission-Line Catalog {\tt FastSpecFit 3.2} (\citealt{Moustakas2023}, Moustakas et al. in prep) and ii) The DESI PRObabilistic Value-Added Bright Galaxy Survey catalog ({\tt PROVABGS}; ~\citealt{Hahn2023b}). {\tt Fastspecfit} is a stellar continuum and emission-line fitting code optimized to jointly model  DESI optical spectra and broadband photometry using physically motivated stellar continuum and emission-line templates. {\tt PROVABGS} also models jointly DESI spectroscopy and photometry using  state-of-the-art Bayesian approach and returns  full posterior distributions of the galaxy properties. We report the metrics for these reference catalogs in Table~\ref{tab:CatalogsComparioson}, however, a more detailed comparison is the subject of a future work. For comparison purposes, all the stellar mass estimates are recalculated, if necessary,  to the cosmology and IMF used to create our VAC. 

\subsection{Metrics}\label{sec:metrics_comparison}
To quantify the degree to which our estimates are different from the ones derived by other catalogs, we adopt the median difference ({\tt $\Delta$}) given as:
\begin{equation}
    \rm \Delta = median(x_{DESI} -  x_{ref}),
\end{equation}
where $\rm x_{DESI}$ and $\rm x_{ref}$ correspond to our logarithmic estimates and estimates in the reference catalog, respectively. 
This metric is followed by the normalized median absolute deviation ({\tt NMAD}) given as:
\begin{equation}
    \rm NMAD = 1.4826 \times median(|x_{DESI} - x_{ref}|).
\end{equation}
NMAD is approximately equivalent to the standard relative deviation, with a reduced impact from extremely outlying errors. 
Finally, we provide the Pearson correlation ({\tt r}).
The table with metrics for all the reference catalogs is provided in Table~\ref{tab:CatalogsComparioson}.

\begin{table}
\centering
        \caption{Comparison of stellar mass estimates between our catalog and reference catalogs. } 
        \label{tab:CatalogsComparioson}
        \footnotesize
        \begin{tabular}{r | r | r | r | r | r}
        catalog & {\tt N} & $\Delta$ & {\tt NMAD} & {\tt r} & {\tt err} \\
        \hline \hline
        {\scriptsize \tt COSMOS2020(FARMER)} &  1,899 &  0.001 &  0.176 &  0.96 & { 0.132} \\ 
        {\scriptsize \tt COSMOS2020(CLASSIC)} & 2,485 & {-0.014} & {0.156} & 0.97 & {0.125} \\
        \hline
        {\scriptsize \tt DEVILS} & 5,080 & -0.100 & 0.207 & 0.82 & 0.142 \\ 
        \hline
        {\scriptsize \tt SDSS(Firefly)} & 24,947 & 0.060 & 0.321 & 0.85 & 0.097 \\ 
        {\scriptsize \tt SDSS(MPA-JHU)} & 18,778 & -0.071 & 0.126 & 0.98 & 0.100 \\ 
        \hline
        {\scriptsize \tt GSWLC} & 17,902 & -0.191 & 0.284 & 0.97 & 0.100 \\ 
        \hline
        {\scriptsize \tt \citealt{Zou2024}} & 761,096 & -0.105 & 0.173 & 0.98 & 0.120 \\ 
        {\scriptsize \tt FastSpecFit 3.2} & 1,121,332 & -0.318 & 0.503 & 0.85 & 0.156 \\ 
        {\scriptsize \tt PROVABGS} & 215,123 & -0.393 & 0.582 & 0.89 & 0.116 \\     

\end{tabular}
\tablefoot{The median difference ($\Delta$), {\tt NMAD}  and Pearson coefficient ({\tt r}; see Sect.~\ref{sec:metrics_comparison} for definitions), and the median error of stellar masses ({\tt err}) for the given sample (with a {\tt N} number of sources) are provided.}
\end{table}

\subsection{COSMOS Catalog}\label{sec:COSMOS_comparison}
Here, we compare our stellar mass estimates with the ones from the {\tt COSMOS2020} catalog~\citep{Weaver2022}. 
The catalog includes sources down to $i\sim27$ observed over a 2~$\rm deg^2$ of the Cosmic Evolution Survey (COSMOS) field. 
The catalog comes in two independent versions: the {\tt CLASSIC}, based on the traditional aperture photometry performed on the PSF-homogenized images, with the exception of IRAC images \citep{Laigle2016}, and  {\tt FARMER}, which uses a new profile-fitting photometric extraction tool based on the tractor~\citep{Lang2016}.
The {\tt COSMOS2020} catalog provides a photo-z accuracy and outlier rate below 1\% for bright galaxies ($ i<22.5$). The photo-z accuracy and outlier rate degrade to $\rm \sim 4\%$, and $\rm \sim 20\%$, respectively, for the faintest galaxies ($ 25< i <27)$. 
The {\tt CLASSIC} version includes 1.7 million galaxies with photometry from optical to NIR, while the {\tt FARMER} version is limited to almost one million galaxies within the UltraVISTA footprint to provide {\it izYJHKs} images used to construct galaxy models. 
For both catalogs, stellar masses are derived with two independent tools: {\tt LePhare}~\citep{Arnouts2002, Ilbert2006} and {\tt Eazy}~\citep{Brammer2008}. 
In this analysis, we compare the {\tt FARMER} version limited to the {\tt LePhare} stellar mass estimates (\citealt{Weaver2022}; the statistical comparison of the stellar masses from {\tt CLASSIC} version is presented in Table~\ref{tab:CatalogsComparioson}). 
{\tt LePhare} is a SED fitting code that uses a set of templates generated using \cite{Bruzual2003} models and assuming a \cite{Chabrier2003} IMF. The SFH is described by an exponentially declining SFH and a delayed SFH assuming solar and half-solar metallicities. The dust attenuation is modeled with the \citealt{Calzetti2000} law and a curve with a slope $\rm \lambda^{0.9}$ (see App. A of \citealt{Arnouts2013}) with color excess limited to 0.7. The AGN templates are not incorporated. 

\begin{figure*}
\centerline{\includegraphics[width=0.99\textwidth]{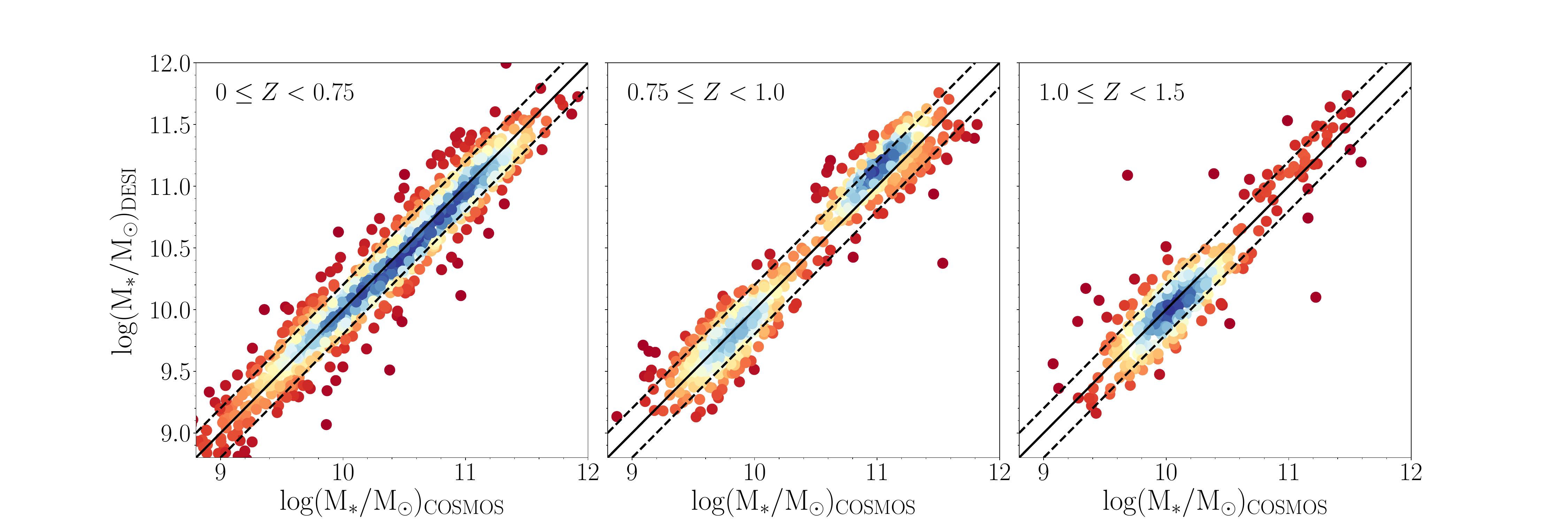}}
        \caption{Comparison of stellar mass estimates from our DESI VAC and {\tt COSMOS2020(FARMER}; \citealt{Weaver2022}) catalog {in three redshift bins}. The 1:1  and $\pm 0.2$ dex lines are marked with black solid and dashed lines, respectively. The linear fit (red line) and the Pearson correlation coefficient are reported in the legend.  }
        \label{fig:COSMOSmasses}
\end{figure*}

We selected {$\sim$ 2,000} galaxies with a high redshift accuracy ($\rm \delta z = | (z_{phot}- z_{spec})/(1+z_{spec})| < 0.01$\footnote{This cut ensures that the comparison between COSMOS2020’s photo-z-based physical parameters and DESI's spec-z-based physical parameters is not dominated by the redshift estimation difference.}) and a stellar mass error  of $\rm \mstar_{err} \le 0.25$ from {\tt COSMOS2020(FARMER)}. We found the negligible differences between our stellar masses and the ones from {\tt COSMOS2020(FARMER)} given by the {$\Delta$} = {0.001} dex and {\tt NMAD} = {0.176} (see Table~\ref{tab:CatalogsComparioson} and  Sect.~\ref{sec:metrics_comparison} for definitions for these metrics). 
The {$\Delta$} value is even smaller than the one found when comparing {\tt COSMOS(FARMER)} with \cite{Zou2024} ($\rm \Delta_{\mstar} = 0.08$). 
We note that such median differences are well below the median error of stellar mass estimates ($\rm 0.13$ dex; see Table~\ref{tab:CatalogsComparioson}).  Finding such a consistency between different codes, parametrization of the SED fitting codes, and different SED coverage confirms the robustness of the stellar mass estimates. 
The comparison of our stellar mass estimates with the ones from {\tt COSMOS2020(FARMER)} catalog {in three redshift bins} is shown in Fig.~\ref{fig:COSMOSmasses}. {There is no clear dependence on redshift, however at $\rm 0.75 \leq Z \leq 1.0$, the clear bimodal distribution in stellar mass reveals smaller offset ({$\Delta$} = -0.02 and {\tt NMAD} = 0.18) for low-mass galaxies ($\mstar \leq 10.5$) than for high-mass galaxies ($\mstar \ge 10.5$, {$\Delta$} = 0.12 and {\tt NMAD} = 0.22). }
A Pearson correlation coefficient found for the comparison of stellar masses from our catalog with the ones from {\tt COSMOS2020(FARMER)} ({\tt r} = 0.96) indicates a very strong positive linear relationship between the stellar mass estimates. 

\begin{figure*}
\centerline{\includegraphics[width=0.99\textwidth]{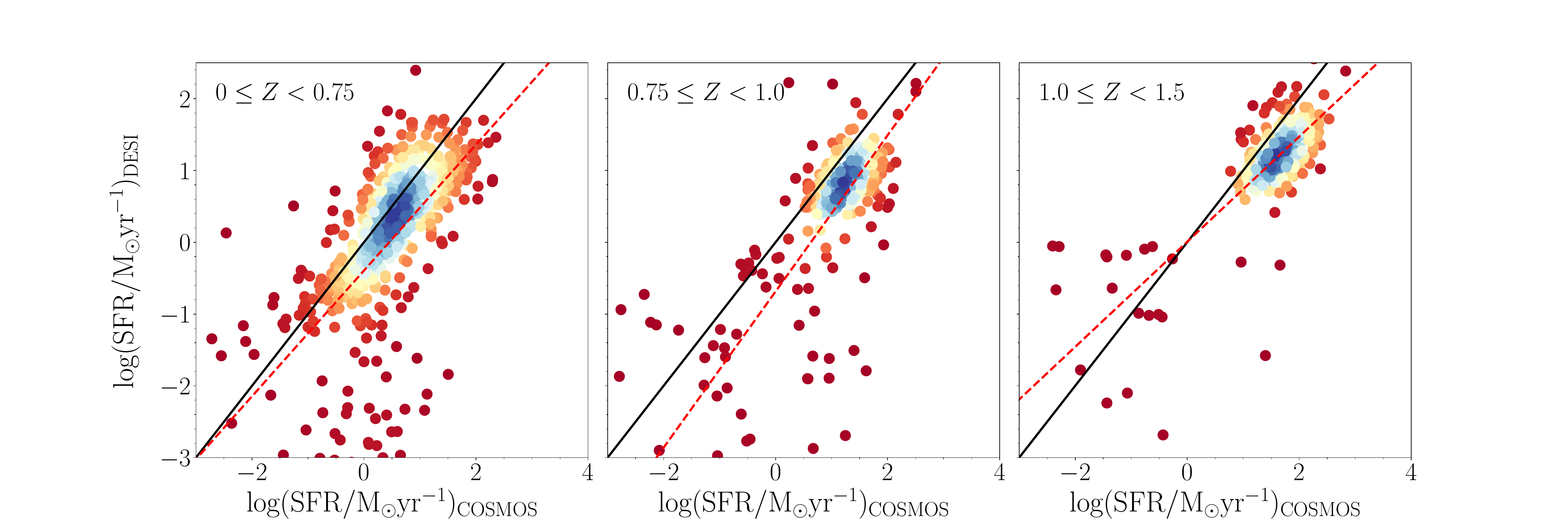}}
        \caption{Comparison of SFR estimates from our DESI VAC and {\tt COSMOS2020(FARMER}; \citealt{Weaver2022}) catalog {in three redshift bins}. The 1:1  and $\pm 0.2$ dex lines are marked with black solid and dashed lines, respectively. The linear fit (red line) and the Pearson correlation coefficient are reported in the legend.  }
        \label{fig:COSMOSFR}
\end{figure*}

We also compared the SFR estimates (see Fig.~\ref{fig:COSMOSFR}) finding  {$\Delta$} of {-0.37} and  {\tt NMAD} of {0.62} with a median error   $\rm log(SFR/M_{\odot}y^{-1})_{err}$ of {0.31}. The rest-frame magnitudes are also in good agreement, namely $\Delta$ is {0.05} and {\tt NMAD} = {0.18} for the rest-frame {\it r} magnitude.

\subsubsection{COSMOS AGN}\label{sec:agn_cosmos}

CIGALE returns several AGN properties such as the AGN fraction ({\tt AGNFRAC}), viewing angle ({\tt AGNPSY}), and AGN luminosity ({\tt AGNLUM}) defined as the total (disk, dust remitted, and scattered) luminosity of the AGN. 
We compared these properties with a sample of 754 X-ray AGN drawn from the {\tt Chandra-COSMOS Legacy Survey}~\citep{Suh2019}. 

\cite{Suh2019} obtained the AGN luminosities and stellar masses (among other properties) using the $\tt AGN_{FITTER}$ SED-fitting code~\citep{CalistroRivera2016} to model near $\rm UV - FIR$ SEDs of X-ray selected COSMOS AGNs. The AGN SEDs were decomposed into a nuclear torus, a host galaxy, and a starburst component with an additional component of the big blue bump template in the UV-optical range for BL AGN. The host galaxy models are generated from \cite{Bruzual2003} SSP assuming solar metallicity, \cite{Chabrier2003} IMF, and a simple exponentially declining SFH. 

\cite{Suh2019} provided stellar masses derived with $\tt AGN_{FITTER}$ that are $\sim 0.15$ dex lower than our estimates with {\tt CIGALE}. A similar underestimation of $\tt AGN_{FITTER}$ stellar masses was found by \cite{Thorne2022} when comparing $\tt AGN_{FITTER}$ stellar masses to DEVILS stellar mass estimates derived with {\tt PROSPECT}~\citep{Davies2021, Thorne2021, Thorne2022, Thorne2022b}. \cite{Thorne2022} explained that this discrepancy is likely driven by the simplification of the host galaxy models (in the prescription of the SFH, metallicity, and dust) by the $\tt AGN_{FITTER}$, which is focused on recovering AGN properties. On the other hand, SED fitting codes such as {\tt PROSPECT} and {CIGALE} are able to recreate the more complex nature of host galaxies.

\begin{figure*}
        \centerline{\includegraphics[width=0.99\textwidth]{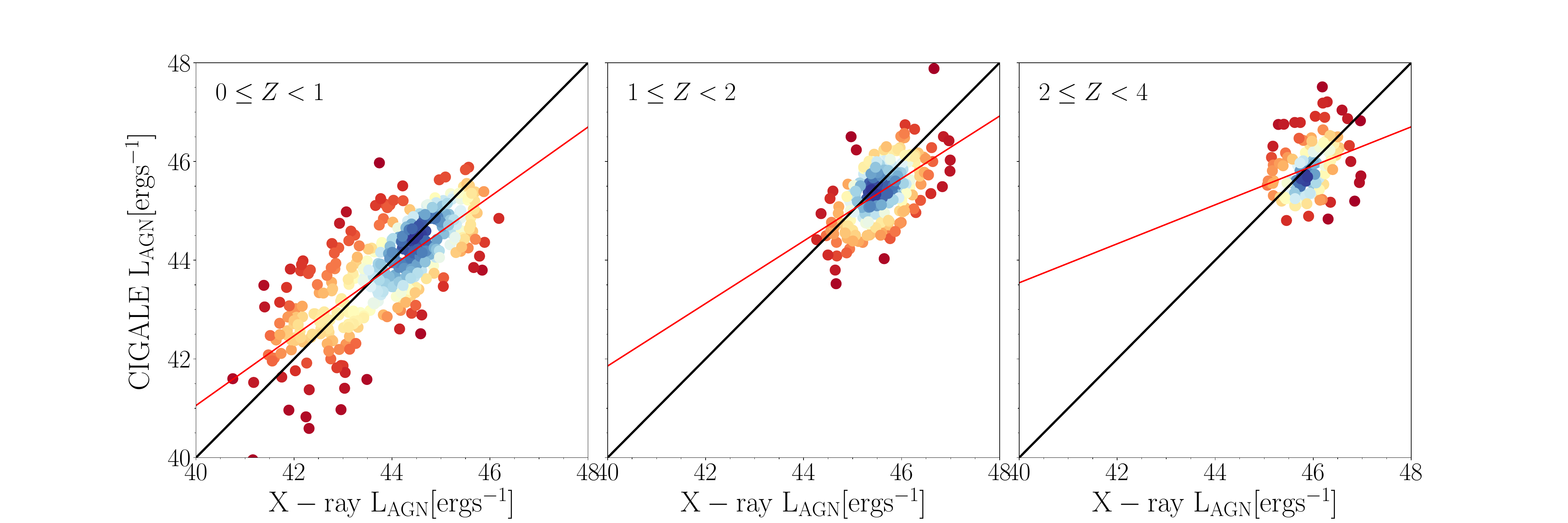}}
        \caption{Comparison of the bolometric AGN luminosity derived from the \textit{Chandra} hard ($\rm 2 - 7$  keV band) X-ray luminosities of~\cite{Suh2019} with those derived with {\tt CIGALE} for a sample of 754 AGNs {in three redshift bins}. The 1:1 correlation is marked with a black solid line. The linear fit (red line) and the Pearson correlation coefficient are reported in the legend.  }
        \label{fig:COSMOSagnluminosity}
\end{figure*}

The AGN luminosities in our catalog stay in good agreement with that estimated by~\cite{Suh2019} using $\tt AGN_{FITTER}$, with a $\Delta = 0.059$ and {\tt NMAD} of 0.378 despite the differences in the SED modeling. We also find close agreement with the bolometric AGN luminosity derived from the \textit{Chandra} hard ($\rm 2 - 7$  keV band) X-ray luminosities (see Fig.~~\ref{fig:COSMOSagnluminosity}) characterized by a $\Delta =-0.106$, {\tt NMAD}  of 0.648 and Pearson correlation of 0.85 {independently of the redshift}.
 
\cite{Suh2019} distinguished the BL/unobscured and NL/obscured AGN types predetermined based on the optical properties (i.e., based on the presence of the BL or NL in their spectra) or their photometric SED (i.e., whether is best fitted by an unobscured or obscured AGN template; see details in  \citealt{Marchesi2016} and \citealt{Suh2019}). We use this information to validate the {\tt AGNFRAC} and {\tt AGNPSY} derived with {\tt CIGALE}. We find that 423 BL/unobscured AGN are characterized by higher AGN fraction (with a median {\tt AGNFRAC} = 0.36) and larger viewing angle ({\tt AGNPSY} = 67$^{\circ}$) than 397 NL/obscured AGN (with a median {\tt AGNFRAC} = 0.27 and {\tt AGNPSY} = 39$^{\circ}$). Only $\rm \sim 40\%$ of BL/unobscured and NL/obscured AGN  are observed with at least 2 MIR bands with S/N > 3 suggesting that {\tt AGNFRAC} and {\tt AGNPSY} have the potential of discriminating between NL and BL AGN based on the {\tt CIGALE} estimates even in the absence of the MIR information (see also Sect.~\ref{sec:AGNFeaturesCompReprSample}). A more detailed discussion about AGN classification based on {\tt CIGALE} is presented in Siudek et al{. (under DESI Collaboration review).}

\subsubsection{DEVILS catalog}\label{sec:devils_cosmos}

\cite{Thorne2022} derived physical properties including stellar masses, SFR and AGN luminosities for $\sim 500,000$ DEVILS galaxies observed in the COSMOS field in the $\rm FUV-FIR$ using the {\tt PROSPECT} SED fitting code incorporating AGN templates from~\cite{Fritz2006} and flexible star formation and metallicity parameters (for details, see \citealt{Thorne2022} and also~\citealt{Thorne2021, Thorne2022b}). {\tt PROSPECT} identified $91\%$ of BPT-selected AGNs and derived AGN luminosities in close agreement with the luminosities derived from Chandra X-ray (\citealt{Marchesi2016}). The AGN identification based on the {\tt PROSPECT} code is based on the AGN fraction, requiring {\tt AGNFRAC} > 0.1 (see also Table 2 in \citealt{Thorne2022} for a comparison of {\tt PROSPECT} AGN identification with standardly used techniques). 
We found {\tt PROSPECT} counterparts for 5080 galaxies and we restricted our comparison to the sample of 1063 AGN and 2313 non-AGN galaxies with FIR photometry ({\tt FIRINPUT = 1}\footnote{\cite{Thorne2022} found that for sources lacking FIR photometry, the AGN fraction is high ({\tt FRACAGN} > 0.8) and {\tt PROSPECT} is not able to resolve properties of the host galaxy.}) and an AGN fraction of {\tt AGNFRAC} > 0.1 and $\le 0.1$, respectively. We find the median difference on the stellar mass estimates to be on the level $\Delta \sim -0.07$ and {\tt NMAD} $\sim 0.17$ for both AGN and non-AGN host galaxies. 
Further restricting the comparison to 386 AGNs with WISE photometry ({\tt FLAGINFRARED} $\ge 3$), we find the same AGN fraction ($\tt AGNFRACTION_{DEVILS}$ = 0.27 and $\tt AGNFRACTION_{DESI}$ = 0.26). This further highlights the independence of the estimates on the code and prescription used for deriving properties of galaxies.

\section{Physical properties: model and photometry dependence}\label{sec:DependenceofPhysProp}

The SED fitting technique introduces several systematic uncertainties in derived properties coming from the assumptions made about the models, which may lead to a disagreement in the estimated stellar masses of up to a factor of $\sim 2$ (e.g., \citealt{Maraston2006,Kannappan2007, Conroy2013, Lower2020,Pacifici2023}). In this section, we consider how our choices of the SED modules (see Table~\ref{tab:SEDParameters}) influence the main physical properties (stellar masses and SFRs). 
We validate the impact of the number of assumptions made about the IMF, fixed metallicity, SSP models, SFH prescription, dust attenuation and emission laws, and AGN models (see Table~\ref{tab:SEDParametersAlternatives} for a list of the changed parameters given in Appendix~\ref{app:SEDParametersAlternatives}). To quantify the effect, we made only one change, namely, in a parameter related to the default configuration (see Table~\ref{tab:SEDParameters}). 
To discriminate the impact of the incorporation of the WISE photometry, we compared the physical properties estimated with or without WISE photometry or only with {\it W1} and {\it W2} bands in Sect.~\ref{sec:DependenceonPhootmetry}. 
Finally, we compared the {\tt CIGALE} derived SFR with the ones estimated from the emission line measurement in Sect.~\ref{sec:sec:SFRSpectra}. 
To preserve the computational time, we validated the influence of the model assumptions on a smaller sample of $\rm \sim50,000$ galaxies representing seven main galaxy classes: 1,750 BL AGNs, 3,900 NL AGNs, 8,393 composite, 8,819 star-forming, 8,526 passive, 8,846 retired, and 9,962 other galaxies. The selection of this representative sample is described in  Appendix~\ref{sec:RepresentativeSample}.

\subsection{Stellar mass: choice of the stellar components}\label{sec:IMFchoice}

Stellar evolution models, such as \cite{Bruzual2003} or \cite{Maraston2005}, under the assumption of the IMF, describe the evolution of the SSP as a function of their stellar ages and metallicities following some fiducial SFH. In this section, we consider the dependence of the stellar masses on the choice of the i) IMF, ii) SSP models, iii) metallicity, and iv) SFH prescription. 

We considered three main IMF choices: i) \cite{Salpeter1955} assuming a power-law distribution with a heavier slope towards high-mass stars, ii) \cite{Kroupa2001} assuming a broken power-law IMF with a shallower slope at higher stellar masses, implying fewer massive stars compared to \cite{Salpeter1955}, iii) \cite{Chabrier2003}, which combines a log-normal distribution for low-mass stars and a power-law distribution for higher masses and thus is considered as a reasonable middle ground between \cite{Salpeter1955} and \cite{Kroupa2001}.  The \cite{Chabrier2003} IMF typically results in lower stellar mass estimates compared to the \cite{Salpeter1955} IMF, because the \cite{Chabrier2003} IMF reduces the contribution of high-mass stars, which are more massive and have a greater contribution  to the total stellar mass. To mitigate the difference in stellar mass estimates we commonly assume a conversion factor between \cite{Chabrier2003} and other IMFs, namely:
\begin{equation} 
\begin{split}\rm
  M_{\star}\_Salpeter = 1.7 \cdot M_{\star}\_Chabrier, \\ 
 \rm M_{\star}\_Kroupa= 1.1 \cdot M_{\star}\_Chabrier,  
\end{split} 
\end{equation}
as shown by, for instance, \cite{Cimatti2008}, \cite{Longhetti2009}, \cite{Bolzonella2010}, and \cite{Ilbert2010}. Here, we take the opportunity, to assess whether a simple scaling is sufficient and whether it varies for different types of galaxies. 
We find a constant median difference of -0.24 dex (see Sect.~\ref{sec:metrics_comparison} for the definition of the median difference and Table~\ref{tab:StMassComparison} for the quantitative comparison of the estimates with different model assumptions) between stellar masses obtained assuming \cite{Chabrier2003} and \cite{Salpeter1955} stellar masses. This median difference is in agreement with the commonly used conversion factor and is independent of the stellar mass and galaxy class. 

 To generate the DESI VAC, we relied on the \cite{Bruzual2003} models. We discriminate the degree to which our estimates are different from the ones derived under the assumption of the \cite{Maraston2005} model. The main difference between \cite{Maraston2005} and \cite{Bruzual2003} relies on the incorporation of the thermally pulsating asymptotic giant branch (TP-AGB) stars in the evolution of galaxies in \cite{Maraston2005} models. TP-AGB stars are evolved stars that can significantly affect the integrated light of galaxies, particularly in the NIR wavelength range. Due to the inclusion of TP-AGB stars, the \cite{Maraston2005} model might yield higher stellar mass estimates compared to \cite{Bruzual2003} for galaxies with significant contributions from these stars, especially at intermediate and old ages (see more details in e.g., \citealt{Maraston2005, Conroy2010, Kriek2010, Maraston2011}). While both models (BC03 and M05) include remnants, the exact mass contributions from remnants might not differ drastically between the two models. However, any differences would stem from the specific stellar evolution prescriptions and the IMF adopted in each model.  The way these effects are included are not the same even for a simple \citealt{Salpeter1955} IMF~\citep{Maraston2005} and some offsets just from that cannot be excluded~\citep[e.g.,][]{Maraston2013}. Typically, remnants constitute a relatively small fraction of the total stellar mass, so while there might be differences, they should not be as pronounced as the differences arising from the TP-AGB treatment.  We note that recent works using {\it James Webb} Telescope spectra reports the spectroscopic detection of the TP-AGB in galaxies at high-z and models with little contribution from that phase do not fit the data well~\citep{Lu2024}. Different works have reported an offset of $\sim 0.15$ dex between stellar masses of \cite{Bruzual2003} and \citealt{Maraston2005} \citep[e.g., ][]{Ilbert2010, Pozzetti2010, Kawinwanichakij2020}. 
The change of the SSP model to \cite{Maraston2005} introduces a median difference of -0.26 dex in our stellar mass estimates with respect to the ones derived with \cite{Bruzual2003} models. We refer to Sect.~\ref{sec:metrics_comparison} for the definition of the median difference and Table~\ref{tab:StMassComparison} for the quantitative comparison of the estimates with different model assumptions. 

Typically, physical properties are derived relying on the parametric SFHs, although they may suffer from strong biases in recovering the proper SFHs due to the assumption of the simplistic prescription (e.g., \citealt{Ciesla2015,Carnall2019,Lower2020,Leja2022}). As a response to the necessity of more advanced SFH prescription, non-parametric SFHs have been proposed (e.g., \citealt{Leja2019, Lower2020,Ciesla2023}). We verified how our stellar mass estimates change when using a non-parametric SFH module ({\tt sfhNlevels}; see  Table~\ref{tab:SEDParametersAlternatives}) implemented in {\tt CIGALE} \citep{Ciesla2023}. The formula of the non-parametric SFH is based on time bins in which the SFRs are constant and linked together by the {\tt bursty continuity} \citep{Tacchella2022}, instead of the assumption of the analytical function. We refer the reader to \cite{ArangoToro2023, Ciesla2023a, Ciesla2023} for more details about the non-parametric SFH module. For DESI galaxies, the SED fitting of the optical-MIR photometry with non-parametric SFH has a negligible impact on the stellar mass estimates across the entire stellar mass range (with a median difference of 0.03). The median difference is two times higher for AGNs and star-forming galaxies ($0.04-0.05$) than for passive and retired galaxies (0.02; see Table~\ref{tab:StMassComparison}). 

Stellar metallicity is poorly constrained from photometric data alone  due to the age-metallicity-dust degeneracy~\citep[e.g.,][]{Worthey1994, Papovich2001}. To overcome this problem, most commonly SED fitting-based approaches rely on fixing the metallictiy to a solar value~\citep[e.g.,][]{Malek2018,Boquien2019} or leaving it as a free but constant value over the  lifetime of the galaxy~\citep[e.g.,][]{Carnall2018,Johnson2021}. These assumptions may affect  
other parameters of interest, such as the stellar mass or SFRs introducing mass-dependent systematics~\citep[e.g.,][]{Pforr2012,Mitchell2013,Thorne2022b}. On the other hand, other works (e.g.,~\citealt{Osborne2024}) do not report the dependence of the stellar mass estimates on the choice of metallicity. We also validated the impact on the stellar mass estimates with models where the metallicity is allowed to vary (see Table~\ref{tab:SEDParametersAlternatives}), instead of assuming the metallicity fixed to a solar value. For our catalog, allowing metallicity to vary introduces a median difference of -0.04 across the entire stellar mass range. The median difference is two times higher for NL AGNs and composite ($-0.07$) than for the remaining galaxy classes ($-0.03$; see Table~\ref{tab:StMassComparison}). 

\begin{table*}
\centering
        \caption{Comparison of stellar mass estimates between our catalog and the ones derived by changing one of the model descriptions outlined in Sect.~\ref{sec:DependenceofPhysProp}. }
        \label{tab:StMassComparison}
        \footnotesize
        \begin{tabular}{c | r | r  | r | r | r | r}       
        module & $\Delta$ & {\tt NMAD} & $\Delta$ & {\tt NMAD} & $\Delta$ & {\tt NMAD} \\
         & \multicolumn{2}{c}{\tt All} & \multicolumn{2}{c}{\tt StarForm}  & \multicolumn{2}{c}{\tt AGN}\\
        \hline \hline
        {\tt IMF} (\citealt{Chabrier2003} vs. \citealt{Salpeter1955})&  -0.244 & 0.362 & -0.248 & 0.367 & -0.246 & 0.364\\ 
        {\tt SSP} (\citealt{Bruzual2003} vs. \citealt{Maraston2005}) &   -0.258 & 0.382 & -0.235 & 0.348 & -0.262 & 0.388\\ 
        {\tt SFH} (delayed with extra burst vs. non parametric) & 0.027 & 0.066 & 0.039 & 0.094 & 0.044 & 0.084 \\ 
        {\tt Z} (metallicity fixed to solar vs. variable) &  -0.039 & 0.076 & -0.040 & 0.073 & -0.046 & 0.101 \\ 
        \hline
        {\tt DustAtt} (\citealt{Calzetti2000} vs. \citealt{CharlotFall2000}) &  -0.031 & 0.063 & -0.024 & 0.063 & -0.109 & 0.167 \\ 
        {\tt DustEm} (\citealt{Draine2014} vs. \citealt{Dale2014}) &  -0.000 & 0.018 & -0.000 & 0.028 & -0.000 & 0.049 \\ 
        \hline
        {\tt AGN} (\citealt{Fritz2006} vs. no AGN) & 0.004 & 0.018 & 0.007 & 0.024 & -0.001 & 0.045 \\ 
        {\tt AGN} (\citealt{Fritz2006} vs. \citealt{Stalevski2012, Stalevski2016}) & 0.001 & 0.006 & 0.003 & 0.007 & 0.005 & 0.030\\ 
       \hline
        {\tt SEDs} ($grzW14$ vs. $grzW12$) &  -0.002 & 0.055 & -0.009 & 0.071 & -0.013 & 0.187\\ 
        {\tt SEDs} ($grzW14$ vs. $grz$) &  0.036 & 0.166 & 0.067 & 0.190 & 0.044 & 0.212\\ 

\end{tabular}
\tablefoot{The median difference ($\Delta$), and {\tt NMAD} (see Sect.~\ref{sec:metrics_comparison} for definitions) are reported for the given change in the model considering the change in the initial mass function ({\tt IMF}), single stellar population models ({\tt SSP}), star formation history prescription ({\tt SFH}), metallicity ({\tt Z}), dust attenuation law ({\tt DustAtt}), dust emission model ({\tt DustEm}), and AGN models ({\tt AGN}).  The influence of the presence of MIR information is also reported. The metrics are derived for the entire representative sample ({\tt All}) and separately for star-forming galaxies ({\tt StarForm}) and {\tt AGN} (including both NLs and BL AGNs). }
\end{table*}

\subsection{Stellar mass: Choice of the dust models}\label{sec:Dustchoice}

We considered an alternate dust attenuation model proposed by \citealt{CharlotFall2000} (see  Table~\ref{tab:SEDParametersAlternatives}) to test the systematics.  
In contrast to \cite{Calzetti2000}, this recipe assumes different attenuation for young ($\rm age<10$ Myr) and old stars ($\rm age>10$ Myr). 
Young stars are attenuated in the birth clouds, while both young and old stars are attenuated in the interstellar medium. In {\tt CIGALE}, the implementation of the \cite{CharlotFall2000} law is more flexible, giving the freedom to choose the values of input parameters (attenuation of the ISM, slopes of power-law attenuation curves for the birth cloud, and the ISM, and the ratio of the total attenuation).
Both attenuation laws are modeled by a power law and normalized to the attenuation in the V band. The main difference in the shape of attenuation curves appears at $\lambda>5000\AA$, where \cite{CharlotFall2000} is flatter than the one given by \cite{Calzetti2000}. For example, \cite{Mitchell2013} estimated that the stellar mass can be underestimated by up to 0.6 dex by assuming the \cite{Calzetti2000} for massive galaxies. 
For DESI galaxies we find a median difference of $-0.03$ dex across the entire stellar mass range, but the median difference is higher for AGN and composite galaxies ($\Delta\sim -0.10$) than for the remaining classes ($\Delta \sim -0.02$; see Table~\ref{tab:StMassComparison}). 

We also considered the prescription of the dust emission model given by \citealt{Dale2014} (see  Table~\ref{tab:SEDParametersAlternatives}), which is much simpler than the complex model of \cite{Draine2014}. 
The star-forming component is described by $\rm dM_{d}\propto U^{-\alpha}dU,$ where $\rm M_{d}$ is the dust mass heated by the radiation field at intensity $\rm U$ and $\alpha$ represents the relative contributions of the different local SEDs~\citep{Dale2014}.
The parameter $\alpha$ is the only free parameter and is tightly connected with the $\rm 60-to-100$ $\mu m$ color. 
Due to a limited variation of the PAH with respect to $\alpha$, the model does have problems with proper modeling of the dust in metal-poor galaxies (e.g., \citealt{Engelbracht2005}). However, when using only optical-MIR SEDs, there is no difference in the used dust emission model, as indicated by the zero median difference in the stellar mass estimates using \cite{Dale2014} and \cite{Draine2014} prescriptions (see Sect.~\ref{sec:metrics_comparison} for the definition of the median difference and Table~\ref{tab:StMassComparison} for the quantitative comparison of the estimates with different model assumptions).

\subsection{Choice of the AGN models}\label{sec:AGNchoice}
Finally, we considered the impact of incorporating AGN templates on the stellar mass estimates by comparing the estimates from our VAC with estimates from {\tt CIGALE} when the AGN contribution is fixed to 0 for the representative sample. As suggested by~\cite{Thorne2022}, the inclusion of the AGN component is argued to be crucial not to overestimate the light coming from the host galaxies. However, our results suggest that the stellar masses for the general sample of galaxies are not affected significantly by the contribution of AGN as the median difference is low ($\Delta$ = 0.004 for the entire representative sample). {On the other hand, the median difference is high for BL AGN ($\Delta = -0.26$; see Fig.~\ref{fig:AGNsubtypesMass}). This implies that the incorporation of the AGN models affects the stellar mass estimates (the median difference is slightly higher than the median stellar mass error; $\mstar_{err} = 0.23$) for the BL AGNs; whereas for the remaining galaxy classes, this effect is negligible from the statistical point of view (i.e., the scatter shown in Fig.~\ref{fig:AGNsubtypesMass} is large if we consider the individual sources). 
{We note that the difference in stellar mass for BL AGNs depends on redshift and stellar mass (see Fig.~\ref{fig:AGNBLAGNMass}), while for the remaining classes, the stellar mass differences are independent of redshift and considered stellar mass range. }

The change of the AGN model to the one proposed by~\citealt{Stalevski2012, Stalevski2016} (for the description of AGN models; see Sect.~\ref{sec:AGNComponentFritz}) does not affect the stellar mass estimates for the entire galaxy population (see Table~\ref{tab:StMassComparison}) and the difference is the highest for the BL AGNs (with a median difference of 0.05 and NMAD of 0.14). However, we note that the scatter for individual galaxies is large;  for individual sources, the masses might be higher even for $2-3$ dex when using \cite{Fritz2006} models than when relying on \cite{Stalevski2012, Stalevski2016}.   

 \begin{figure*}
\sidecaption
  \includegraphics[width=12cm]{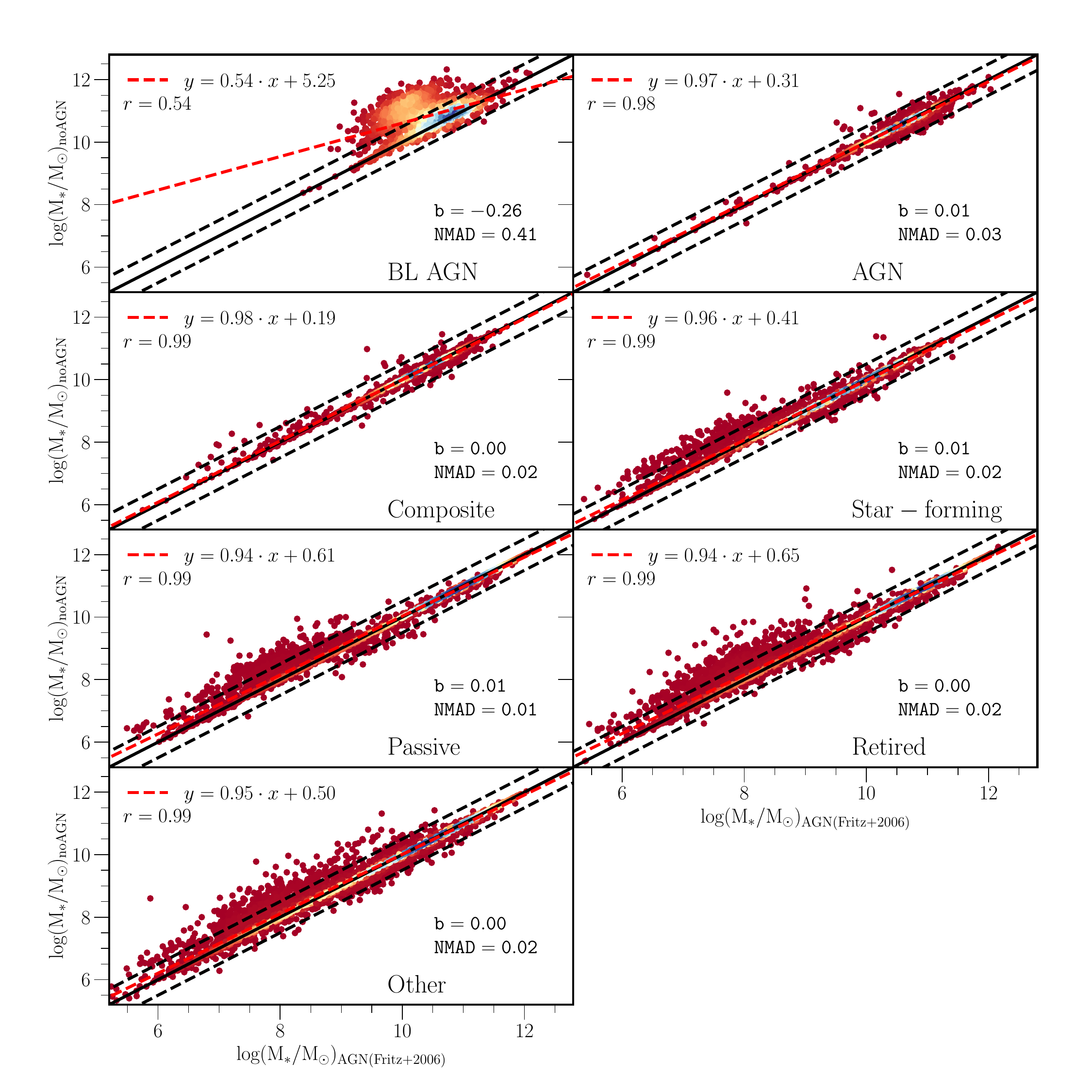}
        \caption{Comparison of stellar mass estimates with or without incorporation of the AGN templates in SED fitting framework for the representative sample composed of seven different galaxy classes. The 1:1  and $\pm 0.5$ dex lines are marked with black solid and dashed lines, respectively. The linear fit (red line) and the Pearson correlation coefficient (r) are reported in the legend. The median difference ($\Delta$) and {\tt NMAD} are reported on the plots for each galaxy class (see Sect.~\ref{sec:metrics_comparison} for the definition of the metric). }    \label{fig:AGNsubtypesMass}
\end{figure*}

\begin{figure}
\centerline{\includegraphics[width=0.49\textwidth]{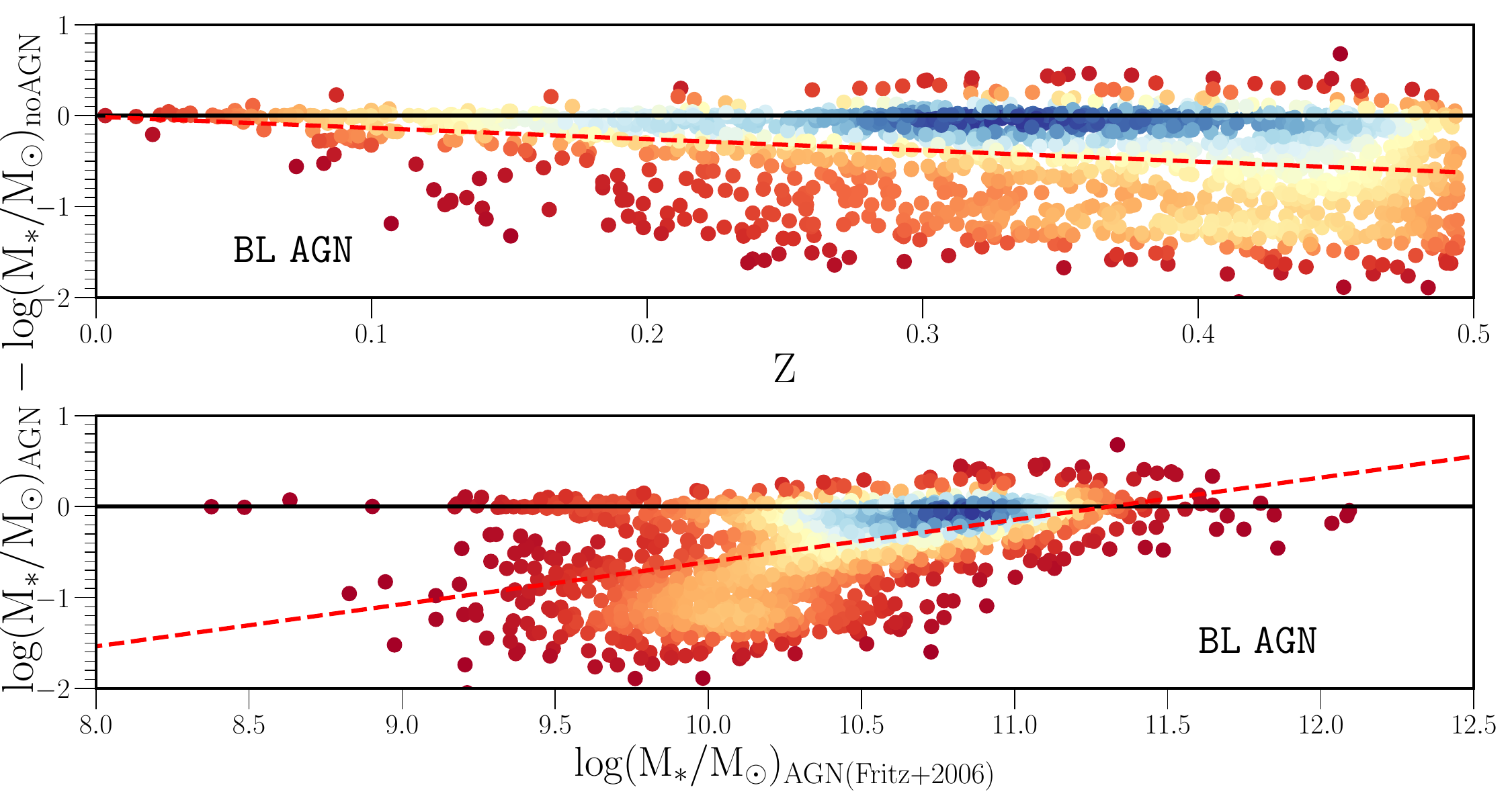}}
        \caption{{Stellar mass difference between estimates obtained with and without incorporating AGN templates in the SED fitting for BL AGNs. The top panel shows the mass difference as a function of redshift, while the bottom panel displays it as a function of stellar mass. The black solid line represents zero mass difference, and the red dashed line corresponds to a linear fit to the data.} }
        \label{fig:AGNBLAGNMass}
\end{figure} 

\subsection{Stellar mass: choice of the photometry}\label{sec:DependenceonPhootmetry}

Aside from the model assumptions discussed in the previous section, the SED coverage is one of the most important ingredients for deciding on the reliability of the SED-derived physical properties. The coverage of the $\rm FUV-FIR$ is highly desired to obtain reliable estimates of the contribution from young and old stars and AGNs (e.g., ~\citealt{Thorne2022, Thorne2022b}). However, numerous works suggest that robust and reliable stellar mass estimates require only optical photometry to find the tight relation between optical color and stellar M/L ratio (e.g., \citealt{Bell2001, Bell2003, Gallazzi2009, Zibetti2009, Taylor2010}). \cite{Gallazzi2009} showed that stellar mass estimates obtained on one optical color are not biased against the estimates based on the optical-NIR SEDs or spectral features. However, \citealt{Gallazzi2009}  based their estimates on the BC03 models that are NIR featureless.  Other works instead find that only optical + NIR photometry can break the age, metallicity, and dust degeneracy~\citep{Maraston2010}. 

In this section, we compare the estimates of stellar mass for DESI galaxies assessing further in the wavelength, namely, using optical colors alone ({\it grz}) with the ones obtained based on the optical-MIR SED fit, adding i) only WISE1 and WISE2 ($grzW12$) and all four WISE photometry ($grzW14$). 
We find a negligible impact when incorporating $W12$ to the SED fit on the stellar mass estimates ($\Delta = - 0.002$; see Table~\ref{tab:StMassComparison}) for the entire galaxy population. The discrepancy between stellar mass estimates increases for AGNs ( $\Delta$ = 0.13 and -0.06 for BL and NL AGNs, respectively, and NMAD 0.26  and 0.16 for BL, and NL AGNs, respectively). Independently of the galaxy class, the scatter is large $\pm 1$ dex, thus the stellar masses for individual sources may differ significantly. 
The stellar masses obtained only based on {\it grz} photometry are still similar to the ones derived based on the {\it grzW14} SED fits ($\Delta$ = 0.04), but the scatter increases ({\tt NMAD} = 0.17; see Table~\ref{tab:StMassComparison}). 
The degree of difference in stellar mass estimates depends on the galaxy types, and we find that this is higher for the BL AGNs (with $\Delta$ = 0.20 and {\tt NMAD} = 0.34) than for the remaining classes  (see Fig.~\ref{fig:WISEsubtypesMass}). 
{The difference in the stellar mass is independent of redshift except for the low redshift ($\rm Z \lesssim 0.1$), where the stellar mass estimated with WISE photometry are higher than the ones estimated only based on the optical bands (see Fig.~\ref{fig:WISEsubtypesMassDifference}). Independently of redshift, the} stellar masses can differ up to $\sim 1$ dex for AGNs and composite galaxies -- and even up to $\sim 2$ dex for star-forming and passive galaxies. 

\begin{figure*}
\sidecaption
\includegraphics[width=12cm]{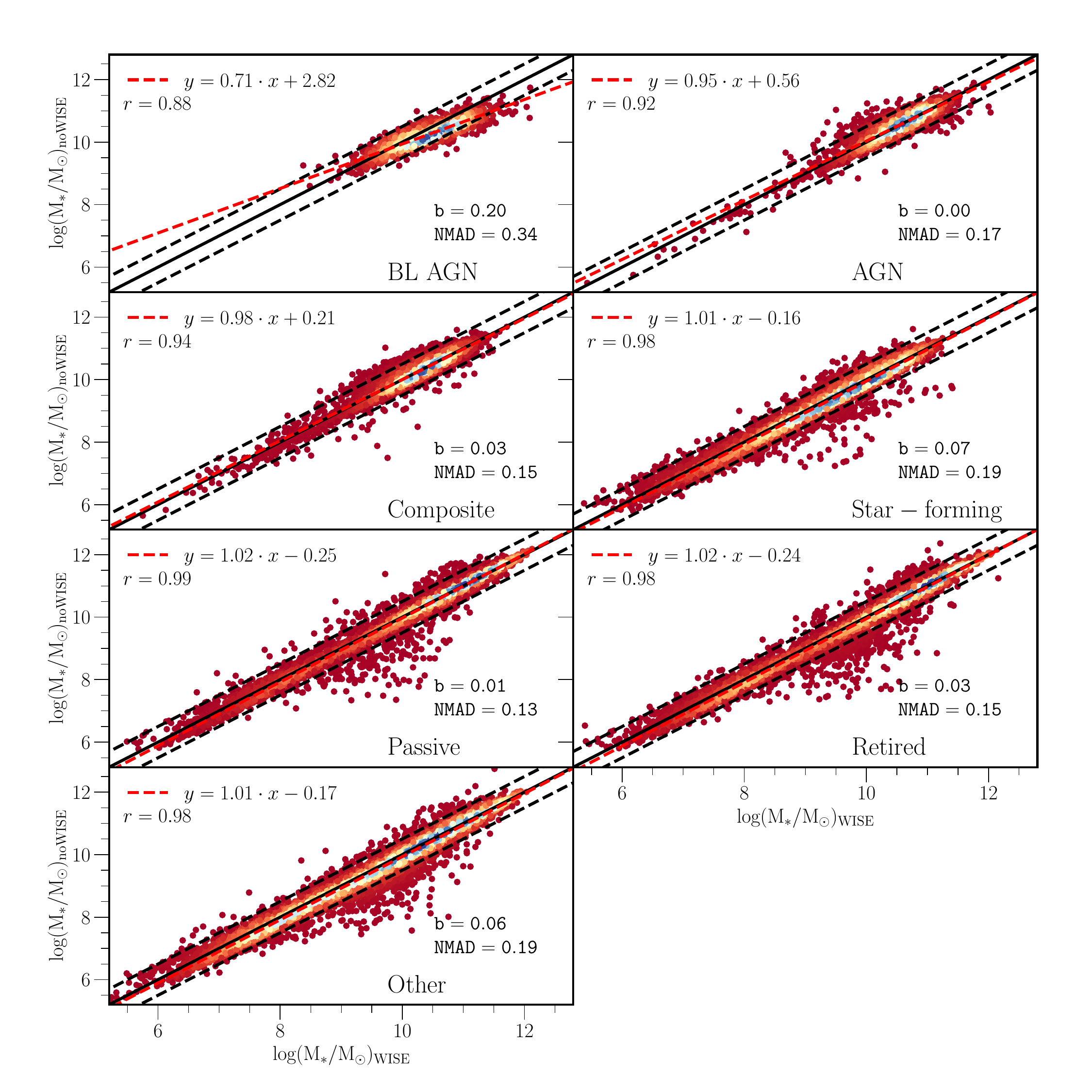}
        \caption{Comparison of stellar mass estimates with or without incorporation of the MIR photometry to the SED fitting for the representative sample composed of seven different galaxy classes. The 1:1  and $\pm 0.5$ dex lines are marked with black solid and dashed lines, respectively. The linear fit (red line) and the Pearson correlation coefficient (r) are reported in the legend. The median difference ($\Delta$) and {\tt NMAD} (see Sect.~\ref{sec:metrics_comparison} for the definition of the metric) are reported on the plots for each galaxy class: BL and NL AGNs, composite objects, star-forming galaxies, passive and retired galaxies, and other sources. }
        \label{fig:WISEsubtypesMass}
\end{figure*} 

\begin{figure*}
\sidecaption
\includegraphics[width=12cm]{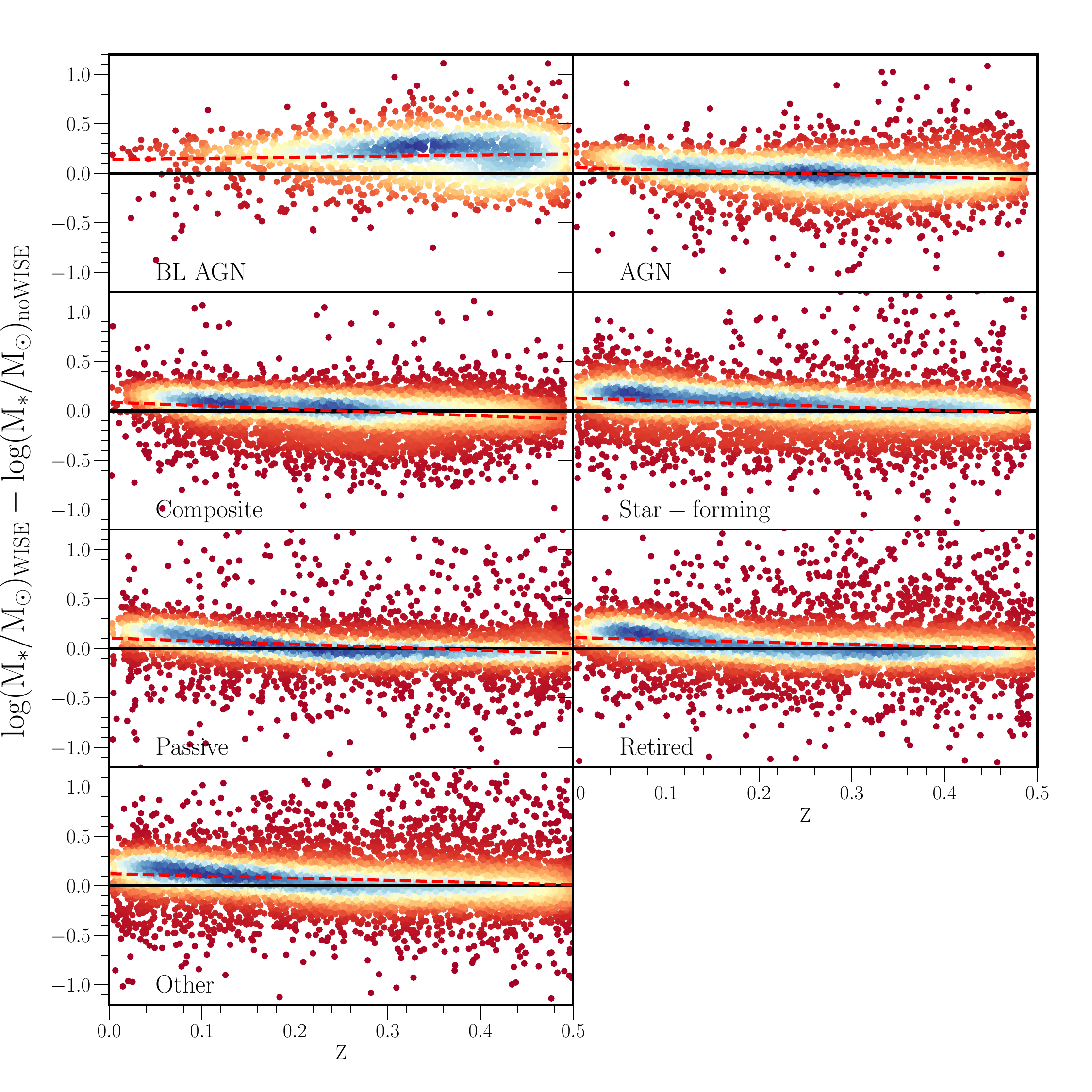}
\caption{{Stellar mass difference between estimates obtained with and without incorporating WISE photometry in the SED fitting for seven different classes. Each panel shows the mass difference as a function of redshift with the black solid line representing zero mass difference and the red dashed line corresponding to a linear fit to the data.} }
        \label{fig:WISEsubtypesMassDifference}
\end{figure*} 

\subsection{SFR: Choice of the models and photometry}\label{sec:sec:SFRSpectra}

In this section, we validate how the SFRs are affected by the aforementioned choices in the SED framework. The SFR derived via SED fitting (SFR(SED)) depends on the wavelength coverage of the SED fits (e.g., the lack of FIR observations may alter its estimations~\citep[]{Ciesla2015}. 
Thus, we used an alternative SFR indicator based on the H$\alpha$ line luminosity ($\rm L(H\alpha)$), following the definition given by~\cite{Kennicutt1998}:
\begin{equation}\label{eq:sfr}
 \rm{   SFR(H\alpha)(M_{\odot}yr^{-1})= 7.9\times10^{-42}L(H\alpha)(ergs^{-1}),}
\end{equation}
where $\rm L(H\alpha)$ is corrected for the dust extinction. 
In this work, we rely on the Balmer decrement method, which is commonly used to correct the line measurements for dust extinction through the comparison of the observed ratio of the H$\alpha$ and  H$\beta$ emission lines ($\rm R_{obs}$) with the dust-free theoretical value ($\rm R_{th} = (H\alpha/H\beta)_{theoretical}  = 2.86$; \citealt{Groves2012}). For the theoretical ratio, we used the case B (the optical thick limit) recombination, corresponding to a temperature of  $\rm T = 10^4$ K and electron density of $\rm N_e = 10^2$ $\rm cm^{-3}$ \citep{Osterbrock1989}. 
The H$\alpha$/H$\beta$ ratio is linked with the attenuation at optical wavelengths by:
\begin{equation}\label{eq:Av}
    \rm A_{V} = R_V \cdot 2.5  log( \frac{R_{obs}}{R_{th}})/(k_{H\beta}-k_{H\alpha}),
\end{equation}
where: 
\begin{equation}
\begin{aligned}
       \rm k_{H\beta} = 2.659 \times [-2.156  + (1.509/\lambda_{H\beta})- \\
    (0.198/\lambda^2_{H\beta}) + (0.011/\lambda^3_{H\beta})] + R_V, 
\end{aligned}
\end{equation}

\begin{equation}
\begin{aligned}
       \rm k_{H\alpha} = 2.659 \times [-1.857  + (1.040/\lambda_{H\alpha})] + R_V, 
\end{aligned}
\end{equation}

where $\rm R_V = 4.05$ (\citealt{Calzetti2000}),  $\rm \lambda_{H\beta} = 0.4861$ $\mu m$ and  $\rm \lambda_{H\alpha} = 0.6563$ $\mu m$. The reddening is then expressed as:
\begin{equation} \label{eq:ebv} \rm
    E(B-V) = A_V/ R_V.
\end{equation}
For the representative sample (see Appendix~\ref{sec:RepresentativeSample} for the description of the representative sample), we selected galaxies with high S/N ({\tt S/N} $\ge3$) of H$\alpha$ and H$\beta$ line fluxes taken from {\tt FastSpecFit v3.2} and correct the H$\alpha$ for extinction following:
\begin{equation}
    \rm F_{corr}  = F_{observed} \times 10^{0.4 \cdot k_{H\alpha} \cdot E(B-V)}.
\end{equation}
We also applied an aperture correction from {\tt FastSpecFit v3.2}. 
 Based on the corrected H$\alpha$ flux and redshift, we derived the H$\alpha$ luminosity; finally, we derived the SFR derived based on the H$\alpha$ emission line (SFR(H$\alpha$)) following Eq.~\ref{eq:sfr}. To preserve the consistency of the IMF with the SFR definition proposed by~\cite{Kennicutt1998}, we compared the SFR estimated with {\tt CIGALE} assuming \cite{Salpeter1955} IMF . We refer to Table~\ref{tab:SFRComparison} for the bias introduced by assumption of the \cite{Salpeter1955} or \cite{Chabrier2003} IMF). 
 We show the comparison of the SFR(SED) and SFR(H$\alpha$) in Fig.~\ref{fig:SFRcomparison} for the entire representative sample as well as for star-forming galaxies and AGNs. The discrepancy between SFR(SED) and SFR(H$\alpha$) given by {$\Delta = 0.119$ and {\tt NMAD} $= 0.567$ (see Sect.~\ref{sec:metrics_comparison} for the definitions of the metrics) are mostly driven by the AGN population. The SFRs for star-forming galaxies are in closer agreement ($\Delta$ = 0.136 and {\tt NMAD} = 0.413) than for AGNs ($\Delta = 0.248$ and {\tt NMAD} = 0.871). 
 {Table~\ref{tab:stats} shows how the $\Delta$ and {\tt NMAD} changes with redshift. There are no clear trends, the consistency of the SFR(SED) and SFR(H$\alpha$) among different redshift bins suggests that there is no dependence on redshift.} 
 The consistency of SFR(SED) and SFR(H$\alpha$) for star-forming galaxies highlights the consistency of the SFRs derived based on the optical-MIR SEDs with the ones derived based on the emission lines.

 The degree of the difference in our SFRs estimates from the ones derived by other prescriptions are quantified in Table~\ref{tab:SFRComparison} (see Sect.~\ref{sec:metrics_comparison} for the definition of the metric). For the star-forming galaxies, the bias introduced by the assumptions of the dust laws or AGN contribution does not  significantly affect the SFR estimates ($\Delta$ and {\tt NMAD} $\lesssim$ 0.1). The moderate differences ($\Delta$ and {\tt NMAD} $\lesssim$ 0.3) are introduced by assumptions over IMF, SSP models, Z, or excluding {\it W3} and {\it W4} from SED fits, while the SFH or SED fits obtained solely on optical bands introduce large median differences and NMAD ($\Delta$ and {\tt NMAD} $\gtrsim$ 0.3). The agreement between SFR(SED) and SFR(H$\alpha$) degrades for AGNs, preserving only small $\Delta$ and {\tt NMAD} for the choice of the AGN model ($\Delta$ and {\tt NMAD} $\lesssim$ 0.1), while introducing large median difference and NMAD ($\Delta$ and {\tt NMAD} $\gtrsim$ 0.3) for SFH module and incorporating MIR information to SED fits. The rest of the assumptions introduced a moderate median difference.

\begin{figure*}
\centerline{\includegraphics[width=0.99\textwidth]{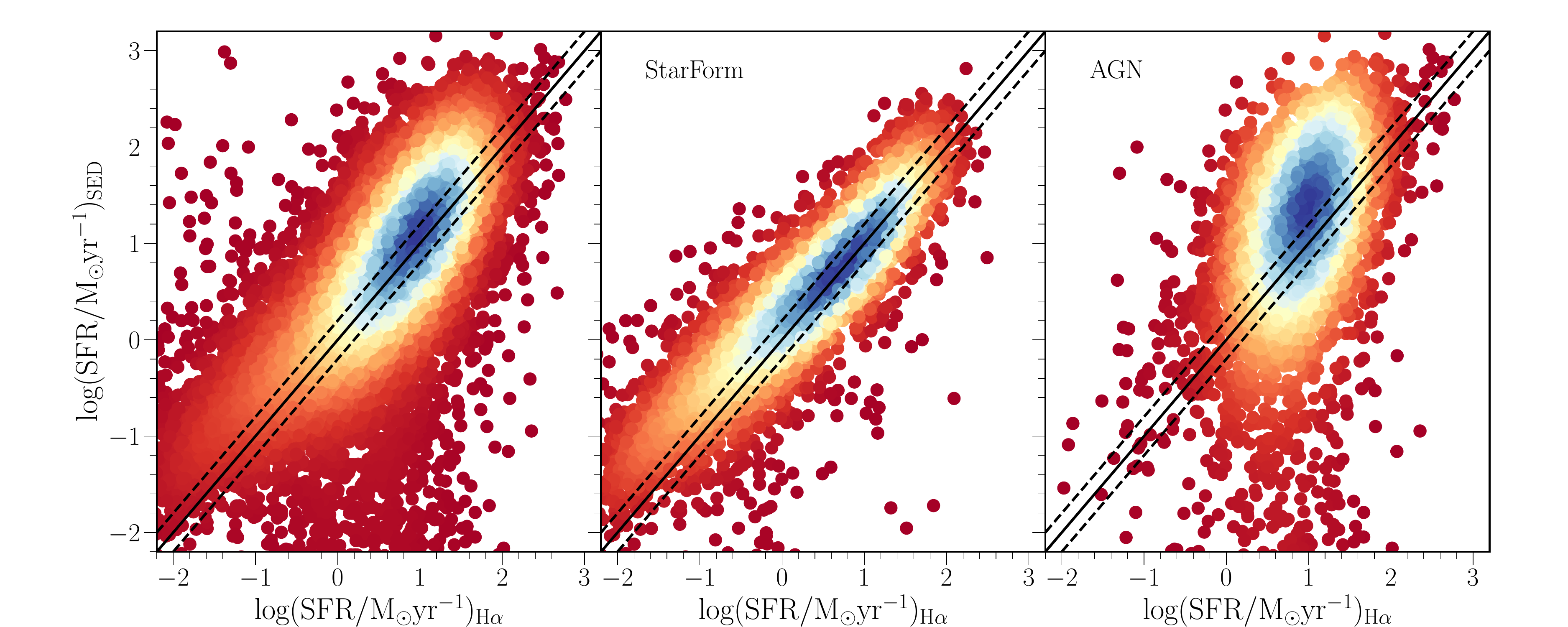}}
        \caption{Comparison of the SFR estimated from the H$\alpha$ line with the SFR obtained from {\tt CIGALE} optical-MIR SED fits for the entire representative sample (left panel), star-forming galaxies (middle panel)m and AGNs (BL and NL together; right panel). The 1:1  and $\pm 0.2$ dex lines are marked with black solid and dashed lines, respectively.  }
        \label{fig:SFRcomparison}
\end{figure*} 

\begin{table*}[ht]
\centering
\caption{{Evolution of $\Delta$ and {\tt NMAD} (see Sect.~\ref{sec:metrics_comparison} for the definitions of the metrics) for the SFR(SED) and SFR(H$\alpha$) for the entire representative sample, star-forming galaxies, and AGNs (BL and NL together). }}
\begin{tabular}{c|c|c|c|c|c|c}
         {\tt Z} & \multicolumn{2}{c|}{\tt All} & \multicolumn{2}{c|}{\tt StarForm}  & \multicolumn{2}{c}{\tt AGN}\\
         & $\Delta$ & {\tt NMAD} & $\Delta$ & {\tt NMAD} & $\Delta$ & {\tt NMAD} \\

        \hline \hline
$0.0-0.5$ & 0.119 & 0.567 & 0.136 & 0.413 & 0.248 & 0.871 \\
\hline
$0.00-0.20$ & 0.203 & 0.604 & 0.263 & 0.488 & 0.216 & 0.777 \\
$0.20-0.35$ & 0.097 & 0.552 & 0.117 & 0.381 & 0.195 & 0.845 \\
$0.35-0.50$ & 0.048 & 0.541 & 0.000 & 0.349 & 0.314 & 0.961 \\
\hline
\end{tabular}
\label{tab:stats}
\end{table*}

\begin{table*}
\centering
        \caption{Comparison of our SFR estimates and the ones derived by changing one of the model descriptions outlined in Sect.~\ref{sec:DependenceofPhysProp}.  }
        \label{tab:SFRComparison}
        \footnotesize
        \begin{tabular}{c | r | r  | r | r | r | r}    
         Module & \multicolumn{2}{c|}{\tt All} & \multicolumn{2}{c|}{\tt StarForm}  & \multicolumn{2}{c}{\tt AGN}\\
         & $\Delta$ & {\tt NMAD} & $\Delta$ & {\tt NMAD} & $\Delta$ & {\tt NMAD} \\

        \hline \hline
        {\tt IMF} (\citealt{Chabrier2003} vs. \citealt{Salpeter1955})& -0.212 & 0.314 & -0.198 & 0.292 & -0.209 & 0.309 \\ 
        {\tt SSP} (\citealt{Bruzual2003} vs. \citealt{Maraston2005}) &  -0.214 & 0.462 & -0.256 & 0.384 & -0.227 & 0.389 \\ 
        {\tt SFH} (delayed with extra burst vs non parametric) &  1.018 & 2.636 & 1.664 & 2.503 & 1.303 & 2.24 \\ 
        {\tt Z} (metallicity fixed to solar vs. variable) &  0.217 & 0.598 & 0.064 & 0.146 & 0.163 & 0.374 \\ 
        \hline
        {\tt DustAtt} (\citealt{Calzetti2000} vs. \citealt{CharlotFall2000}) &   0.047 & 0.171 & 0.012 & 0.096 & 0.052 & 0.198 \\ 
        {\tt DustEm} (\citealt{Draine2014} vs. \citealt{Dale2014}) & -0.009 & 0.097 & -0.007 & 0.051 & -0.003 & 0.184\\ 
        \hline
        {\tt AGN} (\citealt{Fritz2006} vs. no AGN) &  -0.142 & 0.220 & -0.049 & 0.075 & -0.190 & 0.287 \\ 
        {\tt AGN} (\citealt{Fritz2006} vs. \citealt{Stalevski2012, Stalevski2016}) &  0.008 & 0.038 & 0.002 & 0.010 & 0.044 & 0.127 \\ 
       \hline
        {\tt SEDs} ($grzW14$ vs. $grzW12$) &  0.023 & 0.481 & 0.034 & 0.180 & 0.321 & 0.928\\ 
        {\tt SEDs} ($grzW14$ vs. $grz$) &  -0.540 & 1.022 & -0.302 & 0.542 & -0.298 & 0.841\\ 
        \hline
        {\tt SFR} (SED(\citealt{Salpeter1955}) vs. H$\alpha$) & 0.119 & 0.567 & 0.136 & 0.413 & 0.248 & 0.871\\ 
\end{tabular}
\tablefoot{The $\Delta$, and {\tt NMAD}  (see Sect.~\ref{sec:metrics_comparison} for definitions) are reported for the given change in the model considering a change in the initial mass function ({\tt IMF}), single stellar population models ({\tt SSP}), star formation history prescription ({\tt SFH}), metallicity ({Z}), dust attenuation law ({\tt DustAtt}), dust emission model ({\tt DustEm}), and AGN models ({\tt AGN}).  The influence of the presence of MIR information and comparison with SFR derived based on the H$\alpha$ emission line are also reported. The metrics are derived for the entire representative sample ({\tt All}) and separately for star-forming galaxies ({\tt StarForm}) and {\tt AGN} (including both NL and BL AGNs). }
\end{table*}  

\subsection{AGN features}\label{sec:AGNFeaturesCompReprSample}

As shown in the previous sections, the incorporation of the WISE photometry has an influence on the stellar mass and SFR estimates (see Sect.~\ref{sec:DependenceonPhootmetry} and~\ref{sec:sec:SFRSpectra}). Here, we explore how the AGN features derived with {\tt CIGALE} depend on the inclusion of the MIR photometry in SED fits. The change of the AGN fraction ({\tt AGNFRAC}) and viewing angle ({\tt AGNPSY}) for the BL AGNs, NL AGNs, and star-forming galaxies drawn from the representative sample (see Appendix~\ref{sec:RepresentativeSample} for a description of the representative sample) when using 1) solely optical information ({\it grz}), 2) optical and WISE W1 and W2 ({\it grzW12}), or 3) optical and all WISE ({\it grzW14}) is shown in Figs.~\ref{fig:AGNFracCompWISE} and~\ref{fig:AGNPsyCompWISE}, respectively. As expected, optical SED fits cannot distinguish the contribution of AGN to IR, which results in an artificial AGN fraction ({\tt AGNFRAC} $\sim 0.3$) and viewing angle ({\tt AGNPSY} $\sim 35$) for almost all galaxies independent of their galaxy type. The {\tt AGNFRAC} and {\tt AGNPSY} distributions start to differentiate for BL AGNs, NL AGNs, and star-forming galaxies when at least {\it W12} is included in SED fitting. 
When the {\it W12} is incorporated in the SED fits, the BL AGNs are characterized by lower {\tt AGNFRAC} than NL AGNs (with a median {\tt AGNFRAC} increasing from 0.16 to 0.24); whereas star-forming galaxies have an {\tt AGNFRAC} distribution similar to that of NL AGN (with a median 0.22). Interestingly, when including {\it W12},  {\tt AGNPSY} starts to properly identify  BL AGNs based on their high viewing angle ({\tt AGNPSY} $\sim 80^{\circ}$), although there is also a strong peak at {\tt AGNPSY} $\sim 40^{\circ}$. Both star-forming and NL AGNs are characterized by similar viewing angles ({\tt AGNPSY} $\sim 40^{\circ}$).

The incorporation of all four WISE bands ($W14$) results in smoothing the {\tt AGNFRAC} distribution for BL AGNs with a strong peak at {\tt AGNFRAC} $\sim0.15$ and {\tt AGNPSY} $\sim 80^{\circ}$; whereas for NL AGNs and star-forming galaxies, the distribution is characterized by a strong peak at {\tt AGNFRAC} $\sim 0$ and a secondary peak at $\sim 0.25$. The difference in {\tt AGNFRAC} between star-forming and NL AGNs is in the long tail towards high values for NL AGNs, while the fraction of star-forming galaxies exceeding {\tt AGNFRAC} $\sim 0.35$ is negligible. The {\tt AGNPSY} distributions for NL AGNs and star-forming galaxies are similarly characterized by wide peaks at {\tt AGNPSY} $\sim 40^{\circ}$; however, a small fraction is characterized by the viewing angle typical for BL AGNs. 
Even with MIR SED fits, the distribution of {\tt AGNFRAC} and {\tt AGNPSY} for NL AGNs and star-forming galaxies are similar, challenging the usefulness of these AGN features to differentiate between AGN and non-AGN galaxies. Nevertheless, BL AGNs show a clear preference for {\tt AGNFRAC} $\sim 0.15$ and {\tt AGNPSY} $\sim 80^{\circ}$.

Independently of the availability of WISE photometry, 80\% of BL AGNs are identified as AGNs based on the {\tt AGNFRAC} $\ge 0.1$. Following the same criterion, 65\% of NL AGNs are also classified as AGNs. A similar fraction (68\%) of star-forming galaxies is characterized by {\tt AGNFRAC} $> 0.1$, but this is reduced to 19\% if all WISE bands with {\tt S/N} $\ge 3$ are included in the SED fits ({\tt FLAGINFRARED} = 4). This confirms that without the MIR information, the star-forming galaxies overestimate the AGN contribution, although providing measurements in all four WISE bands reduces significantly the fraction of the star-forming galaxies with overestimated AGN fraction (from 68\% to 19\%). 

The physical properties derived for the majority (68\%) of star-forming galaxies are thus derived under the assumption of the false presence of AGN for sources without the MIR information. However, we validated the notion that the stellar mass estimates can still be securely used. For a sample of star-forming galaxies with {\tt AGNFRAC} $\ge 0.1$, we compared the derived stellar masses with the ones estimated without incorporating AGN contribution (i.e., the {\tt AGNFRAC} is fixed to the null value) independently of the availability of  WISE photometry. Both the $\Delta$ and {\tt NMAD} are comparable (with $\Delta$ = 0.014 and  {\tt NMAD}  = 0.037). Similarly, the effect on the SFR is negligible ($\Delta$ = -0.079 and {\tt NMAD} = 0.119). For star-forming galaxies with a low AGN fraction ({\tt AGNFRAC} $\leq 0.1$), there is no impact on stellar mass or SFR estimates (with $\Delta = - 0.001, - 0.016$ and {\tt NMAD} = 0.011, 0.025 for stellar mass and SFR, respectively). To summarize, although the AGN features especially for sources lacking high S/N WISE photometry are not reliable, the overestimation of the AGN fraction for star-forming galaxies does not affect their stellar mass or SFR estimates.

\begin{figure}
\centerline{\includegraphics[width=0.49\textwidth]{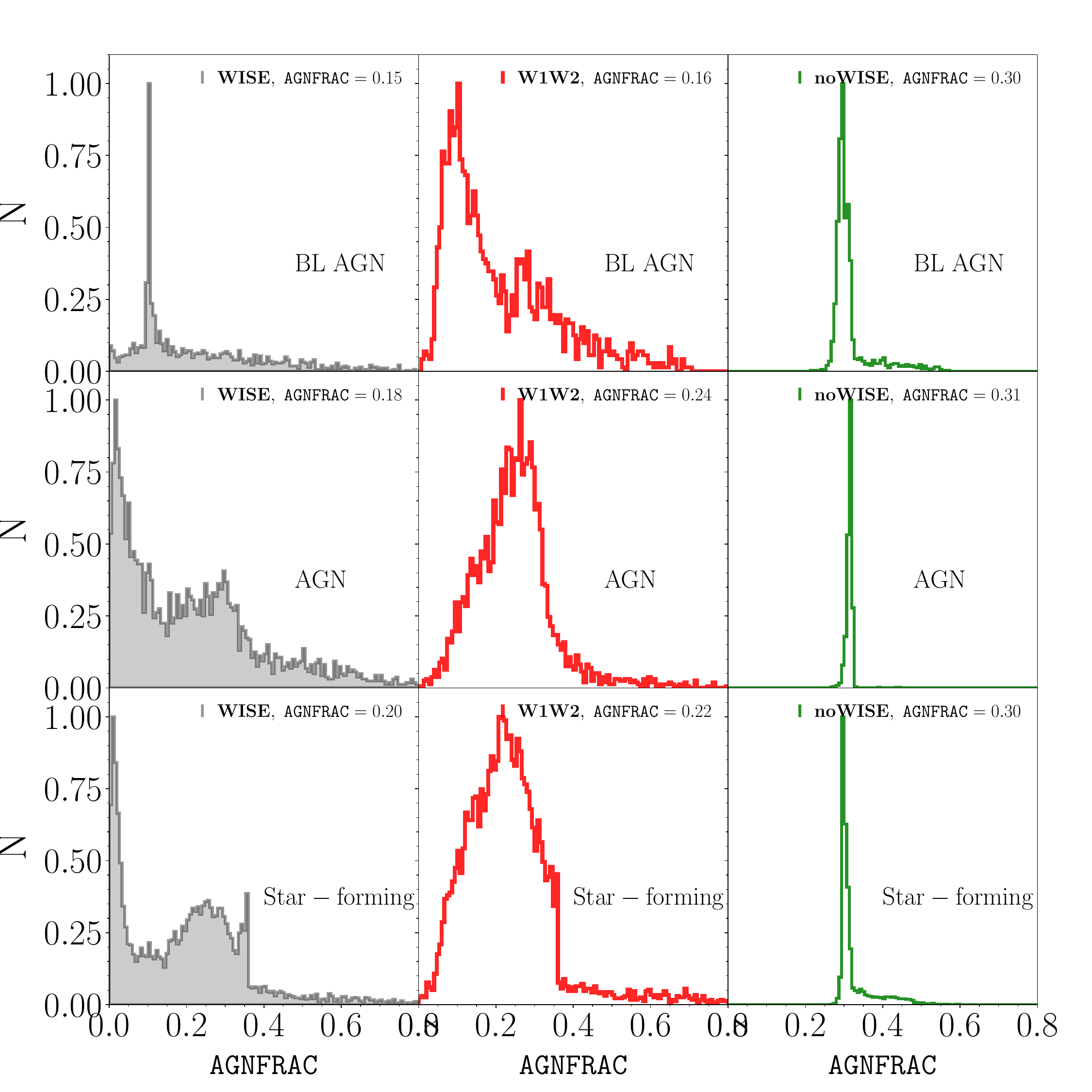}}
        \caption{Comparison of the {\tt AGNFRAC} distribution obtained from SED fits incorporating {\it grzW14} bands (left panels),   {\it grzW12} (middle panels), and solely {\it grz} bands (right panels)  for BL AGNs (top panels), NL AGNs (middle panels), and star-forming galaxies (bottom panels) drawn from the representative sample (see Appendix~\ref{sec:RepresentativeSample}). The number of sources and the median {\tt AGNFRAC} are reported in the plots. }
        \label{fig:AGNFracCompWISE}
\end{figure}

\begin{figure}
\centerline{\includegraphics[width=0.49\textwidth]{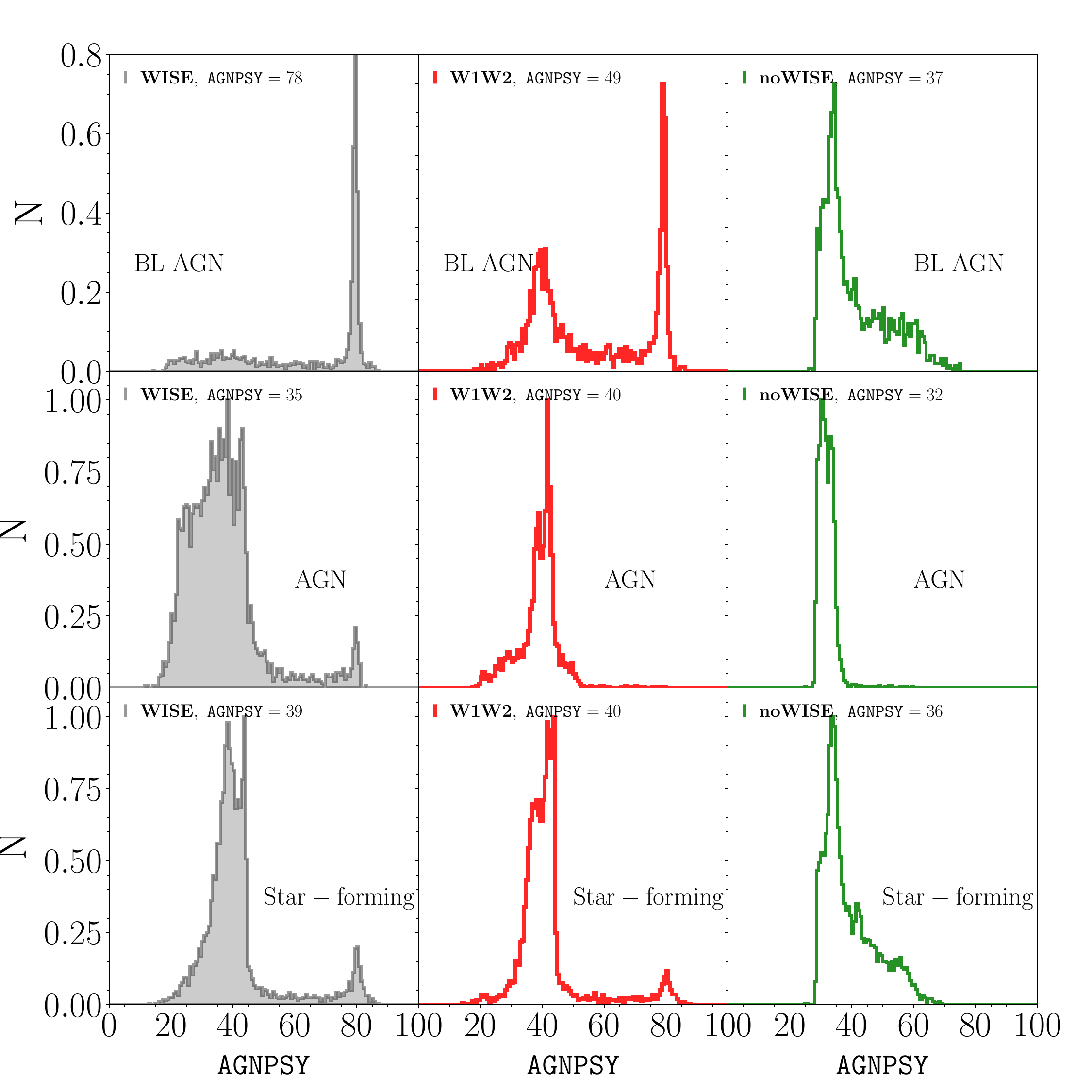}}
        \caption{Comparison of the {\tt AGNPSY} distribution obtained from SED fits incorporating {\it grzW14} bands (left panels),   {\it grzW12} (middle panels), and solely {\it grz} bands (right panels)  for BL AGNs (top panels), NL AGNs (middle panels), and star-forming galaxies (bottom panels) drawn from the representative sample (see Appendix~\ref{sec:RepresentativeSample}). The number of sources and the median {\tt AGNFRAC} are reported in the plots}
        \label{fig:AGNPsyCompWISE}
\end{figure} 

\section{Summary}\label{sec:conclusions}
We created the VAC of physical properties (including stellar masses, SFRs, rest-frame magnitudes, and AGN features) for DESI EDR galaxies. Here, we briefly summarize the VAC:
\begin{itemize}
    \item The catalog includes almost 1.3 million galaxies with optical and NIR photometry ({\it grz}) supplemented
with WISE ({\it W14}) information spanning a broad redshift range ($\rm 0.001 \le z \le 5.968$; for details, see Sect.~\ref{sec:data}).
    \item  The physical properties are derived with the {\tt CIGALE} SED fitting code (\citealt{Boquien2019})  including AGN templates from~\citealt{Fritz2006} (see Sect.~\ref{sec:cigale} and Table~\ref{tab:SEDParameters} for details on the SED fitting prescription).  Tables~\ref{tab:StMassComparison} and~\ref{tab:SFRComparison} elaborate on the extent to which our estimates depend on our model assumptions. 
    \item Several quality flags, such as the quality of the fit expressed by $\chireduced$, the {PDF} of the derived properties, or the {\tt S/N} of available photometry, would allow users to adapt the sample to specific scientific goals (see Sect.~\ref{sec:CleaningVAC}).
    \item  We identified red and blue {\tt BGS} galaxies based on the  UVJ diagram (\citealt{Whitaker2012}) and revised their location on the star-forming MS (see Sect.~\ref{sec:MainSequence}). 
    \item We evaluated the bias and deviations introduced by several model parameters, including SSP, SFH, and dust choice, on stellar mass (see Sect.~\ref{sec:DependenceofPhysProp} and Table~\ref{tab:StMassComparison}) and SFR (see Sect.~\ref{sec:sec:SFRSpectra} and Table~\ref{tab:SFRComparison}). 
    \item We found a very good agreement with {\tt COSMOS} and other VACs that are BC03-based. The relatively good agreement with the {\tt SDSS(Firefly)} VAC, which is obtained by fitting the spectra, rather than the photometry. Using another model prescription (SSP, SFH, metallicities, etc.) showcases the excellent agreement of the stellar mass estimates with other VACs (see Sect.~\ref{sec:comparion} and Table~\ref{tab:CatalogsComparioson}). 
\end{itemize}

The  VAC presented here highlights the importance of incorporating MIR photometry in the SED fits. Without the inclusion of WISE photometry, the AGN fraction would be overestimated for star-forming galaxies, although it does not affect their stellar mass or SFR estimates (see   Sect.~\ref{sec:AGNFeaturesCompReprSample}). We demonstrate that accounting for the AGN contribution is essential for deriving stellar masses for galaxies hosting AGN.  With {\tt CIGALE,} we are able to recover both stellar masses and AGN luminosities in a way that is consistent with other methods. We note that the particular host properties of  DESI AGNs will be explored in detail in a future work. 
 
\begin{acknowledgements}

The presented analysis is conducted on the VAC of physical properties of DESI galaxies {\tt v1.2}, while the latest available version {\tt v1.4} includes an additional 51,126 ($4\%$) that were missing information about the {\tt RELEASE} in the previous version. There is also a supplementary table of 8,422 sources for which physical properties are based on the redshift coming from the {\tt QSO afterbruner} pipeline relying on the {\tt QuasrNet} and the broad Mg II finder pipelines. The {\tt afterburner} shows better performance (94\%) in identifying visually inspected QSOs relative to {\tt Redrock} pipeline (86\%; see \citealt{Chaussidon2023} and \citealt{Alexander2023} for more details).  The VAC for the forthcoming release (DESI Year 1) will be published as part of Data Release 1 similar to the currently public DESI EDR. 

The authors thank the anonymous referee for insightful comments. 
The authors would like to express their appreciation to Laure Ciesla for sharing the {\tt CIGALE} non-parametric SFH module {\tt sfhNlevels}, J.~Thorne, and H. Suh for sharing their catalogs.  

This work has been supported by the Polish National Agency for Academic Exchange (Bekker grant BPN/BEK/2021/1/00298/DEC/1) and the European Union's Horizon 2020 Research and Innovation program under the Maria Sklodowska-Curie grant agreement (No. 754510). M.S. acknowledges financial support from the State Research Agency of the Spanish Ministry of Science and Innovation under the grants 'Galaxy Evolution with Artificial Intelligence' with reference PGC2018-100852-A-I00 and 'BASALT' with reference PID2021-126838NB-I00. 
M.M. acknowledges support from the Spanish Ministry of Science and Innovation through the project PID2021-124243NB-C22. This work was partially supported by the program Unidad de Excelencia Mar\'ia de Maeztu CEX2020-001058-M. H.Z. acknowledges the support from the National Natural Science Foundation of China (NSFC; grant Nos. 12120101003 and 12373010) and  National Key R\&D Program of China (grant Nos. 2023YFA1607800, 2022YFA1602902) and Strategic Priority Research Program of the Chinese Academy of Science (Grant Nos. XDB0550100).

This material is based upon work supported by the U.S. Department of Energy (DOE), Office of Science, Office of High-Energy Physics, under Contract No. DE–AC02–05CH11231, and by the National Energy Research Scientific Computing Center, a DOE Office of Science User Facility under the same contract. Additional support for DESI was provided by the U.S. National Science Foundation (NSF), Division of Astronomical Sciences under Contract No. AST-0950945 to the NSF’s National Optical-Infrared Astronomy Research Laboratory; the Science and Technology Facilities Council of the United Kingdom; the Gordon and Betty Moore Foundation; the Heising-Simons Foundation; the French Alternative Energies and Atomic Energy Commission (CEA); the National Council of Humanities, Science and Technology of Mexico (CONAHCYT); the Ministry of Science, Innovation and Universities of Spain (MICIU/AEI/10.13039/501100011033), and by the DESI Member Institutions: \url{https://www.desi.lbl.gov/collaborating-institutions}. Any opinions, findings, and conclusions or recommendations expressed in this material are those of the author(s) and do not necessarily reflect the views of the U. S. National Science Foundation, the U. S. Department of Energy, or any of the listed funding agencies.

The authors are honored to be permitted to conduct scientific research on Iolkam Du’ag (Kitt Peak), a mountain with particular significance to the Tohono O’odham Nation.

The DESI Legacy Imaging Surveys consist of three individual and complementary projects: the Dark Energy Camera Legacy Survey (DECaLS), the Beijing-Arizona Sky Survey (BASS), and the Mayall z-band Legacy Survey (MzLS). DECaLS, BASS and MzLS together include data obtained, respectively, at the Blanco telescope, Cerro Tololo Inter-American Observatory, NSF’s NOIRLab; the Bok telescope, Steward Observatory, University of Arizona; and the Mayall telescope, Kitt Peak National Observatory, NOIRLab. NOIRLab is operated by the Association of Universities for Research in Astronomy (AURA) under a cooperative agreement with the National Science Foundation. Pipeline processing and analyses of the data were supported by NOIRLab and the Lawrence Berkeley National Laboratory. Legacy Surveys also uses data products from the Near-Earth Object Wide-field Infrared Survey Explorer (NEOWISE), a project of the Jet Propulsion Laboratory/California Institute of Technology, funded by the National Aeronautics and Space Administration. Legacy Surveys was supported by: the Director, Office of Science, Office of High Energy Physics of the U.S. Department of Energy; the National Energy Research Scientific Computing Center, a DOE Office of Science User Facility; the U.S. National Science Foundation, Division of Astronomical Sciences; the National Astronomical Observatories of China, the Chinese Academy of Sciences and the Chinese National Natural Science Foundation. LBNL is managed by the Regents of the University of California under contract to the U.S. Department of Energy. The complete acknowledgments can be found at \url{https://www.legacysurvey.org/}.

\end{acknowledgements}

\section*{Data availability}

The VAC of physical properties of DESI galaxies is publicly available at \url{https://data.desi.lbl.gov/doc/releases/edr/vac/cigale/}. 
The data behind the figures are available at https://doi.org/10.5281/zenodo.13847488.


\bibliographystyle{aa}
\bibliography{VACPaperI}



\begin{appendix}

\section{Catalog structure}\label{app:Catalogcolumns}
A description of the columns in the catalog is given in Table~\ref{tab:CatalogColumns}.

\begin{table*}
\centering
        \caption{Data model of the DESI VAC.}
        \label{tab:CatalogColumns}
        \footnotesize
        \begin{tabular}{r r r l}
\hline
\hline
Name & Format & Unit &  Description\\
\hline
{\tt TARGETID} & int64   & - & Unique DESI target ID \\ 
    {\tt SURVEY}                       & bytes7  & -               & Survey name \\ 
    {\tt PROGRAM}                      & bytes7  & -               & Program name \\ 
    {\tt HEALPIX}                      & int32   & -               & Healpix number \\ 
    {\tt SPECTYPE}                     & bytes7  & -               & Redrock spectral classification \\ 
    {\tt RA}                           & float64 & deg             & Right ascension from target catalog \\ 
    {\tt DEC}                          & float64 & deg             & Declination from target catalog \\ 
    {\tt RELEASE}                      & int16   & -               & Legacy Surveys (LS) Release \\ 
    {\tt Z}                            & float64 & -               & Redshift \\ 
    \hline \\ 
    {\tt CHI2}                         & float64 & -               & reduced chi2 defining the quality of the fit:\\
     &  & & for a more reliable estimations the cut of $\rm chi_{r}^2 \lesssim 17$ is recommended \\ 
    {\tt LOGM} & float64 & log(solMass)    & logarithm of the stellar mass  \\ 
    {\tt LOGM\_ERR} & float64 & log(solMass)    & error on logarithm of the stellar mass \\ 
    {\tt LOGSFR} & float64 & log(solMass/yr) & logarithm of the star formation rate averaged over 10Myr \\ 
    {\tt LOGSFR\_ERR} & float64 & log(solMass/yr) & error on logarithm of star formation rate averaged over 10Myr \\ 
    {\tt AGNLUM} & float64 & W               & total luminosity of the AGN in W \\ 
    {\tt AGNFRAC} & float64 & -               & fraction of the total IR emission coming from the AGN, where 0 means no AGN\\
    & & & contribution, 1 means 100\% AGN contribution \\ 
    {\tt AGNPSY}                       & float64 & deg             & viewing angle, with $\sim 30^{\circ}$ and $\sim70^{\circ}$, for type 1 and type 2 AGN, respectively \\ 
    {\tt LNU\_(U/G/R/I/Z)}              & float64 & W/Hz            & rest-frame luminosity in a given band,\\
    & & & rest-frame magnitudes can by derived using  $\rm -2.5\times log10(Lnu)+34.1$ \\ 
    {\tt NUVR,RK,UV,VJ,GR}           & float64 & AB mag          & rest-frame colors in given bands \\
    {\tt LNU\_(U/G/R/I/Z)\_ERR}          & float64 & W/Hz            & error of the rest-frame luminosity in a given band \\ 
    {\tt (NUVR,RK,UV,VJ,GR)\_ERR}       & float64 & AB mag          & error of the rest-frame colors in given bands \\ 
   {\tt AGE}      & float64 & Myr          & age of the main stellar population \\ 
   $\tt AGE_{ERR}$      & float64 & Myr          & error of the age of the main stellar population \\ 
   {\tt AGEM}      & float64 & Myr          & mass-weighted age of the main stellar population \\ 
  $\tt AGEM_{ERR}$     & float64 & Myr          & error of the mass-weighted age of the main stellar population \\     
   {\tt TAU}      & float64 & Myr          & $\tau$ of the main stellar population \\ 
   $\tt TAU_{ERR}$      & float64 & Myr          & $\tau$ of the main stellar population \\ 
   {\tt FRACYSSP}      & float64 & -          & mass fraction of young stellar population \\ 
   $\tt FRACYSSP_{ERR}$      & float64 & -          & error of the mass fraction of young stellar population \\   
    \hline \\ 
    {\tt FLAG\_MASSPDF}                 & float64 & -               & flag expressed by $\tt M_{best}$/$\tt M_{bayes}$ to reject stellar mass estimates with broad PDF \\ 
    & & & and/or complex likelihood distribution, e.g., $1/5 \lesssim \tt M_{best}$/$\tt M_{bayes} \rm \lesssim 5$ \\ 
    {\tt FLAG\_SFRPDF}                  & float64 & -               & flag expressed by $\tt SFR_{best}$/$\tt SFR_{bayes}$ to reject SFR estimates with broad PDF \\ 
   & &  & and/or complex likelihood distribution, e.g., $1/5 \tt SFR_{best}$/$\tt SFR_{bayes} \rm \lesssim 5$    \\ 
    {\tt FLAGOPTICAL}                  & int64   & -               & flag to select sources observed with high S/N ({\tt S/N} $\ge 10$) in optical bands ({\it grz}); \\
    &  &  & {\tt FLAGOPTICAL} = 3(2/1/0): source is observed in 3(2/1/0) band(s) with {\tt S/N} $\ge 10$ \\ 
    {\tt FLAGINFRARED} & int64   & - & flag to select sources observed with high S/N ({\tt S/N} $\ge 3$) in WISE bands (W1,  \\ 
    & &  & W2, W3, W4); {\tt FLAGINFRARED} = 4(3/2/1/0): source is observed in 4(3/2/1/0) \\
    & &  & band(s) with {\tt S/N} $\ge 3$ \\ 
    \hline \\ 
    {\tt FLUX\_(G/R/Z/W1-4)} & float32 & nanomaggy & flux in a given band; for the reddening corrected flux use: \\
    & & &  {\tt DERED\_FLUX \rm = \tt FLUX\_BAND/MW\_TRANSMISSION\_BAND};  \\ 
    & &  & for magnitude : $\tt MAG\_BAND \rm = -2.5 \times log10(\tt DERED\_FLUX\rm) + 22.5$ \\ 
    $\tt FLUX\_IVAR\_(G/R/Z/W1-4)$       & float32 & $\rm nanomaggy^{-2}$   & inverse variance of the flux in a given band \\ 
    {\tt MW\_TRANSMISSION\_(G/R/Z/W1-4)} & float32 & - & Milky Way foreground dust transmission factor [0-1] in a given band. \\ 
    {\tt S/N\_(G/R/Z/W1-4}) & float32 & - & S/N in a given band calculated as: $ \tt FLUX \rm \times sqrt(\tt FLUX\_IVAR\rm)$ \\ 
\hline \hline
\\

\end{tabular}
\end{table*}  
\section{Stellar mass errors dependence on the quality of the photometry}\label{app:stmass_errors}

The errors in the stellar mass and SFR estimates depend on the model assumptions done for the SED fitting and on the spectral coverage included in the SED fits. Standardly, the logarithmic stellar masses are constrained to the $\rm 0.2-0.4$ dex level~\citep[e.g.,][]{Conroy2013,Comparat2017, Pacifici2023} depending on the assumptions on the SFH, IMF and other model prescriptions (for our stellar mass estimates the dependence on model assumptions is outlined in Sect.~\ref{sec:DependenceofPhysProp}). For our stellar mass estimates,  there is a clear dependence of the uncertainties on the {\tt S/N} of the input photometry (see Fig.~\ref{fig:logMerr}). An accuracy level of $\lesssim 0.25$ dex for galaxies is fulfilled when the {\it g, r, z} observations have $\tt S/N \ge10$. In the case of QSO, an accuracy limit of $\lesssim 0.20$ dex is achieved with at least 3 WISE bands measured with $\tt S/N \ge 7$ (independently of the optical measurements). We note that the errors increase with redshift, implying for QSO at $\rm z>1.5$ that even with high {\tt S/N} the errors may reach $\rm \sim 0.8$ dex (see the blue tail in the left plot in Fig.~\ref{fig:logMerr}).

\begin{figure}  
\centerline{\includegraphics[width=0.49\textwidth]{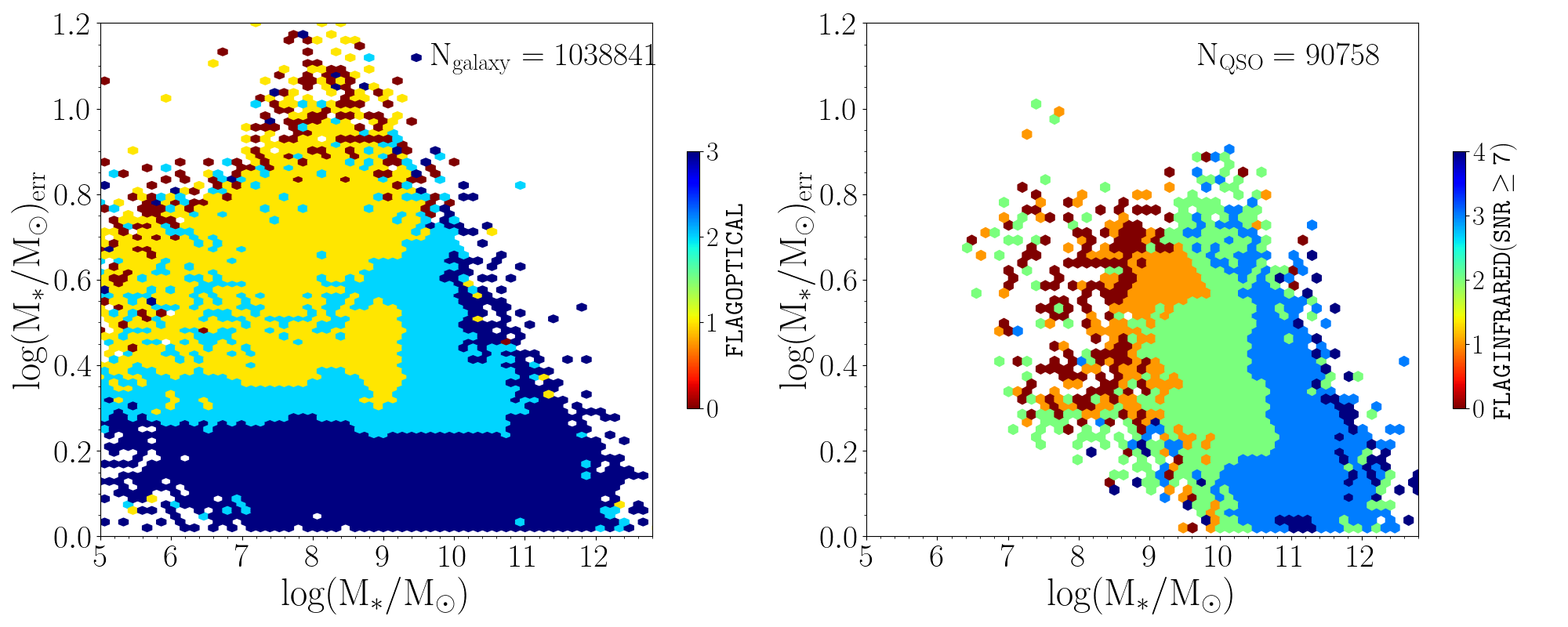}}
        \caption{Dependence of the stellar mass error ($\mstar_{err}$) on the stellar mass ($\mstar$) for two main DESI types: {\tt GALAXIES} (left panel) and {\tt QSOs} (right panel). The stellar mass uncertainties depend on the {\tt S/N} of the optical ({\it igrz}) photometry (left panel) as well as on the S/N of the WISE photometry (right panel). The low errors ($\mstar_{err} \lesssim0.25$) are achieved when at least 3 optical bands are observed with  $\tt S/N \ge10$ and at least 3 WISE bands are measured with $\tt S/N \ge 7$. }
        \label{fig:logMerr}
\end{figure} 

\section{Comparison with other catalogs}\label{sec:DESISTMassCatComparison}

In this appendix, we further explore the comparison of the stellar mass obtained with {\tt CIGALE} for DESI EDR galaxies with the reference catalogs. In Sect.~\ref{sec:comparion}, we  offer a comparison of stellar masses drawn from the {\tt COSMOS2020} catalog and AGN and host properties based on the samples overlapping with {\tt AGN-COSMOS} catalogs. Here, we expand the comparison of the stellar masses with the {\tt SDSS(FIREFLY)} and {\tt SDSS(MPA-JHU)} catalogs (Sect.~\ref{sec:sdss_comparison}), with the {\tt GSWLC} (Sect.~\ref{sec:gswlc_comparison}) and with the DESI catalog presented by \citealt{Zou2024}  (Sect.~\ref{sec:desi_comparison}). 

\subsection{SDSS}\label{sec:sdss_comparison}
To date, the largest spectroscopic survey, SDSS, had provided several VACs including estimates of the stellar properties. In this work, we compare our stellar mass estimates derived with {\tt CIGALE} with two commonly used SDSS catalogs: i)  eBOSS SDSS FIREFLY DR16 {\tt SDSS(FIREFLY)}{\footnote{\url{https://live-sdss4org-dr16.pantheonsite.io/spectro/eboss-firefly-value-added-catalog}}} \citep{Comparat2017}, and ii) {\tt MPA-JHU DR8}\footnote{\url{https://www.sdss4.org/dr17/spectro/galaxy_mpajhu/}}  \citep{Kauffmann2003b, Brinchmann2004,Tremonti2004}. 
The {\tt SDSS(Firefly)} catalog provides stellar properties (age, metallicity, dust reddening, stellar mass, and the SFH) obtained through a comprehensive analysis of the galaxy spectra from SDSS, Baryon Oscillation Spectroscopic Survey (BOSS), and eBOSS. The physical properties were inferred via full spectral fitting with the chi-squared minimization fitting code {\tt FIREFLY}~\citep{Wilkinson2017}. The catalog relies on the stellar population models from~\cite{Maraston2011} utilizing different stellar libraries and IMFs. In this work, we use the stellar masses obtained assuming a \cite{Chabrier2003} IMF and using the MILES libraries \citep{Sanchez2006, Beifiori2011, Falcon2011}. 
The {\tt SDSS(MPA-JHU)} catalog is a VAC of physical properties of SDSS DR8 galaxy derived from fitting their photometry. The catalog provides line measurements, Lick, and other indices as well as the physical properties. The stellar masses and SFRs are derived from the SED fitting (covering {\it ugriz}) using the models and methodology outlined in~\cite{Kauffmann2003}. To avoid the aperture effect on the stellar mass estimates,  these are derived based on the photometry rather than the spectral indices as originally proposed by~\cite{Kauffmann2003}. The SED fitting relies on templates generated with \cite{Bruzual2003} SSP models and a \cite{Kroupa2001} IMF. The catalog includes the stellar mass measurements corresponding to the median and 2.5\%, 16\%, 84\%, and 97.5\% of the PDF. 
 
The comparison of the DESI stellar masses derived with {\tt CIGALE} with {SDSS} stellar masses for the galaxies in the overlap between the samples is shown in Fig.~\ref{fig:SDSSmasses} and the metrics are presented in Table~\ref{tab:CatalogsComparioson}. There is a good agreement between our stellar masses with {\tt SDSS(MPA-JHU)}, with a small median difference and NMAD ($\Delta = -0.071$; {\tt NMAD} = 0.126) that increases when comparing to the {\tt SDSS(FIREFLY)} estimates, which is the only one characterized by a positive median difference. This comparison suggests that the {\tt SDSS(FIREFLY)} stellar mass might be underestimated by an average of 0.2 dex for massive galaxies ($\mstar \sim 11$; see the left panel in Fig.~\ref{fig:SDSSmasses}) as their stellar masses are derived from spectra with apertures that do not cover the full galaxies. On the contrary, for the {\tt SDSS(MPA-JHU)} we find a close agreement across the entire mass range.

\begin{figure}
\centerline{\includegraphics[width=0.49\textwidth]{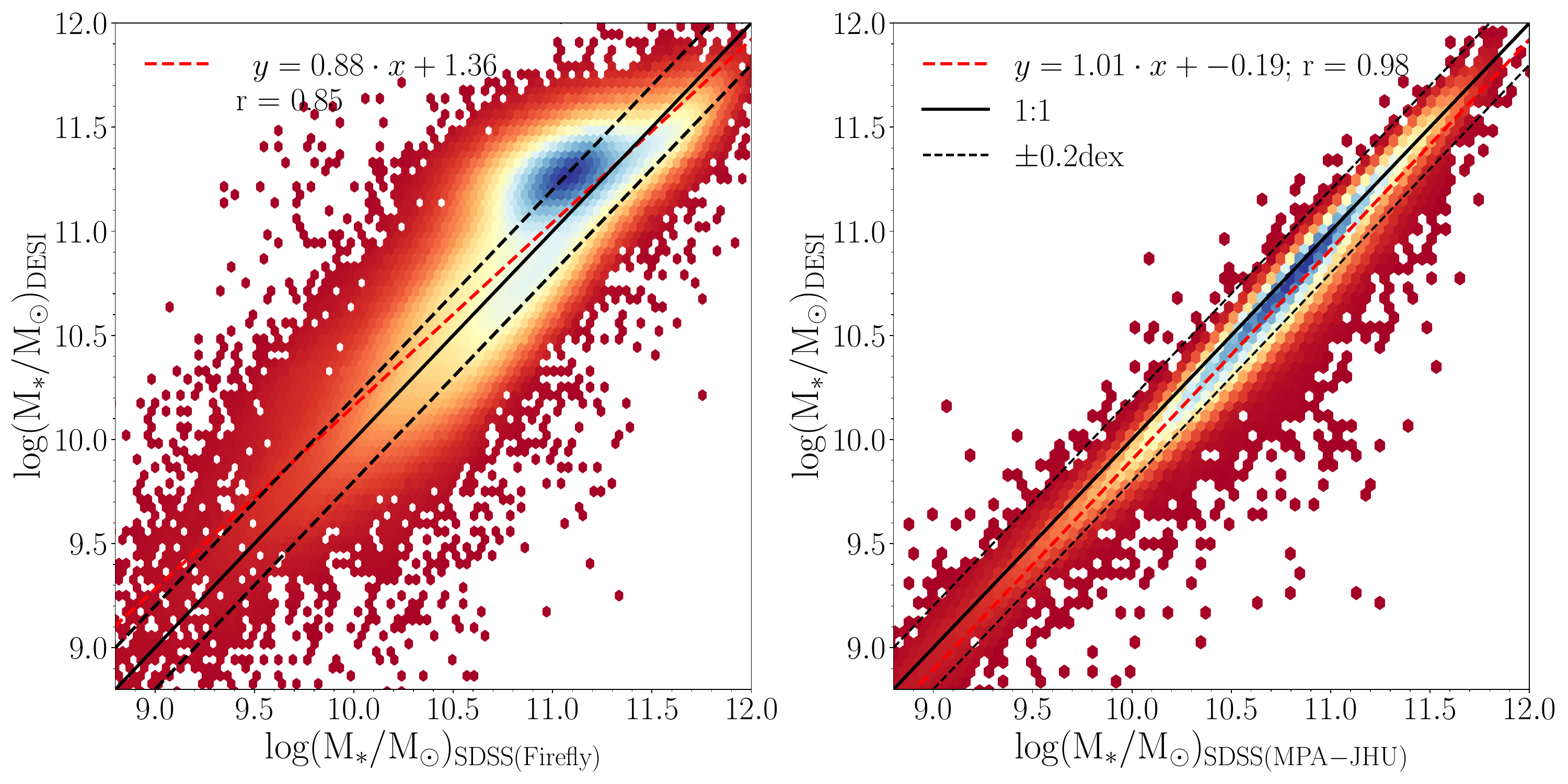}}
        \caption{Comparison of our DESI stellar mass estimates with 24,947 and 18,778 stellar mass estimates from the {\tt SDSS(FIREFLY)} (left panel) and the {\tt SDSS(MPA-JHU)} catalogs (right panel), respectively. The 1:1 and $\rm 0.2$ dex lines are marked with black solid and dashed lines, respectively. The linear fit (red line) and the Pearson correlation coefficient are reported in the legend. }
        \label{fig:SDSSmasses}
\end{figure} 

\subsection{GSWLC}\label{sec:gswlc_comparison}
In this work, we use the {\tt GSWLC X2} catalog (\citealt{Salim2018A}), an updated version of the {\tt GSWLC 1} version~\citep{Salim2016}. The catalog includes physical properties derived for more than 650,000 galaxies covering 90\% of SDSS with the {\tt CIGALE} SED fitting code. The updated version modifies the energy balance in the SED fitting by using luminosity-dependent IR templates to derive the total IR luminosity from {\it W3} or {\tt W4} photometry to derive more accurate SFRs. For SDSS sources classified as AGN based on the BPT diagram \citep{Tremonti2004}, the IR luminosity is corrected before SED fitting (see details in \citealt{Salim2018A}). The {\tt GSWLC 2} catalog is based on the \cite{Chabrier2003} and \cite{Bruzual2003} SSP models.
For the sample of 17,902 DESI galaxies matched with the {\tt GSWLC 2} catalog, we find a good agreement between the stellar masses across the full stellar mass range with an offset of the $\rm 0.2$ dex towards lower estimates for our measurements (see Fig.~\ref{fig:GSWLCmasses}). Although both catalogs rely on the {\tt CIGALE}, the dust attenuation and energy balance are different and could explain the offset. Interestingly, the scatter of the stellar mass estimates is correlated with the quality of the WISE photometry (see lower panel in Fig.~\ref{fig:GSWLCmasses}). Sources observed with at least 2 WISE bands with {\tt S/N} > 3 are characterized by a larger spread. The investigation of this relation is left for future work on the comparison of different stellar mass estimates for DESI galaxies.

\begin{figure}
\centerline{\includegraphics[width=0.49\textwidth]{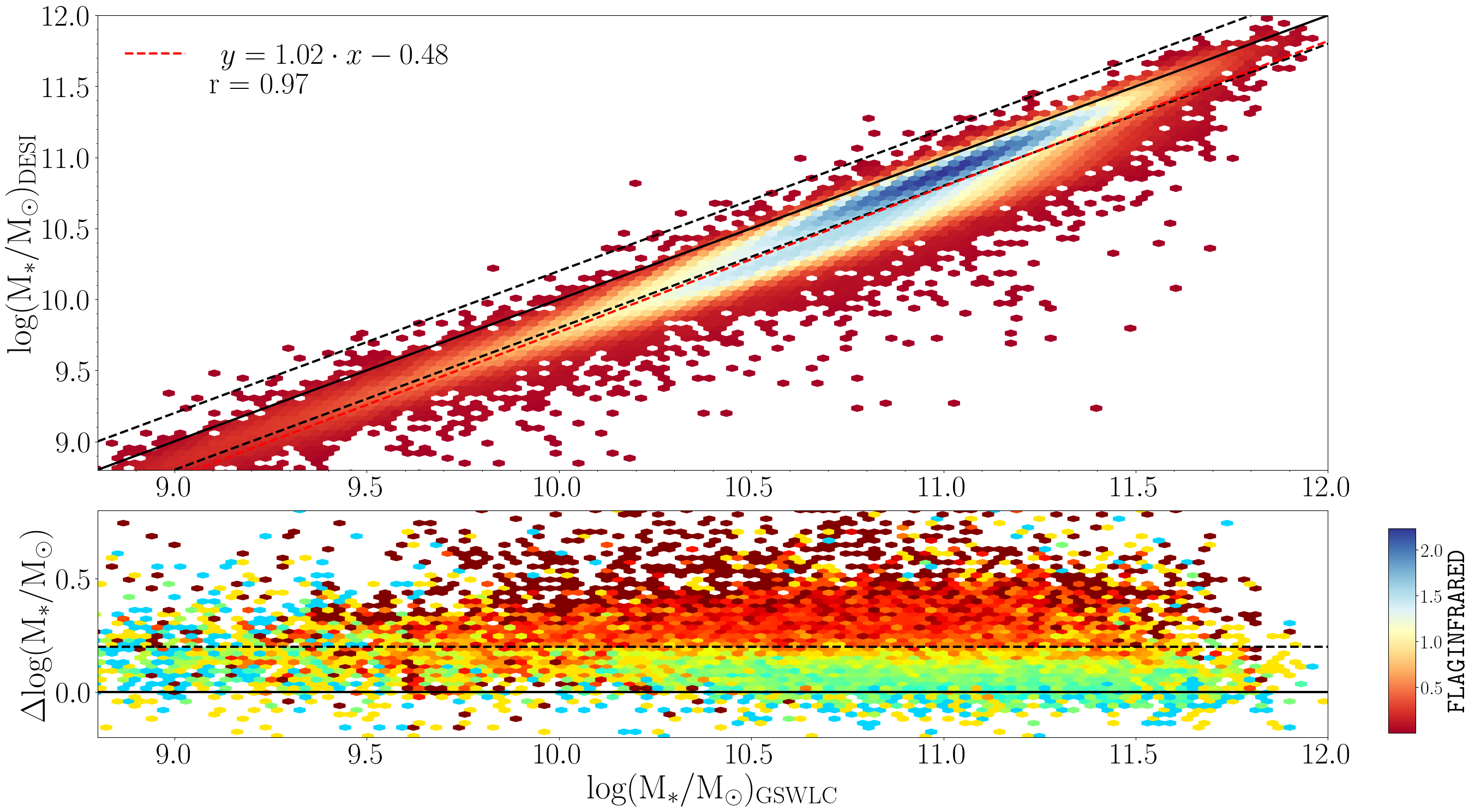}}

        \caption{Comparison of $\sim$18,000 stellar mass estimates from our VAC and {\tt GSWLC 2} catalogs (top panel) and difference in stellar mass estimates (bottom panel). The 1:1 and $\rm 0.2$ dex lines are marked with black solid and dashed lines, respectively. The linear fit (red line) and the Pearson correlation coefficient are reported in the legend. The bottom panel shows the dependence of the stellar mass residuals on the availability of  WISE photometry expressed by {\tt FLAGINFRARED}: the scatter increases when WISE photometry is readily available. }
        \label{fig:GSWLCmasses}
\end{figure} 

\subsection{DESI catalogs}\label{sec:desi_comparison}
The comparison of stellar masses across different VACs built for DESI galaxies will be discussed in future work. We restrict here the comparison to the stellar mass estimates obtained by~\cite{Zou2024}. Both VACs rely on {\tt CIGALE} estimates with the same SSP models and IMF, however, they differ in the prescription, e.g: \cite{Zou2024} leave stellar metallicity as a free parameter, while in our catalog it is fixed to a solar value and our catalog accounts for the AGN templates. We compare separately stellar masses for 125,077 star-forming and 9,852 AGN galaxies selected based on the [N II] Baldwin, Phillips, and Terlevich diagram~\citep[BPT][]{Baldwin1981}. The discrepancy between stellar mass measurements is higher for AGNs than for star-forming galaxies (median difference increases from -0.136 to -0.199 and {\tt NMAD} from 0.211 to 0.303). The comparison of the stellar mass estimates is shown in Fig.~\ref{fig:HuMassComparison}.

\begin{figure}
        \centerline{\includegraphics[width=0.49\textwidth]{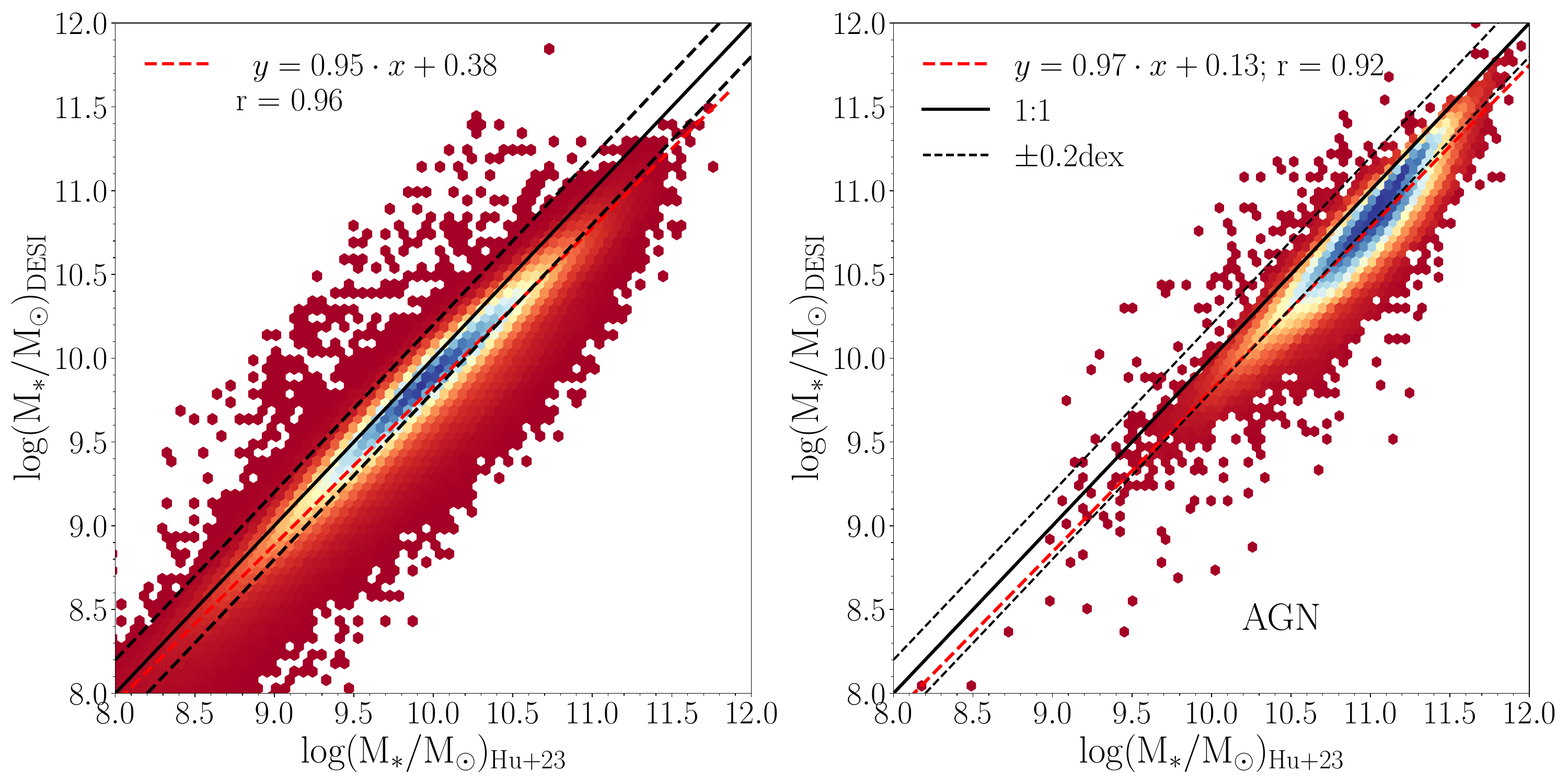}}
        \caption{Comparison of our stellar masses with the estimates from \cite{Zou2024} for   [N II] BPT-selected star-forming galaxies (left panel) and AGNs(right panel). The 1:1 and $\rm 0.2$ dex lines are marked with black solid and dashed lines, respectively. The linear fit (red line) and the Pearson correlation coefficient are reported in the legend. The discrepancies between stellar masses are larger for AGNs than for star-forming galaxies. }
        \label{fig:HuMassComparison}
\end{figure}  

\section{SED fitting modules}\label{app:SEDParametersAlternatives}
As outlined in Sect.~\ref{sec:DependenceofPhysProp}, we quantified the effect of the model assumptions for the stellar mass estimates on the representative sample of EDR galaxies described in the following App.~\ref{sec:RepresentativeSample}. The priors of the SED fitting modules we analyzed are summarized in  Table~\ref{tab:SEDParametersAlternatives}.

\begin{table*}
\centering
        \caption{Alternative parameters used in SED fitting with {\tt CIGALE}.}
        \label{tab:SEDParametersAlternatives}
        \footnotesize
        \begin{tabular}{r r r}
\hline
\hline
Parameter & Symbol &  Values\\
\hline
\multicolumn{3}{c}{Stellar population models:  M05 (\citealt{Maraston2005})}\\
\hline \hline
Initial mass function & IMF & \cite{Salpeter1955}\\
Metallicity & $\rm Z$ & 0.0004, 0.004, 0.008, 0.02, 0.05\\
\hline
\hline
\multicolumn{3}{c}{SFH: non-parametric SFH}\\
\hline \hline
Age of the oldest stars in the galaxy (Gyr) & age & 0.1, 0.5, 1, 3, 4.5, 6, 8, 10, 13 \\
Seeds of different SFH  & $\rm N\_SFH$ &  1000 \\
\hline
\hline
\multicolumn{3}{c}{Dust attenuation: \cite{CharlotFall2000}}\\
\hline \hline
V-band attenuation in the interstellar medium & $\rm AV\_ISM$ & 0.01, 0.1, 0.4, 0.7, 1.1,1.5, 2,2.5, 3, 3.5, 5\\
\hline
\hline
\multicolumn{3}{c}{Dust emission: \cite{Dale2014}}\\
\hline \hline
AGN fraction & $\rm fracAGN$ &  0\\
Alpha slope & $\rm \alpha$ & 0.0625, 0.6875, 1.2500, 2.0, 3.0, 4.0\\
\hline
\multicolumn{3}{c}{AGN: \cite{Stalevski2012, Stalevski2016}}\\
\hline \hline
inclination, i.e., viewing angle & AGNPSY [deg] & 0, 10, 20, 40, 70, 90 \\
Contribution of the AGN to the total LIR & AGNFRAC &  0, 0.01, 0.1, 0.3, 0.5, 0.7, 0.9 \\
\end{tabular}
\end{table*}   
\subsection{Representative sample selection}\label{sec:RepresentativeSample}
To analyze the influence of the assumptions made to generate models and the incorporation of the MIR photometry to the SED fitting on the main physical properties of DESI galaxies (stellar masses and SFRs), we select a representative sample of 50,196 sources. 
The sample consists of seven different galaxy types: BL AGNs, NL AGNs, composite objects, star-forming galaxies, passive galaxies, retired galaxies, and others.  
To select the representative sample we restrict the sample of 1,286,124 EDR galaxies (see Sect.~\ref{sec:data}) to 512,002 sources ({\tt GALAXIES} and {\tt QSOs}) spanning out to redshift $\rm z=0.5$ where H$\alpha$ is visible in DESI spectra. 
The galaxy selection is based on the emission line measurements coming from {\tt FastSpecFit v2.0} (\citealt{Moustakas2023}; Moustakas in prep.). 
The selection of 1,750 BL AGNs is based on the H$\alpha$ line requiring that the amplitude of the narrow and broad components have high S/N ($\tt S/N_{HALPHA\_BROAD\_AMP}$ $\rm \ge 3$,  $\tt S/N_{HALPHA\_NARROW\_AMP}$ $\rm \ge 3$) and that the 2-component Gaussian fit is favored over a single fit by $\rm \chi^2$ ($\tt \Delta\chi^{2} \ge 0$). From the remaining set, NL AGNs, composite, and star-forming galaxies are identified based on the  [N II] BPT diagram requiring high {\tt S/N} for the amplitudes of the lines used in the  [N II] BPT  diagram ($\tt S/N_{HALPHA\_AMP}$ $\rm \ge 3$, $\tt S/N_{HBETA\_AMP}$ $\rm \ge 3$, $\tt S/N_{[OIII]\lambda5007\_AMP}$ $\rm \ge 3$). Following the selection proposed by \cite{Kewley2001} and \cite{Kauffmann2003}, we select 3,900 NL AGNs, 8,393 composite, and 101,757 star-forming galaxies. Among the remaining set of galaxies (396,202) we select 47,797 passive galaxies following the criterion based on the equivalent width of the H$\alpha$ line:
\begin{equation}
    0 < \tt HALPHA\_EW < 0.5 \AA,
\end{equation}
and 63,486 retired galaxies based on the cut:
\begin{equation}
    0.5 \le \tt HALPHA\_EW \le 5 \AA.
\end{equation}
Sources that did not match any of the above criteria (284,919) are kept as others. Among all the galaxy classes, except for AGNs and composite, we randomly select sources in different bins of absolute z magnitude ($\tt ABSMAG\_SDSS\_Z$)-redshift ({\tt Z}) space.  
The summary of the composition of the representative sample is given in Table~\ref{tab:RepresentativeSample}. 

\begin{table}
    \centering
    \caption{Composition of the representative DESI sample drawn from 1,286,124 EDR galaxies.}
    \label{tab:RepresentativeSample}
    \begin{tabular}{|c|c|c|}
    \hline
    \hline
    galaxy class  &  N &  criterion \\
    \hline
    BL AGN   &    1,750   &         $\tt S/N_{H\alpha_{broad}}$ $\ge$ 3 \& $\tt S/N_{H\alpha_{narrow}}$ $\ge$ 3 \\
    NL AGN   &    3,900   &     [N II] BPT  diagram \\
    composite   &    8,393   &     [N II] BPT  diagram \\
    star-forming   &    8,819   &     [N II] BPT  diagram \\
    passive   &    8,526   &   $0 < \rm EW\_H\alpha < 0.5$ \AA \\
    retired   &    8,846   &   $0.5 \le \rm EW\_H\alpha \le  3$ \AA \\
    other   &    9,962   &  - \\
    \hline
    all sources & 50,196 & \\
    \hline
    \end{tabular}
\end{table}

\subsection{The reliability of the fit}\label{sec:ReliabilityofFit}
The goodness of the fit is quantified by the reduced $\chi^2$ of the best-fitting model ($\chi^2_r$). 
In principle, one would expect a good fit with $\chi^2_r = 1$, as $\chi^2_r > 1$ indicates issues in fully capturing the data, and $\chi^2_r < 1$  points to overfitting.
However, since the models are highly non-linear, it is almost impossible to correctly quantify the number of degrees of freedom for each galaxy (\citealt{Andrae2010, Chevallard2016}), and thus $\chi^2_r$ in {\tt CIGALE} corresponds to the $\chi^2$ divided by the total number of input fluxes (instead of number of freedom in its statistical definition).
Therefore, we follow the commonly used approach of removing the bad fits by sigma-clipping the $\chireduced$ (e.g., \citealt{Malek2018}) and keep only those galaxies for which $\chireduced \le 17$, which corresponds to $\sim 95\%$ of the sample. This threshold is also dictated by our visual inspection of the quality of the SED fitting, where we find a degradation of the fitting of the MIR part of the spectrum. We note that in the literature more restrictive cuts ($\chireduced \le 5$) are adapted for the scientific analysis and we encourage the user to use the one more appropriate for their scientific case (e.g., \citealt{Buat2021}). 

\end{appendix}

\end{document}